\documentclass[aps,prx,onecolumn,superscriptaddress,nofootinbib,floatfix,longbibliography]{revtex4-2}
\usepackage[utf8]{inputenc}
\usepackage{amsmath,amssymb}
\usepackage{amsthm}
\usepackage{bm}
\usepackage{mathtools}
\usepackage{graphicx,float}
\usepackage{algorithm}
\usepackage{algpseudocode}
\usepackage{subcaption}
\usepackage{booktabs}
\usepackage{multirow}
\usepackage{enumitem}
\usepackage{xcolor}
\usepackage{float}
\usepackage[normalem]{ulem}
\usepackage{url}
\usepackage{tikz}
\usetikzlibrary{arrows.meta,positioning,fit,backgrounds}
\usepackage{geometry}
\usepackage{hyperref}

\hypersetup{colorlinks=true, linkcolor=blue!50!black, urlcolor=blue!50!black, citecolor=blue!50!black}

\newif\ifshownotes
\shownotestrue

\DeclareMathOperator{\Var}{Var}
\DeclareMathOperator{\Cov}{Cov}
\DeclareMathOperator{\tr}{tr}
\DeclareMathOperator{\diag}{diag}
\DeclareMathOperator{\sgn}{sgn}
\DeclareMathOperator{\XOR}{XOR}

\newcommand{\opnorm}[1]{\lVert #1 \rVert_{\mathrm{op}}}
\providecommand{\E}{}\renewcommand{\E}{\mathbb{E}} \providecommand{\PP}{}\renewcommand{\PP}{\mathbb{P}} \providecommand{\R}{}\renewcommand{\R}{\mathbb{R}}
\newcommand{\F}{\mathbb{F}}
\newcommand{\SNR}{\mathrm{SNR}}

\newcommand{\Wcal}{\mathcal{W}}
\newcommand{\Scal}{\mathcal{S}}
\newcommand{\Hcal}{\mathcal{H}}
\newcommand{\Ccal}{\mathcal{C}}
\newcommand{\Kcal}{\mathcal{K}}
\newcommand{\Ical}{\mathcal{I}}

\newcommand{\norm}[1]{\left\lVert #1 \right\rVert}

\theoremstyle{plain}
\newtheorem{theorem}{Theorem}
\newtheorem{proposition}[theorem]{Proposition}
\newtheorem{lemma}[theorem]{Lemma}
\newtheorem{corollary}[theorem]{Corollary}
\theoremstyle{definition}
\newtheorem{definition}[theorem]{Definition}
\newtheorem{remark}[theorem]{Remark}

\newcommand{\figpath}{Figures/}
\newlength{\resultfigwd}
\newcommand{\resultfig}[2][\linewidth]{%
  \setlength{\resultfigwd}{#1}%
  \IfFileExists{\figpath#2}%
    {\includegraphics[width=\resultfigwd]{\figpath#2}}%
    {\fbox{\parbox[c][0.40\resultfigwd][c]{0.90\resultfigwd}{\centering \resizebox{0.86\linewidth}{!}{\ttfamily\detokenize{Figures/#2}}\\[4pt]
       \normalfont\itshape\footnotesize placeholder: figure not present in this
       repository snapshot}}}%
}

\begin{document}
\title{Hypergraph cycle enumeration for efficient inference of planted $k$-XOR with~smooth trade-off}

\author{Arman Babakhani}
\affiliation{Global Technology Applied Research, JPMorganChase, New York, NY 10017, USA}
\author{Enrico Fontana}
\affiliation{Global Technology Applied Research, JPMorganChase, New York, NY 10017, USA}
\author{Junhyung~Lyle~Kim}
\affiliation{Global Technology Applied Research, JPMorganChase, New York, NY 10017, USA}
\author{Shouvanik Chakrabarti}
\affiliation{Global Technology Applied Research, JPMorganChase, New York, NY 10017, USA}
\author{Ruslan Shaydulin}
\affiliation{Global Technology Applied Research, JPMorganChase, New York, NY 10017, USA}

\begin{abstract}
We introduce \emph{hypergraph cycle enumeration} (HCE), a simple spectral framework for detecting and recovering planted signals in noisy hypergraph inference problems. We study HCE on one representative problem from this class, namely planted noisy $k$-XOR, also known as sparse Learning Parity with Noise (LPN). Both theoretically and empirically, HCE achieves comparable recovery behavior to Kikuchi method, matching Kikuchi hierarchy. We identify parameter regimes where HCE outperforms Kikuchi method, showing that neither algorithm is universally better. We also show a subquadratic quantum speedup for HCE by replacing one of the algorithm's steps with a quantum walk. Our classical and quantum algorithms do not refute the quartic quantum speedup of quantum Kikuchi method, which remains valid for the practically relevant regime of large size and high noise strength.
\end{abstract}

\maketitle

\section{Introduction}

Many problems in high-dimensional statistics and average-case complexity ask the same basic question: when is a weak, noisy signal hidden in random data detectable, and when can it be recovered \emph{efficiently}? In many settings there is a striking separation between what is possible with unlimited computation and what is possible with known polynomial-time algorithms. Understanding such informational--computational gaps has become a central goal in the study of random inference problems.~\cite{Hopkins2018,KWB2019} 

A particularly clean setting in which these questions arise is that of noisy parity measurements, such as Learning Parity with Noise (LPN)~\cite{BKW2003}. In the $\mathbb{F}_2^n$ setting, suppose an unknown vector $x$ is observed only through a set $\{s_i \cdot x_i+n_i\}$ with each $s_i$ revealing a parity of a set of coordinates, and each measurement being corrupted by noise $n_i$. Natural questions are how many such measurements are needed to identify the hidden vector, and how many are needed for an \emph{efficient} algorithm? In some natural regimes, this task appears to be computationally difficult, and is indeed a canonical model for the algorithmic limits of extracting a planted discrete signal from sparse, noisy, high-order data.~\cite{schmidhuber2025quartic,mao2026near}

Here we examine the planted noisy $k$-XOR problem, which is also known as sparse LPN, since we require the observations $s_i$ to be Hamming weight $k$.~\cite{schmidhuber2025quartic,gupta2026classical} This problem is closely related to tensor PCA \cite{HSS2015,HSS2019,RM2014} and to the refutation of random sparse CSPs \cite{AOW2015,Feige2002,GK2001,RRS2017}. It is more conveniently framed in terms of signed vectors with entries in $\{\pm 1\}$ instead of vectors in $\mathbb{F}^2$. A $k$-XOR instance $\Ical$ consists of clauses $(S_a,b_a)$, $a=1,\dots,m$, where each support $S_a\subseteq[n]$ has size $k$ and each label $b_a\in\{\pm1\}$. Given an assignment $z\in\{\pm1\}^n$, the clause is satisfied when $z_{S_a}:=\prod_{v\in S_a}z_v=b_a$. In the planted noisy model there is a hidden $x\in\{\pm1\}^n$ such that
\begin{equation}\label{eq:model}
   b_a \;=\; \eta_a\prod_{v\in S_a}x_v,\qquad a=1,\dots,m,
\end{equation}
where the supports $S_a$ are i.i.d.\ uniform $k$-subsets of $[n]$ and the noise variables $\eta_a\in\{\pm1\}$ are independent with $\E[\eta_a]=\rho$. The \emph{detection} task is to decide whether the data has a planted solution, i.e., to distinguish $\rho=0$ null case from planted case with some $\rho>0$. The \emph{recovery} task is to reconstruct $x$. The noiseless case $\rho=1$ is solved by Gaussian elimination whenever $m\gtrsim n$. Information-theoretically, weak recovery (non-trivial correlation with $x$) is possible already at $m=\Theta(n)$ clauses for any constant $\rho\in(0,1]$, while exact recovery requires every variable to appear in at least one clause, necessitating $m=\Omega(n\log n)$ clauses~\cite{HS2026}. No polynomial-time algorithm is known below $m=\widetilde\Theta(n^{k/2})$ for even $k\ge4$, and this threshold is conjectured optimal for efficient algorithms \cite{AOW2015,RRS2017,KMOW2017}. The noisy case is conjectured to exhibit a sharp statistical--computational gap \cite{AOW2015,RRS2017,KMOW2017,WAM2019}.

Historically this and related problems have been approached with methods such as tensor unfolding~\cite{RM2014} and the sum-of-squares (SoS) hierarchy~\cite{Hopkins2018,HSS2019,HKPRSS2017}. We review these algorithms in Section~\ref{sec:related}. The dominant family of algorithms in the computational threshold regime is the \emph{Kikuchi hierarchy} \cite{WAM2019,GKM2022,HS2026}: spectral methods indexed by a degree parameter $\ell$ that trade runtime $n^{O(\ell)}$ for a sample complexity that decreases smoothly with $\ell$. This \emph{smooth trade-off} is precisely the property that makes the Kikuchi and SoS hierarchies interesting, since it interpolates between the polynomial-time and the information-theoretic thresholds.

This work started as an attempt to answer the following question: \textit{can Gaussian elimination produce a $2$-XOR from which the planted solution of the original $k$-XOR problem can be extracted in the noisy case?} The algorithm we propose is inspired by Gaussian elimination. More formally, we introduce a practical hypergraph-cycle-enumeration hierarchy for planted noisy $k$-XOR that retains the smooth trade-off of Kikuchi-type methods, while reducing recovery at every level to spectral estimation on an $n\times n$ matrix. Our contributions can be summarized as follows:
\begin{itemize}[leftmargin=*]
    \item We introduce \textbf{hypergraph cycle enumeration} (HCE), a hierarchical algorithmic approach to planted noisy $k$-XOR (Section~\ref{sec:hce}). On level $l$, HCE aggregates sets of $l$ clauses, each giving a weak parity signal, into an $n\times n$ recovery matrix. Crucially, unlike Kikuchi, whose level-$\ell$ operator acts on a lifted space of dimension $\binom{n}{\ell}$, HCE keeps the spectral problem in the original variable space, with matrix size independent of $l$.
    \item We provide \textbf{detection} and \textbf{recovery guarantees} based respectively on collecting size-$l$ even covers, and using size-$l$ sets of clauses with symmetric difference of size 2 to generate a dense $n\times n$ matrix, the principal component of which aligns with the solution. For both detection and recovery, the parameter $l$ traces a similar threshold curve as in the Kikuchi and SoS hierarchies: increasing $l$ combines more clauses at higher computational cost. Specifically, HCE succeeds at the same threshold number of clauses per variable, $m/n=\widetilde{\Omega}(\rho^{-2}(n/l)^{k/2-1})$. Thus, the HCE hierarchy exhibits a \textbf{smooth trade-off}, joining SoS, Kikuchi and the gadget-based methods reviewed in Section~\ref{sec:related} as hierarchies with this property.
   \item We compare the \textbf{complexity} of HCE to the best competitor, the Kikuchi method. At matched even levels and $k=4$, their stated runtime bounds have the same leading exponent with the improved HCE probing rule, while HCE uses a smaller matrix representation. We run \textbf{experiments} for $k=4,5,6$ (Section~\ref{sec:experiments}) using the original all-pairs probe, showing that HCE matches Kikuchi recovery overlap at equal level while using substantially less memory and runtime, especially in the ultra-sparse regimes (e.g., at $n=50$, $k=4$, level $6$, HCE completes in ${\approx}300$\,s versus ${\approx}5.5$\,h per instance for Kikuchi, a ${\sim}60\times$ speed-up; Appendix~\ref{app:MITM}). Thus HCE presents a practical advantage, being able to solve some instances beyond the capability of Kikuchi, and the advantage is largest in the hardest (sparsest) regime.
\end{itemize}

\textit{Note:} for recovery, we use smooth trade-off in the soft-asymptotic sense common in work on the Kikuchi hierarchy, where thresholds are compared after suppressing logarithmic factors \cite{WAM2019}. Our detection result (Theorem~\ref{thm:detection}) is stronger: it has no logarithmic loss. We conjecture that a normalized or suitably trimmed HCE recovery operator removes the logarithmic slack, uniformly over a non-trivial range of growing levels $l=l(n)$; proving this remains open.

\begin{figure}[h]
  \centering
  \resultfig[0.85\linewidth]{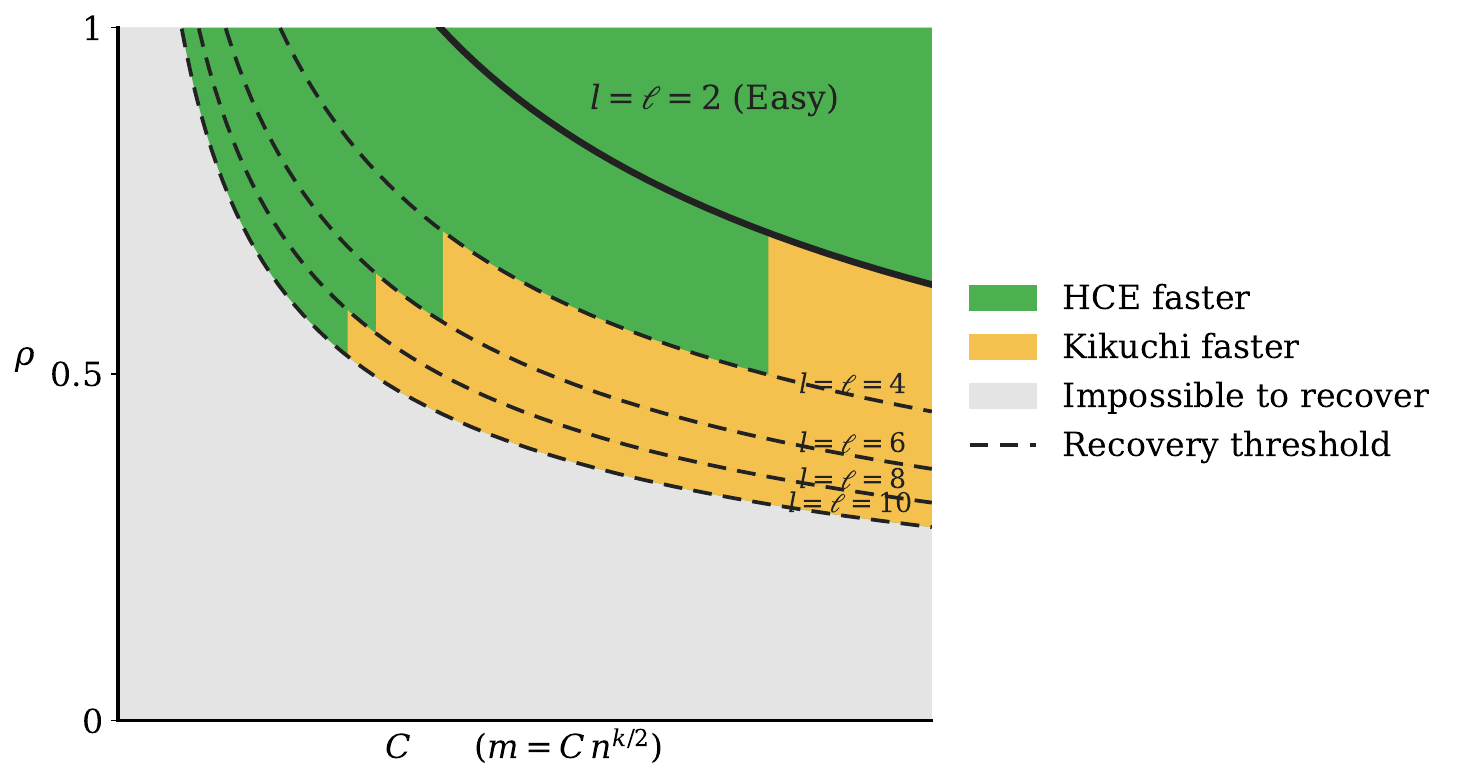}
  \caption{\textbf{Where HCE is faster than Kikuchi} for fixed large $n$ and $k=4$. Each curve marks the recovery threshold at level $l=\ell$; deeper levels allow recovery at smaller $C$ or $\rho$. Above the solid curve, level $2$ suffices (easy to recover); the gray region is beyond the levels shown. At the lowest level that permits recovery, green or yellow indicate that HCE or Kikuchi is faster, respectively. HCE generally wins when shallow levels suffice or $C$ is small. Kikuchi is faster when low $\rho$ requires deeper levels and large $C$ makes HCE's clause enumeration costly. The figure is schematic and illustrates the finite-size behavior in Section~\ref{sec:experiments}, not the asymptotic scaling in Table~\ref{tab:complexity}.
  }
  \label{fig:hce-kikuchi-cartoon}
\end{figure}

Figure~\ref{fig:hce-kikuchi-cartoon} summarizes where each method is preferable. At $m=Cn^{k/2}$ the level-$l$ threshold reduces to an $n$-independent condition $\rho^2C\gtrsim c_l$ with $c_l$ decreasing in $l$ (Section~\ref{sec:theory}), so the $(C,\rho)$ plane splits into nested regions indexed by the minimum usable level. Easy instances with large $C$ and $\rho$ are already solved at $l=\ell=2$; for weaker signals deeper levels are needed, and HCE remains preferable in the ultra-sparse regime $C\ll1$, whereas Kikuchi overtakes it at larger densities, where the cost of enumerating $l$-clause products outweighs the benefit of HCE's smaller matrix (Section~\ref{sec:experiments}).

\section{Preliminaries}
\label{sec:prelim}

\subsection{Kikuchi method}
\label{sec:kikuchi}
The Kikuchi hierarchy \cite{WAM2019,GKM2022,HKM2023} is the benchmark spectral method for planted noisy $k$-XOR, and the main baseline for HCE. For even $k$ and a level $\ell\ge k/2$, the symmetric level-$\ell$ Kikuchi matrix is indexed by $\ell$-subsets of variables,
\begin{equation}\label{eq:kikuchi}
   K_{T,U}=\sum_{a=1}^{m} b_a\,\mathbf 1[T\triangle U=S_a],
   \qquad T,U\in\binom{[n]}{\ell},
\end{equation}
so each clause links two $\ell$-sets that differ exactly by its support. Its top eigenvector correlates with the planted signal once $m\gtrsim\widetilde\Theta\bigl(n(n/\ell)^{k/2-1}\bigr)$, at cost $n^{O(\ell)}$; increasing $\ell$ gives the smooth trade-off discussed in the introduction. Odd $k$ requires an unbalanced lifting \cite{WAM2019}.

HCE is similar in spirit: both methods multiply clause labels along chains of symmetric differences, so that planted signs cancel. The difference is where the spectral step takes place. Kikuchi works on a $\binom{n}{\ell}$-dimensional lifted space, whereas HCE compresses all such chains onto an $n\times n$ matrix. The two hierarchies follow the same threshold curve (Section~\ref{sec:cycles}), so we compare them at matched level in Table~\ref{tab:complexity} and Section~\ref{sec:experiments}. In the numerical experiments, $K$ is stored sparsely, solved by Lanczos and decoded by a voting matrix. HCE uses the meet-in-the-middle construction of Section~\ref{sec:complexity} and the post-processing of Appendix~\ref{app:decoding} (details in Appendix~\ref{app:protocol}).

\subsection{Hypergraph Even covers and witnesses}
Throughout, $[n]=\{1,\dots,n\}$ and $\Kcal:=\binom{[n]}{k}$ denotes the complete $k$-uniform hypergraph, with $M:=|\Kcal|=\binom{n}{k}$ the number of possible clauses. The instance hypergraph is $\Hcal\subseteq\Kcal$ with $|\Hcal|=m$, and we parameterize the clause density by
\begin{equation}
   m = C\,n^{k/2}.
\end{equation}
For a set of clauses $\Scal\subseteq\Hcal$ we write $\triangle(\Scal):=\triangle_{S\in\Scal}S\subseteq[n]$ for the symmetric difference (equivalently, the $\F_2$ sum of indicator vectors) of their supports, and $X_\Scal:=\prod_{a\in\Scal}b_a$ for the product of their labels. A full notation table is given in Appendix~\ref{app:notation}.

Because we work with hypergraphs, the word ``cycle'' is ambiguous; we fix the following standard notion, which is the correct analogue for $\F_2$ linear algebra.

\begin{definition}[Even cover / hypergraph cycle]\label{def:evencover}
A \emph{proper even $l$-cover} is a set $F=\{S_1,\dots,S_l\}$ of $l$ \emph{distinct} hyperedges such that every vertex is contained in an even number of them, i.e.\ $\triangle(F)=\emptyset$. We write $\Ccal_l$ for the set of proper even $l$-covers contained in $\Hcal$ and $Z_l:=|\Ccal_l|$.
\end{definition}

Since $|\triangle(F)|\equiv kl \pmod 2$ for any $l$ sets of size $k$, a level $l$ can only carry even covers or $(i,j)$-witnesses (defined below) when $kl$ is even; we call such levels \emph{admissible}. For odd $k$ and odd $l$ this \emph{parity obstruction} forces $Z_l=0$ and makes the recovery matrix of Section~\ref{sec:matrix} vanish identically. Proper even covers need $l\ge3$, and we write $\pi_k$ for the smallest level carrying one: $\pi_k=3$ for even $k$ and $\pi_k=4$ for odd $k$.

An estimator achieves \emph{weak recovery} if $|\langle\hat x,x\rangle|/n=\Omega(1)$, \emph{almost-exact recovery} if this overlap tends to one, and \emph{exact recovery} if $\hat x=\pm x$. The recovery theorem below is a weak-recovery statement.

\begin{definition}[Witness]\label{def:witness}
For $i\neq j$, an \emph{$(i,j)$-witness of level $l$} in the recovery model is an index set $I\in\binom{[m]}l$ with $\triangle_{a\in I}S_a=\{i,j\}$. We write $\Wcal_l(i,j)$ for the set of such index sets, $W_l(i,j):=|\Wcal_l(i,j)|$, and $N_l:=\E\,W_l(i,j)$ (which by symmetry does not depend on $i,j$).
\end{definition}

In simple terms, a witness is an even cover ``with two loose ends''. HCE, as will be introduced below, is an algorithm that traces these ``loose ends'' in order to detect if there is a planted instances, and if so, recover it. Throughout this work, we reserve $l$ for the \emph{HCE level} and $\ell$ for the \emph{Kikuchi level}; the two are distinct parameters and are only approximately equal by coincidence at small $k$ (Section~\ref{sec:cycles}).

\section{Related Work}
\label{sec:related}
We first discuss general algorithms for planted inference and random-CSP refutation.

\paragraph{Sum of squares.} The sum-of-squares (SoS) hierarchy is a common benchmark for planted inference and random-CSP refutation. In tensor PCA, low-degree SoS relaxations and spectral algorithms derived from their moment matrices attain the conjectured polynomial-time recovery scale \citep{HSS2015,HSS2019,HKPRSS2017}. For random $k$-XOR, the spectral/SoS method of \citet{RRS2017} gives strong-refutation certificates, while \citet{KMOW2017} prove complementary SoS degree lower bounds. Kikuchi was introduced as a simpler spectral hierarchy attaining the same degree/runtime--signal trade-off as SoS \citep{WAM2019}, and Kikuchi matrices also underlie SoS certificates for random CSPs \citep{GKM2022}. The normalized analysis of \citet{HS2026} makes this trade-off log-free for planted noisy $k$-XOR detection and weak recovery while retaining degree-$O(\ell)$ SoS refutation certificates. HCE is a low-degree spectral statistic in this lineage, but is not itself an SoS relaxation.

\paragraph{Tensor unfolding and spectral baselines.} A simpler and historically important approach is \emph{tensor unfolding} (matricization): represent the observations as a sparse order-$k$ tensor $T$ supported on the observed clauses with values $b_a$, so that $\E[T]\propto x^{\otimes k}$, and partition the $k$ modes into two groups of size $k/2$ to obtain a matrix whose leading singular vectors estimate tensor powers of the signal \cite{RM2014,HSS2015,HSS2019}. Spectral separation occurs at $m\gtrsim\widetilde\Theta(n^{k/2})$, matching the polynomial-time threshold for constant-degree methods; for odd $k$ one uses an unbalanced unfolding, and related square constructions give the corresponding spectral relaxations. This baseline is simple and efficient but provides \emph{no} smooth sample--time trade-off, and it treats the data as noisy tensor entries without exploiting the algebraic cancellation structure of XOR constraints, exactly the structure that HCE uses.

\paragraph{Trace methods, even covers and low-degree polynomials.} The matrix $A^{(l)}$ is closely related to the $l$-th step of the trace method for refuting random CSPs \cite{AOW2015,RRS2017}, whose spectral antecedents go back to \citet{GK2001} and \citet{FGK2005}, and to the ``square'' spectral construction of \citet{HSS2019} for tensor PCA. The witness counts of Appendix~\ref{app:N_ell} are an explicit version of the $\F_2$-cycle counts of \citet{AOW2015} and \citet{RRS2017}; even-cover counting also underlies the hypergraph Moore bound \cite{HKM2023} and Feige-type refutation \cite{Feige2002,GKM2022}. Closest to our work, \citet{HS2026} show that the Kikuchi hierarchy is \emph{sharp} for random $k$-XOR, without polylogarithmic losses. Accordingly, our novelty claim is not a new statistical regime, a proved equivalence with one Kikuchi slice, or uniform runtime superiority; it is the exact final-boundary compression, its cycle interpretation, the $n\times n$ spectral endpoint at every level, the explicit witness-count constants, and the empirical memory frontier.

\medskip
We now discuss algorithms that are directly related to or share ideas with our approach.

\paragraph{Classical speedup by finding short cycles.~\cite{gupta2026classical}}
Gupta \textit{et al.} gave a classical detection algorithm for planted $k$-XOR that runs in roughly the square root of the time of the Kikuchi power method when $k$ is a large constant, reducing the quartic quantum speedup of \citet{schmidhuber2025quartic} to a quadratic one in that regime~\cite{gupta2026classical}. Like HCE, their algorithm finds even covers by meet-in-the-middle: a closed walk in the Kikuchi graph yields an even cover, and by the birthday paradox two short random walks from a common start soon meet. HCE instead finds every cover of a given size rather than a random sample. This lets us compute the mean and variance of $X(l)$ exactly and obtain the explicit constant $\gamma_k$ of Theorem~\ref{thm:detection}, whereas their condition keeps the logarithmic factor of the Kikuchi method. Their result covers detection only, while HCE also gives recovery.

\paragraph{Sub-hypergraph enumeration.~\cite{li2025smooth, mao2026near}}
Where Gupta \textit{et al.} sample covers at random from the Kikuchi matrix, \citet{li2025smooth} and \citet{mao2026near} sample from a subset of covers with a specific shape. Li proved a smooth trade-off between running time and signal strength, with no logarithmic loss, for Gaussian tensor PCA, and Mao adapted the method to planted noisy $k$-XOR. Both sum products of labels over rings of small \emph{gadgets} glued end to end for detection, and over open chains of gadgets between two variables $i$ and $j$ for recovery. A ring is an even cover in which every variable appears exactly twice, and a chain is an $(i,j)$-witness with the same property, so their statistics are restricted parts of our sums $X(l)$ and $A^{(l)}$. Notably however they use a different method for recovery, which may explain why they have no $\log(n)$ factor. We leave open whether HCE recovery can avoid this logarithmic loss. Mao also obtains exact recovery and a matching low-degree lower bound.

HCE, the works by Gupta \textit{et al.}~\cite{gupta2026classical} and Mao~\cite{mao2026near}, and implicitly the Kikuchi method~\cite{WAM2019}, all rest on the same object: an even cover, that is, a set of clauses whose variables cancel in pairs, so that the product of its labels is correlated with the planted assignment. They differ in which covers they collect and how: Gupta \textit{et al.} sample covers through random-walk collisions, Li and Mao count covers of one restricted shape, and HCE enumerates all covers of a given size. These choices determine the threshold and the running time and explain the different thresholds achieved. The same view applies to the Kikuchi method. The trace of a power of the Kikuchi matrix is a sum over closed walks in the Kikuchi graph, and each closed walk yields an even cover, so the Kikuchi method can also be read as a particular weighting of even covers. Comparing these weightings is a natural way to understand where the thresholds of spectral methods come from.

\section{Hypergraph Cycle Enumeration}
\label{sec:hce}
\begin{figure}[t]
\centering
\captionsetup{font=small,skip=5pt}
\begin{minipage}[c]{0.615\linewidth}
\centering
\resizebox{\linewidth}{!}{%
\begin{tikzpicture}[
    font=\footnotesize, >=Latex,
    box/.style={draw, rounded corners, align=center, inner sep=4pt,
                minimum height=0.9cm, fill=white},
    inputbox/.style ={box, fill=blue!6,   draw=blue!50!black,   text width=4.0cm},
    stepbox/.style  ={box, fill=orange!8, draw=orange!60!black, text width=4.0cm},
    outbox/.style   ={box, fill=purple!6, draw=purple!55!black, text width=4.0cm},
    arrow/.style={->, thick}
]

\node[inputbox] (data) at (3.0,0) {%
\textbf{Noisy $k$-XOR clauses}\\[2pt]
$(S_1,b_1),\dots,(S_m,b_m)$\\[2pt]
$b_a=\eta_a\prod_{v\in S_a}x_v$\\[2pt]
{\scriptsize sparse noisy entries of $x^{\otimes k}$}
};

\node[stepbox] (enum) at (9.0,0) {%
\textbf{1. Enumerate level-$l$ witnesses}\\[2pt]
$\Scal\subseteq[m]$, $|\Scal|=l$, with
$\triangle_{a\in\Scal}S_a=\{i,j\}$
};

\node[stepbox] (mult) at (9.0,-2.15) {%
\textbf{2. Multiply the $l$ labels}\\[2pt]
$\prod_{a\in\Scal}b_a=x_ix_j\prod_{a\in\Scal}\eta_a$
};

\node[stepbox] (acc) at (3.0,-2.15) {%
\textbf{3. Accumulate}\\[2pt]
$A^{(l)}_{ij}\mathrel{+}=\prod_{a\in\Scal}b_a$
};

\node[outbox] (spec) at (3.0,-4.3) {%
\textbf{4. Spectral recovery}\\[2pt]
$A^{(l)}\approx\rho^{l}N_l\,xx^\top+\text{noise}$\\[2pt]
$\hat x=\sgn(\text{top eigenvector})$
};

\begin{scope}[shift={(8.3,-3.6)}]
    \def\s{0.28}
    \foreach \r in {0,...,4}{\foreach \c in {0,...,4}{
        \draw[fill=purple!10, draw=purple!45!black, thin]
            ({\c*\s},{-\r*\s}) rectangle ++(\s,-\s);}}
    \foreach \d in {0,...,4}{
        \fill[purple!45!black, opacity=0.55]
            ({\d*\s},{-\d*\s}) rectangle ++(\s,-\s);}
    \draw[draw=purple!55!black, thick] (0,0) rectangle ({5*\s},{-5*\s});
    \node[font=\scriptsize] at ({5*\s/2},{0.25}) {$n$};
    \node[font=\scriptsize, rotate=90] at ({-0.25},{-5*\s/2}) {$n$};
    \node[align=center, font=\scriptsize, text width=3.4cm] at ({5*\s/2},{-5*\s-0.42})
        {$A^{(l)}$: rank one $+$ noise;\\ size independent of $l$};
\end{scope}

\draw[arrow] (data.east) -- (enum.west);
\draw[arrow] (enum.south) -- node[right, font=\scriptsize] {if collapse} (mult.north);
\draw[arrow] (mult.west)  -- (acc.east);
\draw[arrow] (acc.south)  -- (spec.north);
\draw[arrow] (spec.east)  -- (7.9,-4.3);
\end{tikzpicture}%
}
\end{minipage}\hfill
\begin{minipage}[c]{0.365\linewidth}
\begin{algorithm}[H]
\caption{HCE at level $l$}
\label{alg:hce}
\scriptsize
\begin{algorithmic}[1]
   \Require Clauses $\{(S_a,b_a)\}_{a=1}^{m}$ with $|S_a|=k$; level $l$ with $kl$ even.
   \Ensure $\hat x\in\{\pm1\}^{n}$, estimating $x$ up to a global sign.
   \State $r\gets\lfloor l/2\rfloor$, $s\gets\lceil l/2\rceil$, $G^{(l)}\gets0$.
   \State Hash the $s$-sets and $r$-sets by boundary, storing aggregate label products; reuse the table when $r=s$.
   \State Probe same-size partners by replacing one element in each $r$-boundary.
   \State Probe size-difference-two partners by deleting two elements from keys in either table.
   \State Accumulate each match in $G^{(l)}$ with its multiplicity.
   \State Recover $A^{(l)}$ from $G^{(l)}$ and the lower-level matrices using~\eqref{eq:mitm-recurrence}.
   \If{$k$ is odd and $l\ge6$}
      \State $M\gets\bar A^{(l)}$ as in Definition~\ref{def:trim}.
   \Else
      \State $M\gets A^{(l)}$.
   \EndIf
   \State $\hat u\gets$ a top eigenvector of $M$.
   \State \Return $\hat x=\sgn(\hat u)$.
\end{algorithmic}
\end{algorithm}
\end{minipage}
\caption{\textbf{Overview of hypergraph cycle enumeration (HCE) algorithm}. \emph{Left:} the clauses are sparse noisy entries of the rank-one order-$k$ tensor $x^{\otimes k}$; HCE multiplies $l$ clauses whose supports XOR to a pair $\{i,j\}$, so that every variable cancels modulo two except $i$ and $j$, leaving an effective noisy $2$-XOR constraint on $x_ix_j$. The case $l=2$ is elementary: two clauses with $S_a\triangle S_b=\{i,j\}$ share a common $(k-1)$-set. \emph{Right:} Algorithm~\ref{alg:hce}, the meet-in-the-middle implementation, analyzed in Section~\ref{sec:complexity}.}
\label{fig:hce-schematic}
\end{figure}

\subsection{Cycles carry the signal}
\label{sec:cycles}

The core idea is to use the clause structure of the $k$-XOR instance directly. Recall that $\Ccal_l$ denotes the set of length-$l$ even covers, with $Z_l = |\Ccal_l|$. Consider an even cover $F\in\Ccal_l$. Every variable in $F$ appears an even number of times, so all planted signs cancel and
\begin{equation}
   X_F \;=\; \prod_{S\in F}b_S \;=\; \prod_{S\in F}\eta_S,
   \qquad \E[X_F]=\rho^{l},\qquad \Var(X_F)=1-\rho^{2l}.
\end{equation}
Under the null model ($\rho=0$) the mean vanishes, so even covers are exactly the local structures that certify planting. Collecting enough of them therefore enables detection; hence the name \emph{hypergraph cycle enumeration}. Define the \emph{level-$l$ discriminator}
\begin{equation}\label{eq:discriminator}
   X(l) \;:=\;
   \begin{cases}
      Z_l^{-1}\sum_{F\in\Ccal_l}X_F, & Z_l>0,\\
      0, & Z_l=0.
   \end{cases}
\end{equation}
If the summands were uncorrelated --- as they are for edge-disjoint covers, and asymptotically after the overlap terms are controlled through Proposition~\ref{prop:covariance} --- then detection with constant probability requires
\begin{equation}
   \frac{(\E X(l))^2}{\Var X(l)}\;\approx\;\rho^{2l}Z_l\;\gg\;1
   \iff Z_l=\Omega(\rho^{-2l}).
\end{equation}
For random uniform hypergraphs $Z_l$ concentrates around its expectation (Corollary~\ref{cor:concentration}), whose asymptotics follow from an exact Kravchuk-polynomial count of even covers in the complete hypergraph (Appendix~\ref{app:complete}) combined with the sampling step (Subsection~\ref{app:expected}, Theorem~\ref{thm:EZell}). Writing $Z^{\Kcal}_l=(\alpha_{k,l}+o_n(1))\,n^{kl/2}$ for the count on the complete hypergraph, sampling gives $\mu_l:=\E[Z_l]=(\alpha_{k,l}+o_n(1))(k!\,C)^{l}$, and for large admissible $l$
\begin{equation}\label{eq:EZasymp}
   \E[Z_l]=\bigl(L_k+o_l(1)\bigr)\,l^{-1/2}
   \left(\gamma_k^{-1}\,\frac{m\,l^{k/2-1}}{n^{k/2}}\right)^{l},
   \quad
   \gamma_k:=k^{-k/2}e^{k/2-1},
   \quad L_k:=\tfrac{e^{-(k-1)/2}}{\sqrt\pi}.
\end{equation}
Detection is therefore possible for large $n$, constant $k$ and large admissible $l$ whenever
\begin{equation}\label{eq:detthreshold}
   m\;\gg\;\gamma_k\,\rho^{-2}\,n\left(\frac{n}{l}\right)^{k/2-1}.
\end{equation}
Up to logarithmic factors, the Kikuchi detection threshold \cite{WAM2019,GKM2022,HKM2023} is the same condition with $\gamma_k$ replaced by $\binom{k}{k/2}^{-1}$ and $l$ replaced by the Kikuchi level $\ell$. Thus HCE exhibits the same smooth trade-off, and at $k=4$ the two constants agree to within $2\%$.

\subsection{From cycles to a matrix: the recovery matrix \texorpdfstring{$A^{(l)}$}{A(l)}}
\label{sec:matrix}

Recovery proceeds by reducing the problem to standard (matrix) PCA. Here the clauses are the i.i.d.\ sample $S_1,\dots,S_m$, and witnesses are indexed by clause indices: for $I\subseteq[m]$ put $\partial I:=\triangle_{a\in I}S_a$ and define the \emph{recovery matrix}
\begin{equation}\label{eq:Adef}
   A^{(l)}_{ij}\;:=\;\sum_{\substack{I\in\binom{[m]}{l}\\ \partial I=\{i,j\}}}
   \ \prod_{a\in I}b_a \quad(i\ne j),
   \qquad A^{(l)}_{ii}:=0 .
\end{equation}
Every $I$ in the sum has the property that all variables cancel modulo two except $i$ and $j$. If $x$ is the planted assignment, $\prod_{a\in I}b_a = x_ix_j\prod_{a\in I}\eta_a$ by problem definition. Conditionally on the supports,
\begin{equation}\label{eq:condmean}
   \E\bigl[A^{(l)}_{ij}\;\big|\;S_1,\dots,S_m\bigr]
   \;=\; \rho^{l}\,x_ix_j\,W_l(i,j),
\end{equation}
where $W_l(i,j)$ is the number of $(i,j)$-witnesses of level $l$. Hence $A^{(l)}$ is a weighted noisy estimator of the rank-one matrix $xx^\top$. 
Recall that $N_l$ is the expected number of witnesses for a fixed pair
$i\ne j$. In the i.i.d.\ support model, where clauses are drawn with independent equal probabilities
\begin{equation}\label{eq:normalization}
   N_l=\E[W_l(i,j)]
   =\binom ml\,\PP\bigl[\partial\{1,\dots,l\}=\{i,j\}\bigr].
\end{equation}
Averaging~\eqref{eq:condmean} also over the supports gives
$\E[A^{(l)}]=\rho^lN_l(xx^\top-I)$. We therefore write
\begin{equation}\label{eq:decomposition}
   A^{(l)}=\rho^lN_l\,xx^\top+R^{(l)},
\end{equation}
where the residual $R^{(l)}$ includes both the random fluctuations
and the deterministic diagonal correction.

Since $xx^\top x=nx$, the rank-one signal has eigenvector $x$
and eigenvalue $\rho^l nN_l$. Therefore, for recovery it suffices to show that $\opnorm{R^{(l)}}$ is suitably
smaller than $\rho^l nN_l$, so that the top eigenvector of $A^{(l)}$
remains close to the planted direction. HCE estimates $x$, up to
a global sign, by reading off the signs of this eigenvector;
see Figure~\ref{fig:hce-schematic}. Comparing this condition across
levels $l$ gives the recovery trade-off.

Recovery is, thus, a signal--noise comparison: it suffices that $\opnorm{R^{(l)}}$ be suitably smaller than $\theta_l$. Comparing this condition across fixed levels gives the formal trade-off curve. HCE estimates $x$ from the sign pattern of the top eigenvector of $A^{(l)}$; see Figure~\ref{fig:hce-schematic}.

\section{Theoretical Guarantees}
\label{sec:theory}

Theorems \ref{thm:detection} and \ref{thm:recovery} lay the theoretical foundation of detection and recovery via HCE. These are the two main results that we will state in this section, and defer complete proofs to Appendices~\ref{app:detection} and~\ref{app:recovery}.

The literature on planted inference problems uses a number of different models for how the clauses are sampled. For convenience and clarity of derivation, we consider two models. For \emph{detection}, $\Hcal$ is a uniformly random $m$-edge subhypergraph of $\Kcal$: the supports are drawn uniformly \emph{without replacement}, so $\Hcal$ is a simple hypergraph and even covers are sets of distinct edges. This model was used in Ref.~\cite{gupta2026classical}. For \emph{recovery}, $S_1,\dots,S_m$ are \emph{i.i.d.}\ uniform $k$-subsets, so repeated supports may occur and witnesses are indexed by clause indices rather than by edges; repeated supports are retained rather than discarded. This model or its Poissonization were used in Refs.~\cite{WAM2019,fontana2025end,schmidhuber2025quartic,HS2026}. 
In practice the two models can be seen to be equivalent up to constants factors. Given a Poissonized instance with $\hat m$ variables, $\E [\hat m] = m$, with high probability $\hat m = (1+o(1))m$ and so asymptotically the regimes are equivalent. The Poissonized instance may have repeated clauses; in that case one can deduplicate the clauses, again with a subleading loss in the regime of interest $m = \Theta(n^{k/2})$.

\subsection{Detection}
The heuristic of Section~\ref{sec:cycles} used two unproven simplifications: that $Z_l$ concentrates around $\mu_l=\E_\Hcal[Z_l]$, and that the variance of $X(l)$ is dominated by edge-disjoint pairs of covers. Both are controlled by a single combinatorial estimate. Let $n_s$ be the number of ordered pairs of even $l$-covers of $\Hcal$ sharing exactly $s$ hyperedges. Lemma~\ref{lem:cover-intersect-bound} computes the corresponding count on the complete hypergraph \emph{with its constant}, and after sampling this yields
\begin{equation}\label{eq:momentratio}
   \frac{\Var_\Hcal[Z_l]}{\E_\Hcal[Z_l]^2}
   =\sum_{s=1}^{l}\frac{\mu_s\mu_{l-s}^{2}}{\mu_l^{2}}+o_n(1)
   =\Theta_{k,C}\bigl(l^{-(k-2)\pi_k}\bigr)+o_n(1),
   \qquad \mu_q:=\alpha_{k,q}\,(C\,k!)^{q},
\end{equation}
with $\pi_k$ as in Section~\ref{sec:prelim}. The rate $l^{-(k-2)\pi_k}$ is the mechanism behind the trade-off: at fixed density, deeper levels make overlapping covers relatively rarer, and the discriminator behaves like an average of independent $\pm1$ variables of mean $\rho^{l}$. This makes the error probabilities explicit.

\begin{theorem}[Smooth trade-off for detection]\label{thm:detection}
Fix $k>2$, $\rho\in(0,1]$ and a target error probability $\delta\in(0,1)$. There is a constant $c_k>0$, depending only on $k$, such that the following holds for every fixed admissible $l\ge c_k\delta^{-2/((k-2)\pi_k)}$ and every fixed $C>0$. Set $m=Cn^{k/2}$ (assumed integral). Then, for all sufficiently large $n$, the test that declares the instance planted when $Z_l>0$ and $X(l)>\rho^{l}/2$ has false-positive and false-negative probabilities each at most $\delta$ provided
\begin{equation}\label{eq:detexact}
   m\;\ge\;\left(\frac{32}{\delta}\right)^{1/l}
   \frac{\rho^{-2}}{k!\,\alpha_{k,l}^{1/l}}\;n^{k/2} .
\end{equation}
For large $l$ this reads $m\ge\theta_{l,k,\delta}\,\rho^{-2}\,n\,(n/l)^{k/2-1}$ with $\theta_{l,k,\delta}=(1+o_l(1))\bigl(32\sqrt l/(\delta L_k)\bigr)^{1/l}\gamma_k$, i.e.\ exactly the threshold~\eqref{eq:detthreshold} up to an explicit factor tending to one as $l$ grows. The same sample condition gives a test whose decision rule does not use $\rho$: declare the instance planted when $Z_l>0$ and
\begin{equation}\label{eq:adaptive-detector}
   X(l)>\sqrt{\frac{2}{\delta Z_l}}.
\end{equation}
Its false-positive and false-negative probabilities are again at most $\delta$ for all sufficiently large $n$.
\end{theorem}

\begin{table}[t]
\centering
\small
\begin{tabular}{lcc}
\toprule
Method & Memory & Runtime \\
\midrule
HCE, level $l$
  & $\widetilde O\!\left(n^2+m^{\lceil l/2\rceil}\right)$
   & $\widetilde O\!\left(n^3+m^{\lceil l/2\rceil}+nm^{\lfloor l/2\rfloor}\right)$ \\
HCE, even $l$, $m=Cn^{k/2}$
  & $\widetilde O\!\left(n^2+C^{l/2}n^{kl/4}\right)$
   & $\widetilde O\!\left(n^3+C^{l/2}n^{kl/4+1}\right)$ \\
HCE, odd $l$, $m=Cn^{k/2}$
  & $\widetilde O\!\left(n^2+C^{(l+1)/2}n^{k(l+1)/4}\right)$
   & $\widetilde O\!\left(n^3+C^{(l+1)/2}n^{k(l+1)/4}+C^{(l-1)/2}n^{k(l-1)/4+1}\right)$ \\
\midrule
Kikuchi, level $\ell$
  & $\widetilde O\!\left(n^{\ell}+mn^{\ell-k/2}\right)$
  & $\widetilde O\!\left(mn^{\ell-k/2}+n^{\ell+1}\right)$ \\
Kikuchi, even $\ell$, $m=Cn^{k/2}$
  & $\widetilde O\!\left((C+1)\,n^{\ell}\right)$
  & $\widetilde O\!\left(n^{\ell+1}\right)$ \\
\midrule
Quantum HCE, even $l$, $m=Cn^{k/2}$
  & ---
  & $\widetilde O\!\left(C^{l/3}\,n^{kl/6+2}\right)^{\ast}$ \\
Quantum Kikuchi, even $\ell$, $m=Cn^{k/2}$
  & ---
  & $\widetilde O\!\left(C\,n^{\ell/4+k/2+1}\right)$ \\
\bottomrule
\end{tabular}
\caption{\textbf{Asymptotic runtime and memory for level-$l$ HCE and level-$\ell$ Kikuchi spectral recovery} on planted noisy $k$-XOR with $m=Cn^{k/2}$ clauses and even $k$. $\widetilde O(\cdot)$ suppresses logarithmic factors and constants depending on $k$, $l$ and $\ell$. Classical HCE uses the memory tricks discussed in Appendix~\ref{app:MITM}. Quantum rows: HCE with quantum-walk claw finding \cite{Tani2009} for witness search, and the quartic quantum speedup of Kikuchi \cite{schmidhuber2025quartic} (Section~\ref{sec:complexity}).}
\label{tab:complexity}
\end{table}

\subsection{Recovery}
It is convenient to define expected witness count, its $n$-scaled and noise-normalized versions below,
\begin{equation}\label{eq:normalization}
   N_l=\binom ml\,\PP\bigl[\partial\{1,\dots,l\}=\{i,j\}\bigr],
   \qquad d_l:=nN_l,
   \qquad \theta_l:=\rho^{l}d_l .
\end{equation}
Writing $D_x=\diag(x)$ and $\widehat A^{(l)}=D_xA^{(l)}D_x$, using the i.i.d.\ support model, we obtain
\begin{equation}\label{eq:decomposition}
   \E\widehat A^{(l)}=\rho^{l}N_l\bigl(\mathbf 1\mathbf 1^\top-I\bigr),
   \qquad
   A^{(l)}=\rho^{l}N_l\,xx^\top+R^{(l)} ,
\end{equation}
so the planted eigenvalue is $\rho^{l}N_l(n-1)$, while every orthogonal eigenvalue of the mean is $-\rho^{l}N_l$. 

The exactly centered conjugated residual $\widetilde R^{(l)}:=\widehat A^{(l)}-\E\widehat A^{(l)}$ differs from $R^{(l)}$ only by the deterministic diagonal shift,
\begin{equation}\label{eq:shift}
   D_xR^{(l)}D_x=\widetilde R^{(l)}-\rho^{l}N_lI,
   \qquad
   \bigl|\opnorm{R^{(l)}}-\opnorm{\widetilde R^{(l)}}\bigr|\le\theta_l/n .
\end{equation}

 By~\eqref{eq:decomposition}, $A^{(l)}$ is a rank-one spike of strength $\theta_l$ along $xx^\top$ plus the residual $R^{(l)}$. By Weyl's and the Davis--Kahan inequalities, we show that its top eigenvector aligns with $x$ once $\opnorm{R^{(l)}}$ is small compared with $\theta_l$. For recovery we must control $\opnorm{R^{(l)}}$ globally rather than entrywise. The exact overlap formula in Appendix~\ref{app:recovery} shows that the entry variance contains both the $N_l$ term and correlation contributions from overlapping witnesses. Entrywise information alone cannot exclude their localized effects; we therefore bound high trace moments of the exactly centered residual $\widetilde R^{(l)}$ of~\eqref{eq:shift}.

Splitting witnesses according to whether $i$ and $j$ lie in the same support component (\emph{boundary-connected}) or in different ones (\emph{boundary-disconnected}) gives $\widetilde R^{(l)}=R^{(l)}_{\mathrm{bc}}+R^{(l)}_{\mathrm{bd}}$. A Hoeffding decomposition controls the first sector (Appendix~\ref{app:recovery}), and for even $k$ a parity argument gives $R^{(l)}_{\mathrm{bd}}=0$ identically.

\begin{theorem}[HCE weak recovery for even $k$]\label{thm:recovery} 
Fix even $k\ge4$, admissible $l\ge2$, $\rho\in[0,1]$, $0<C\le n^{k/2-1}$ with $d_l\ge1$, and set $a=(k+1)l+3$. There are constants $B=B_{k,l}$ and $n_0=n_0(k,l)$ such that for all $n\ge n_0$ and even $q\ge2$,
\begin{equation}\label{eq:recovery-moment}
   \bigl(\E\tr|\widetilde R^{(l)}|^{q}\bigr)^{1/q}
   \le n^{1/q}(Bq)^{a}\Bigl[d_l^{1/2}+\rho^{l-2}d_l^{1-1/l}\Bigr].
\end{equation}
Hence, for every fixed $\gamma>0$, with probability at least $1-n^{-\gamma}$,
\begin{equation}\label{eq:recovery-radius}
   \opnorm{R^{(l)}}\le\mathfrak R_{n,\gamma}
   :=B_\gamma(\log n)^{a}\Bigl[d_l^{1/2}+\rho^{l-2}d_l^{1-1/l}\Bigr]
   +\frac{\theta_l}{n}.
\end{equation}
If $\theta_l>(3+\varepsilon)\mathfrak R_{n,\gamma}$ for some $\varepsilon>0$, put $\delta=\mathfrak R_{n,\gamma}/(\theta_l-\mathfrak R_{n,\gamma})$. Then the top eigenvalue of $A^{(l)}$ is isolated, and its unit eigenvector $\hat u$ and $\hat x=\sgn(\hat u)$ satisfy, with $u=x/\sqrt n$ and up to a global sign,
\begin{equation}\label{eq:recovery-correlation}
   |\langle\hat u,u\rangle|\ge\sqrt{1-\delta^{2}}>0, \quad\text{and}\quad \frac{|\langle\hat x,x\rangle|}{n}\ge1-4\delta^2>0.
\end{equation}
\end{theorem}

The exact signal-to-noise ratio is $\SNR_l=\theta_l(1-1/n)/\opnorm{\widetilde R^{(l)}}$, and, ignoring fixed constants, it diverges as soon as
\begin{equation}\label{eq:snr}
   \rho^{2l}d_l\gg(\log n)^{al} .
\end{equation}
It remains to count witnesses. Appendix~\ref{app:N_ell} shows $N_l=(1+o(1))\kappa_{l,k}\,m^{l}/n^{kl/2+1}$ with $\kappa_{l,k}^{1/l}\sim\frac{e}{l}\bigl(\frac{kl}{e}\bigr)^{k/2}$, so that $d_l=nN_l=(1+o(1))\kappa_{l,k}C^{l}$ when $m=Cn^{k/2}$. Thus the separation condition becomes a condition on the clause density alone.

\begin{corollary}[Smooth trade-off for recovery]\label{cor:recovery-threshold}
Fix even $k\ge4$, $0<\rho\le1$, $\gamma>0$, and $\varepsilon>0$. There is a constant $K_{k,\gamma,\varepsilon}>0$, independent of $l$, such that for every fixed admissible $l\ge2$, with $a=(k+1)l+3$, the following holds for all sufficiently large $n$ (where the lower bound on $n$ may depend on $l$). If $m=Cn^{k/2}$, $C\le n^{k/2-1}$, and
\begin{equation}\label{eq:mainthreshold}
   C\;\ge\;K_{k,\gamma,\varepsilon}\,
   \rho^{-2}\kappa_{l,k}^{-1/l}(\log n)^{(k+1)l+3},
\end{equation}
then, for all sufficiently large $n$, the binary output of HCE satisfies, with probability at least $1-n^{-\gamma}$,
\[
   \frac{|\langle\hat x,x\rangle|}{n}
   \ge 1-\frac{4}{(2+\varepsilon)^2}>0.
\]
\end{corollary}

For fixed $k$ and large $l$, $\kappa_{l,k}^{-1/l}\sim\gamma_k l^{1-k/2}$, so \eqref{eq:mainthreshold} traces the Kikuchi curve up to an $l$-independent constant; the results are stated for fixed $l$ only (Appendix~\ref{app:recovery}). For odd $k$, the same curve holds at every even $l$ after trimming a sublinear, support-determined set $\mathcal B_l$ of coordinates, which is empty for $l\in\{2,4\}$ (Theorem~\ref{thm:recovery-odd} and Corollary~\ref{cor:recovery-threshold-odd} in Appendix~\ref{app:rec-odd}).

\begin{remark}
Since the raw HCE matrix can localize on high-degree vertices, removing the $(\log n)^{a}$ slack in~\eqref{eq:recovery-radius} will likely require degree normalization or trimming, as in the normalized Kikuchi construction of \citet{HS2026}.
\end{remark}

\subsection{Complexity}
\label{sec:complexity}

A direct implementation of~\eqref{eq:Adef} enumerates all $l$-tuples of clauses in $O(m^{l})=O(n^{kl/2})$ time when $m=Cn^{k/2}$, using only $O(n^2)$ matrix memory. MITM splits each $l$-set into halves of sizes $s=\lceil l/2\rceil$ and $r=\lfloor l/2\rfloor$ and hashes the $s$-sets by XOR boundary using $\widetilde O(m^s)$ memory. The original all-pairs probe takes $\widetilde O(m^s+n^2m^r)$ construction time. Instead, aggregate both half sizes by boundary. Scan the $r$-boundary keys for same-size partners (one-element replacements), and scan keys in each table for smaller partners (two-element deletions). Every pair of boundaries with XOR $\{i,j\}$ is then accounted for. Since each key has at most $ks$ elements, this takes $\widetilde O_{k,l}(m^s+nm^r+n^2)$ deterministic construction time with the same asymptotic memory; for even $l$, $r=s=l/2$, giving $\widetilde O_{k,l}(nm^{l/2}+n^2)$. For odd $l$, the $m^s$ cost of building the larger table remains and cannot in general be dropped from this bound. An exact lower-level recurrence removes pairs of halves sharing clauses, and the final $n\times n$ eigen-decomposition costs $O(n^3)$. Appendix~\ref{app:MITM} gives the construction and Table~\ref{tab:complexity} its resource bounds. The experimental implementation uses the original all-pairs probe; its measured times do not benchmark the improved key scan.
\paragraph{Quantum complexity and implications.} The enumeration underpinning HCE is an instance of the $l$-XOR problem: find $l$-sets of clause supports, viewed as vectors in $\F_2^{n}$, with prescribed total XOR. Classically, MITM is the standard generic baseline \cite{Wagner2002}. Quantumly, the problem maps to claw finding: split the clauses into two halves, let $f=\XOR$ on $\lceil l/2\rceil$-sets of the first half and $g=\XOR(\cdot)\triangle\{i,j\}$ on $\lfloor l/2\rfloor$-sets of the second, so that a claw $f(\Scal)=g(\Scal')$ is exactly an $(i,j)$-witness. Tani's quantum-walk algorithm \cite{Tani2009}, built on the element-distinctness walk of \cite{Ambainis2007} and Szegedy's framework \cite{Szegedy2004}, finds a claw in $\widetilde O(m^{l/3})$ time instead of $O(m^{\lceil l/2\rceil})$ for a fixed pair; repeating over all pairs gives the quantum HCE row of Table~\ref{tab:complexity}. This is the cost of witness search, not an end-to-end speedup over the improved classical key scan: building $A^{(l)}$ requires the signed sum over all witnesses, so sampling or enumerating many of them adds output-dependent factors. For Kikuchi, \citet{schmidhuber2025quartic} give a quartic quantum speedup of the level-$\ell$ spectral method. The two speedups matter in different regimes. Where classical Kikuchi is already preferable (large $C$ with deep levels, yellow region of Figure~\ref{fig:hce-kikuchi-cartoon}), its quantum speedup can play a significant role. In the ultra-sparse regime $C\ll1$, where HCE is empirically preferable at the sizes we test, theoretical quantum speedups for Kikuchi do not naturally translate into end-to-end recovery speedups over HCE, and the relevant quantum question becomes how far the HCE enumeration itself can be accelerated.
\section{Experiments}
\label{sec:experiments}

\begin{figure}[t]
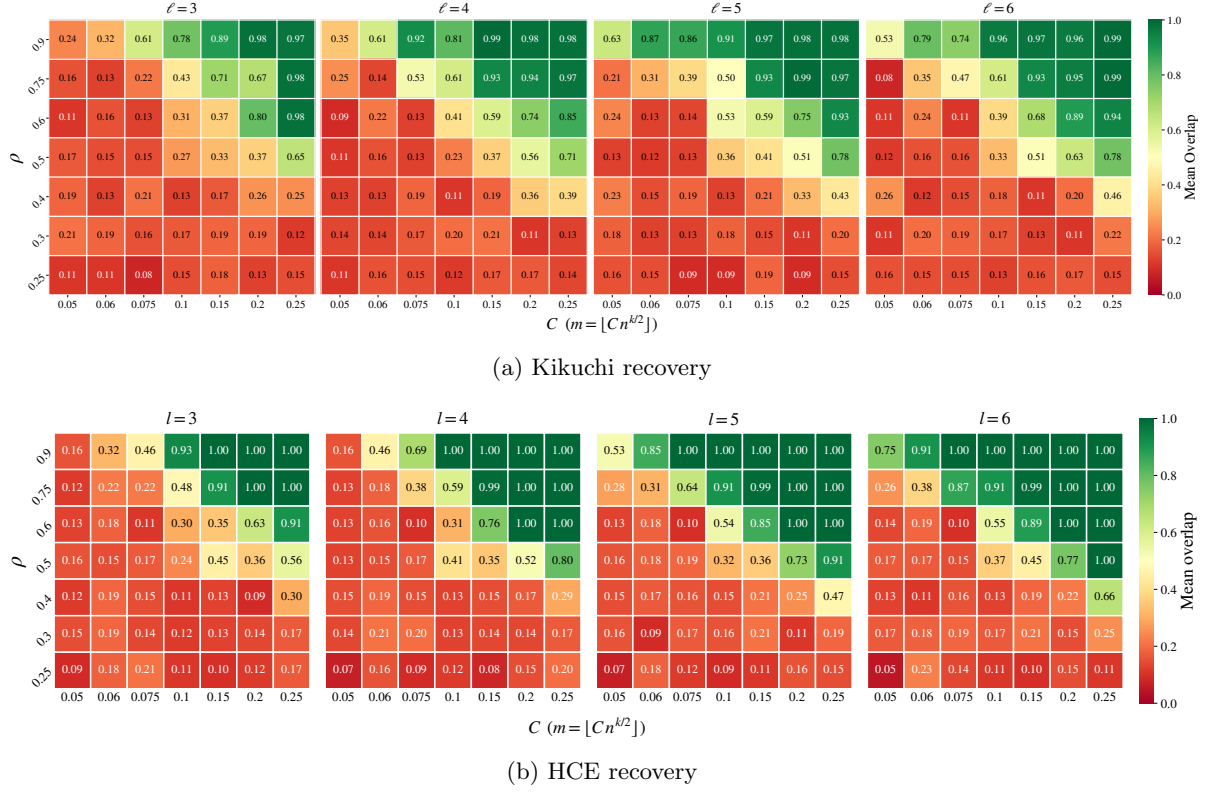

  \centering
  \begin{subfigure}{\linewidth}
    \centering
    \resultfig{Kikuchi_n30_k4_heatmap_pub.pdf}
    \caption{Kikuchi recovery}
    \label{fig:k4-kikuchi}
  \end{subfigure}\\[0.6em]
  \begin{subfigure}{\linewidth}
    \centering
    \resultfig{HCE_n30_k4_heatmap_pub.pdf}
    \caption{HCE recovery}
    \label{fig:k4-hce}
  \end{subfigure}
  \caption{\textbf{Recovery of HCE and Kikuchi is comparable when $l=\ell$}. Recovery overlap with the planted bitstring for $n=30$, $k=4$, averaged over $10$ random $k$-XOR instances. The HCE level $l$ and the Kikuchi level $\ell$ are matched, $l=\ell$. The corresponding heatmaps for $k=5$ and $k=6$ are reported in Appendix~\ref{app:extra-experiments}, Figures~\ref{fig:k5-overlap}--\ref{fig:k6-overlap}.}
  \label{fig:k4-overlap}
\end{figure}

Full specifications of both methods and the resource metrics are given in Appendix~\ref{app:protocol}. The figure of merit for recovery is the \emph{overlap} $|\langle \hat x, x\rangle|/n$ between the recovered and planted bitstrings. Unless stated otherwise, results are averaged over random instances generated from identical seeds for all methods, so that HCE and Kikuchi see exactly the same instances. We compare the methods at matched level $l=\ell$. At $k=4$ their detection-threshold constants agree to within $2\%$ (Section~\ref{sec:cycles}), making this a natural point of comparison.

\paragraph{Recovery quality at $k=4$.} In Figure ~\ref{fig:k4-overlap}, we see that at matched level, HCE attains overlap comparable to Kikuchi, consistent with the constant-level comparison of Section~\ref{sec:cycles}. Since the HCE pipeline uses the post-processing described in Appendix~\ref{app:decoding}, whereas the recovery theorem concerns the bare top eigenvector of one layer, these heatmaps compare the practical pipelines rather than directly testing Theorem~\ref{thm:recovery}. We also noted that tensor power iteration did not appreciably boost the overlaps for either methods, for reasonably large number of iteration steps.


\paragraph{Odd and larger $k$.} At $k=6$, the $k=4$ comparison repeats, with HCE matching Kikuchi recovery quality at equal level ($l=\ell$). At $k=5$, HCE can be utilized directly for recovery, unlike symmetric Kikuchi; by the parity obstruction of Section~\ref{sec:prelim}, the $l=3,5$ matrices vanish, giving no additional recovery contributions, as seen by plateaus from $l=2$ to $l=3$, and $l=4$ to $l=5$. Every reported odd-$k$ run thus has signal only at $l\in\{2,4\}$, where $\mathcal B_l=\varnothing$ and trimmed HCE (Appendix~\ref{app:rec-odd}) coincides with HCE; trimming at $l\ge6$ is left for future work. Both heatmaps are in Appendix~\ref{app:extra-experiments} (Figures~\ref{fig:k5-overlap} and~\ref{fig:k6-overlap}).

\paragraph{Scaling in the problem size.} Figure~\ref{fig:scaling-n} shows runtime and the size of the largest persistent data structure, as defined in Appendix~\ref{app:protocol}, for a ultra-sparse instances ($C=0.15$). We observe that HCE is consistently faster than Kikuchi at matched level, and the memory gap grows with $n$: Kikuchi stores a lifted sparse operator, whereas HCE diagonalizes an $n\times n$ matrix after constructing its hash table.

\begin{figure}[t]
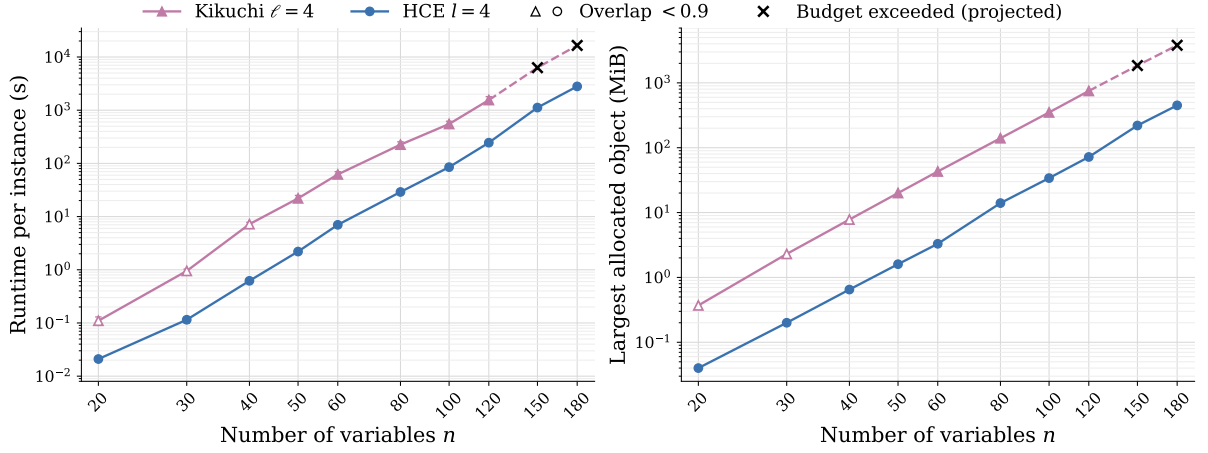

  \centering
  \begin{subfigure}{\linewidth}
    \centering
\resultfig{Runtime_memory_vs_n_pub.pdf}
  \end{subfigure}
   \caption{\textbf{HCE is faster than Kikuchi at the same level.} Scaling of runtime and largest memory object with the number of variables $n$ for noisy $k$-XOR with $k=4$, $C=0.15$, $\rho=0.75$ at fixed level $l=\ell=4$, averaged over $3$ random instances. At these finite sizes HCE is much faster than Kikuchi for $l=\ell$, although at $k=4$ the asymptotic bounds of Table~\ref{tab:complexity} have the same leading exponent.}
  \label{fig:scaling-n}
\end{figure}

\begin{figure}[t]
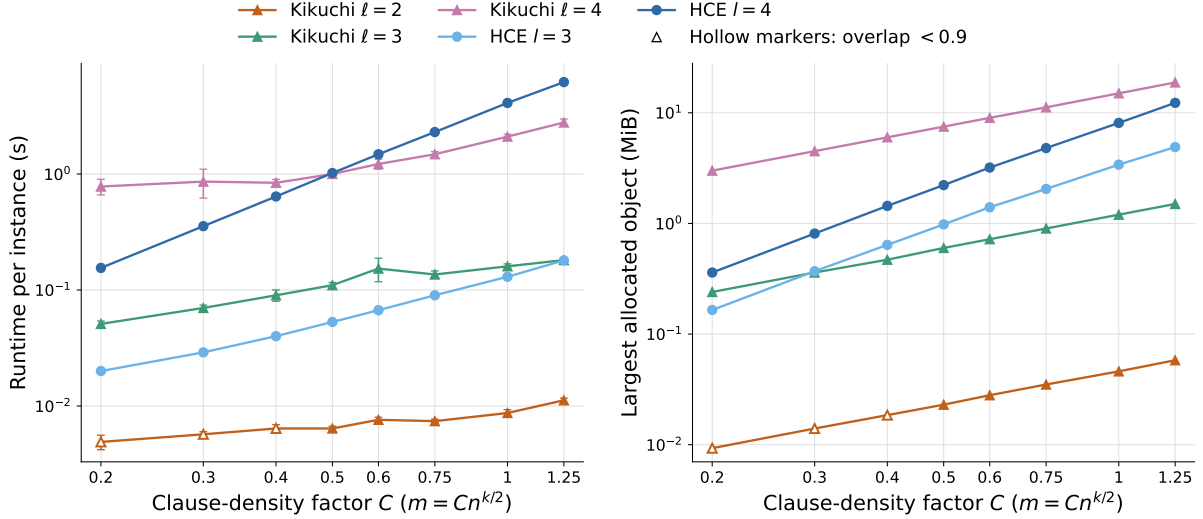

  \centering
  \begin{subfigure}{\linewidth}
    \centering
    \resultfig{Runtime_memory_vs_C_pub.pdf}
  \end{subfigure}
   \caption{\textbf{The advantage over Kikuchi is largest in the hardest regime, where $C$ is small.} Scaling of runtime and largest memory object with clause density $C$ for $k=4$, $n=30$, $\rho=0.75$; HCE at $l=4$ and Kikuchi at $\ell=2,3,4$, averaged over $3$ random instances. Note that for $C<0.5$ (ultra-sparse), $\ell=2$ Kikuchi does not achieve recovery overlap $>0.9$.}
  \label{fig:scaling-C}
\end{figure}

\paragraph{Scaling in the clause density.} As can be noted in Figure~\ref{fig:scaling-C}, the advantage of HCE is largest in the hardest regime, i.e.\ when $C$ is small: there, few clause products collapse to a pair, so the enumeration that HCE pays for is cheap, while the Kikuchi lifted operator remains $\binom{n}{\ell}$-dimensional irrespective of density. Conversely, at large $C$ the enumeration cost of HCE grows and Kikuchi becomes preferable. This delineates three regimes: for $C\gtrsim1$ (`sparse-but-dense') the instance is already solved at level $2$; near $C\sim1$ (sparse) the two methods are comparable at matched level; and for solvable $C\ll1$ (ultra-sparse) HCE is empirically preferable.

\section{Discussion and Conclusion}
\label{sec:discussion}

We introduced hypergraph cycle enumeration (HCE), an elementary spectral hierarchy for planted noisy $k$-XOR. At level $l$, HCE multiplies $l$ clauses whose supports XOR to a pair $\{i,j\}$, similar to a Gaussian elimination procedure on finite field $\mathbb{F}_2$. This reduces the problem, and hence sparse rank-one tensor recovery, to PCA on an $n\times n$ matrix at every level. 

The mechanism for HCE is transparent: every hypergraph cycle contributes an explicit parity signal, aligned with $x_ix_j$ for recovery. The threshold~\eqref{eq:detthreshold} is exactly where the noise cost $\rho^{2l}$ balances the witness gain $N_l$. Unlike symmetric Kikuchi, HCE applies directly to odd $k$.
At $k=4$ and matched even levels, the improved HCE bound has the same leading runtime exponent as the stated Kikuchi bound. In practice, even the original all-pairs implementation solves $k=4$ instances at a fraction of Kikuchi's memory and runs considerably faster in the ultra-sparse regime. 

Three directions remain open. The first is removing the polylogarithmic slack in~\eqref{eq:recovery-radius}, which would match the sharp Kikuchi result of \citet{HS2026}. The second is lowering the enumeration cost below the generic meet-in-the-middle baseline by exploiting planted structure. The third is extending HCE to smoothed and semi-random CSPs, since even covers underlie both refutation and the Moore bound.

These results place HCE among several recent methods for planted noisy $k$-XOR that rest on counting even covers, each method differing by how the covers are weighted. The Kikuchi hierarchy sums closed walks in a lifted graph on $\binom n\ell$ vertices. \citet{HS2026} normalize this matrix and count its closed walks, which gives the conjectured trade-off without logarithmic loss and a quartic quantum speedup. \citet{gupta2026classical} find even covers through collisions of short random walks in the same Kikuchi graph, giving a classical quadratic speedup for detection when $k$ is a large constant. \citet{li2025smooth} and \citet{mao2026near} count only thin rings and chains of small gadgets achieving detection and recovery. HCE conversely counts every even cover and every $(i,j)$-witness of a fixed size. We have shown numerically that HCE, despite its asymptotically worse runtime compared to some of these methods, in practice is very efficient and superior to the Kikuchi method in sparse regimes. Whether some other algorithm of this class has a practical advantage compared to HCE remains open.

There is a further connection between the HCE matrix and the Kikuchi matrix, which we show in Appendix~\ref{app:operator-comparison}: both algorithms are restrictions of the same operator on the Boolean cube. Define for all $a \in [m]$ the operators $\mathsf A_a = b_a \mathsf X_a$, where $\mathsf X_a$ flips the bits corresponding to clause $S_a$. Then the operator of interest is
$$
\mathsf H = \sum_{a=1}^m \mathsf A_a.
$$
In quantum language, this is the Hadamard transform of the $k$XOR Hamiltonian. Both HCE and Kikuchi can be directly related to this object. HCE applies the degree-$l$ Kravchuk polynomial of this operator first and then either evaluates its Rayleigh quotient on the all-zero basis vector for detection, or restricts it to the $n$ Hamming weight 1 states to form the recovery matrix. The Kikuchi matrix restricts the operator to one $\ell$-slice first, obtaining the Kikuchi matrix, and then takes powers to find the top eigenvector, so every intermediate state must stay on that slice.
The relation between HCE and the Kikuchi method reveals interesting possibilities for future quantum algorithms. The quartic speedups of \citet{schmidhuber2025quartic} and \citet{HS2026} come from the lifted Kikuchi space, whereas HCE gives up that space and instead operates on the underlying space of even covers. It is open whether this interpretation allows a similarly large quantum speedup to be obtained for HCE.


\subsection*{Acknowledgments}
We thank Rob Otter for the executive support of the work and invaluable feedback on this project.
We thank the colleagues at the Global Technology Applied Research center of JPMorganChase for support.

\bibliographystyle{apsrev4-2}
\bibliography{ref}

\section*{Disclaimer}
This paper was prepared for informational purposes by the Global Technology Applied Research center of JPMorgan Chase \& Co. This paper is not a product of the Research Department of JPMorgan Chase \& Co. or its affiliates. Neither JPMorgan Chase \& Co. nor any of its affiliates makes any explicit or implied representation or warranty and none of them accept any liability in connection with this paper, including, without limitation, with respect to the completeness, accuracy, or reliability of the information contained herein and the potential legal, compliance, tax, or accounting effects thereof. This document is not intended as investment research or investment advice, or as a recommendation, offer, or solicitation for the purchase or sale of any security, financial instrument, financial product or service, or to be used in any way for evaluating the merits of participating in any transaction.

\clearpage
\appendix
\renewcommand{\thetheorem}{\thesection.\arabic{theorem}}

\paragraph*{Organization of the appendices.} The appendices follow the logical flow of the paper. Appendix~\ref{app:notation} collects the notation. Appendix~\ref{app:detection} proves Theorem~\ref{thm:detection}. It uses Appendix~\ref{app:complete}, which counts even covers in the complete hypergraph and then computes their expected number in a random hypergraph (Subsection~\ref{app:expected}). Appendix~\ref{app:recovery} proves Theorem~\ref{thm:recovery} and Corollary~\ref{cor:recovery-threshold} for even $k$, and the trimmed odd-$k$ result in Subsection~\ref{app:rec-odd}. Appendix~\ref{app:N_ell} counts the witnesses $N_l$ that set the recovery threshold. Appendix~\ref{app:operator-comparison} relates $A^{(l)}$ to the Kikuchi and square constructions, and Appendix~\ref{app:MITM} gives the meet-in-the-middle construction behind Table~\ref{tab:complexity}. The remaining appendices concern the experiments: the decoding pipelines (Appendix~\ref{app:decoding}), the experimental protocol (Appendix~\ref{app:protocol}), and additional results for $k=5,6$ (Appendix~\ref{app:extra-experiments}).

\section{Notation}
\label{app:notation}

\begin{center}
\small
\begin{tabular}{p{0.16\linewidth}p{0.34\linewidth}p{0.42\linewidth}}
\toprule
Symbol & Type & Meaning / property \\
\midrule
$n,m,k$ & positive integers & problem size, number of sampled clauses,
   clause size \\
$l$ & positive integer & \textbf{HCE level}: even-cover size / witness size;
   admissible when $kl$ is even \\
$\ell$ & positive integer & \textbf{Kikuchi level}: degree of the lifted
   Kikuchi matrix; used only in comparisons \\
$\pi_k$ & $3$ ($k$ even), $4$ ($k$ odd) & minimal admissible cover level \\
$[n]$ & finite set $\{1,\ldots,n\}$ & vertex set \\
$\binom{[n]}{k}$ & finite set of $k$-subsets of $[n]$ & possible supports \\
$M$ & $=\binom{n}{k}$ & number of possible clauses \\
$C$ & positive real & clause density, $m=Cn^{k/2}$ \\
$\rho$ & real number in $[0,1]$ & noise mean (planted advantage) \\
$x$ & element of $\{\pm1\}^{n}$ & planted sign vector \\
$\Kcal$ & $=\binom{[n]}{k}$ & complete $k$-uniform hypergraph \\
$\Hcal$ & subset of $\Kcal$ & instance clause set, size $m$ \\
$\Scal$ & subset of $\Kcal$ & any subset of $\Hcal$ \\
$S_a$ & element of $\Kcal$ & clause support indexed by $a$ \\
$\eta_a$ & r.v.\ in $\{\pm1\}$ & independent noise, $\E[\eta_a]=\rho$ \\
$b_a$ & r.v.\ in $\{\pm1\}$ & observation $b_a=\eta_a\prod_{v\in S_a}x_v$ \\
$\triangle$ & operation & symmetric difference \\
$\triangle(\Scal)$, $\partial I$ & subset of $[n]$ & $\triangle_{S\in\Scal}S$;
   $\triangle_{a\in I}S_a$ for a set $I$ of clause indices \\
$X_\Scal$ & r.v.\ in $\{\pm1\}$ & $\prod_{a\in\Scal}b_a$ \\
$F$ & subset of $\Kcal$ & even cover: $|F|=l$, $\triangle(F)=\emptyset$ \\
$\Ccal_l$, $Z_l$ & set / integer & even $l$-covers in $\Hcal$ and their number \\
$\Ccal^{\Kcal}_l$, $Z^{\Kcal}_l$ & set / integer & even $l$-covers in $\Kcal$ and their number \\
$E_{n,k}(l)$ & integer & $=Z^{\Kcal}_l$, the exact Kravchuk count \\
$\alpha_{k,l}$ & non-negative real & $Z^{\Kcal}_l=(\alpha_{k,l}+o_n(1))n^{kl/2}$; zero for inadmissible $l$; $\alpha_{k,0}=1$ \\
$\mu_q$ & non-negative real & $\alpha_{k,q}(C\,k!)^{q}$, so $\E_\Hcal[Z_q]=\mu_q+o_n(1)$; $\mu_0=1$ \\
$n_s$, $n^{\Kcal}_s$ & integers & ordered pairs of even $l$-covers of $\Hcal$ (resp.\ $\Kcal$) meeting in exactly $s$ edges \\
$\Wcal_l(i,j)$, $W_l(i,j)$ & set / integer & $(i,j)$-witnesses in $\Hcal$ and their number \\
$N_l$ & non-negative real & $\E\,W_l(i,j)$ \\
$d_l$, $\theta_l$ & positive reals & $d_l=nN_l$; signal scale $\theta_l=\rho^{l}d_l$ \\
$\widehat A^{(l)}$, $\widetilde R^{(l)}$ & $n\times n$ matrices & $D_xA^{(l)}D_x$ and its exact centering $\widehat A^{(l)}-\E\widehat A^{(l)}$ \\
$q$, $q_n$ & even integers & trace-moment order; $q_n=2\lceil\log n\rceil$ \\
$\mathcal M_q$, $\mathcal N_q$ & functionals & $\mathcal M_q(M)=(\E\tr|M|^{q})^{1/q}$, $\mathcal N_q(M)=(\E\opnorm{M}^{q})^{1/q}$ \\
$a$, $a_0$ & positive integers & $a_0=(k+1)l+2$ (pattern-count exponent); $a=a_0+1$, the polylogarithmic exponent \\
$p_r(w)$ & probability & $\PP[\partial\{1,\dots,r\}=w]$, the prescribed-boundary probability \\
$R^{(l)}_{\mathrm{bc}}$, $R^{(l)}_{\mathrm{bd}}$ & $n\times n$ matrices & boundary-connected and boundary-disconnected sectors of $\widetilde R^{(l)}$ \\
$\mathcal B_l$, $\mathcal G_l$ & subsets of $[n]$ & singleton-boundary trimming set and its complement, \eqref{eq:bad-set} \\
$\Pi_l$, $\bar A^{(l)}$ & $n\times n$ matrices & projection onto $\mathcal G_l$ and trimmed recovery matrix, \eqref{eq:trimmed} \\
$N_{l,\mathrm{bc}}$, $d_{l,\mathrm{bc}}$, $\theta_{l,\mathrm{bc}}$ & non-negative reals & boundary-connected witness count, degree scale, and signal scale \\
$\kappa^{\mathrm{bc}}_{l,k}$ & positive real & boundary-connected restriction of the full i.i.d.\ witness coefficient \\
$T_l,Q_l(i,j)$ & integers & distinct-support universe count and ordered i.i.d.\ support-tuple count \\
$\kappa_l,\kappa_{l,k}$ & positive reals & normalized ordered-tuple and sampled i.i.d.\ witness constants; $\kappa_{l,k}=(k!)^l\kappa_l$ \\
$\gamma_k$, $a_k$, $L_k$ & positive reals & $\gamma_k=k^{-k/2}e^{k/2-1}$; $a_k=1/(\gamma_kk!)$; $L_k=e^{-(k-1)/2}/\sqrt\pi$ \\
\bottomrule
\end{tabular}
\end{center}

Two conventions are worth restating. First, the HCE level is always $l$ and the Kikuchi level is always $\ell$; the two coincide only in the matched-level experiments, where we write $l=\ell$ explicitly. Second, $C$ denotes the clause density throughout, so the even-cover count present in the instance is written $Z_l$ (and not $C_l$), the witness constants are $\kappa_l,\kappa_{l,k}$ (and not $C_l,C_{l,k}$), and the generic constants appearing in the moment bounds are written $B$. The trace-moment order is $q$, which is unrelated to the inclusion probability $p=m/\binom nk$ of Appendix~\ref{app:N_ell}.

\section{Proof of the Detection Theorem}
\label{app:detection}

Throughout this appendix $\Hcal$ is a uniformly random $m$-edge subhypergraph of $\Kcal$ with $m=Cn^{k/2}$, and $X(l)$ is the discriminator~\eqref{eq:discriminator}, with $X_F=\prod_{S_a\in F}b_a$. We keep the two sources of randomness apart by a subscript: $\PP_\Hcal,\E_\Hcal,\Var_\Hcal$ refer to the draw of the clause supports, while $\PP_\rho,\E_\rho,\Var_\rho$ and $\Cov_\rho$ refer to the signs at planted advantage $\rho$, the null model being $\rho=0$.

We only consider \emph{admissible} levels, i.e.\ $l\ge3$ with $kl$ even, for which $\Kcal$ possesses proper even $l$-covers; the minimal admissible level is $\pi_k$ of Section~\ref{sec:prelim}. By Theorem~\ref{thm:Enk}, $Z^{\Kcal}_l=\alpha_{k,l}n^{kl/2}+o_n(n^{kl/2})$ for an $n$-independent $\alpha_{k,l}$, with $\alpha_{k,l}=0$ when $l$ is inadmissible and $\alpha_{k,0}=1$ (the empty family is the unique even $0$-cover). For $k>2$ and large admissible $l$,
\begin{equation}\label{app:eq:alpha-asymp}
   \alpha_{k,l}=\bigl(L_k+o_l(1)\bigr)\,l^{-1/2}
   \bigl(a_k\,l^{k/2-1}\bigr)^{l},
   \qquad a_k:=\frac{k^{k/2}e^{1-k/2}}{k!}=\frac{1}{\gamma_k\,k!},
   \qquad L_k:=\frac{e^{-(k-1)/2}}{\sqrt\pi}.
\end{equation}
Finally set
\begin{equation}
   n_s:=\bigl|\{(F,F')\in\Ccal_l\times\Ccal_l : |F\cap F'|=s\}\bigr|,
   \qquad
   n^{\Kcal}_s:=\bigl|\{(F,F')\in\Ccal^{\Kcal}_l\times\Ccal^{\Kcal}_l : |F\cap F'|=s\}\bigr| ,
\end{equation}
for the instance and for the complete hypergraph respectively.

\begin{lemma}[Pairs of covers meeting in $s$ edges]\label{lem:cover-intersect-bound}
Fix $k,l$ and $0\le s\le l$, and set $r:=l-s$. Then
\begin{equation}
   n^{\Kcal}_s=\bigl(\alpha_{k,s}\,\alpha_{k,r}^{2}+o_n(1)\bigr)\,n^{kl-ks/2} .
\end{equation}
If $\Kcal$ has no proper even cover of size $s$ or of size $r$, the corresponding coefficient $\alpha_{k,s}$ or $\alpha_{k,r}$ vanishes and the statement is to be read as $n^{\Kcal}_s=o_n(n^{kl-ks/2})$.
\end{lemma}

\begin{proof}
Define $J:=F\cap F'$, $A:=F\setminus J$ and $B:=F'\setminus J$. Every triple $(J,A,B)$ corresponds to a unique pair $(F,F')$, so we count triples with $|J|=s$ and $|A|=|B|=r$. Since $F$ and $F'$ are even covers, $\triangle(J)=\triangle(A)=\triangle(B)$; call this common vertex set $U$ and put $u:=|U|$.

A family of $q$ hyperedges with symmetric difference $U$ uses at most $u+(kq-u)/2$ vertices: the vertices of $U$ consume at least one incidence each, while every other used vertex has positive even degree and hence consumes at least two. Counting the vertices of $U$ only once across $J,A,B$, the whole triple uses at most
\begin{equation}
   u+\frac{ks-u}{2}+2\cdot\frac{kr-u}{2}=kl-\frac{ks}{2}-\frac{u}{2}
\end{equation}
distinct vertices. Therefore $U\ne\emptyset$ contributes at a strictly lower order in $n$. If $U=\emptyset$, then $J,A,B$ are themselves even covers, of sizes $s,r,r$, and the leading contribution comes from triples with pairwise disjoint vertex sets, since sharing a vertex again reduces the number of free labels. The number of such triples is
\begin{equation}
   Z^{\Kcal}_s\bigl(Z^{\Kcal}_r\bigr)^{2}
   =\bigl(\alpha_{k,s}\alpha_{k,r}^{2}+o_n(1)\bigr)n^{ks/2+kr},
\end{equation}
and removing the vertex-intersecting triples does not affect the leading term. Since $ks/2+kr=kl-ks/2$, the claim follows.
\end{proof}

\begin{proposition}\label{prop:EZ}
Under the detection sampling model, $\E_\Hcal[Z_l]=Z^{\Kcal}_l\,\dfrac{(m)_l}{(M)_l}$, where $(y)_j$ denotes the falling factorial and $M=\binom{n}{k}$. Consequently, for $m=Cn^{k/2}$ and fixed $k,l$, $\E_\Hcal[Z_l]=\mu_l+o_n(1)$ with $\mu_l=\alpha_{k,l}(C\,k!)^{l}$.
\end{proposition}
\begin{proof}
The first identity is linearity of expectation: a fixed $l$-set of edges lies in $\Hcal$ with probability $(m)_l/(M)_l$. The second follows from $Z^{\Kcal}_l=(\alpha_{k,l}+o_n(1))n^{kl/2}$, $(m)_l/(M)_l=(1+o_n(1))(m/M)^l$ and $M=(1+O(k^2/n))n^k/k!$.
\end{proof}

\begin{lemma}[Normalized second moment]\label{lem:Zvar}
Fix $k,l$ with $\alpha_{k,l}>0$ and let $m=Cn^{k/2}$ for a constant $C>0$. With $\mu_q:=\alpha_{k,q}(C\,k!)^{q}$ and $\mu_0=1$,
\begin{equation}\label{app:eq:Zvar}
   \frac{\Var_\Hcal[Z_l]}{\E_\Hcal[Z_l]^{2}}
   =\sum_{s=1}^{l}\frac{\mu_s\mu_{l-s}^{2}}{\mu_l^{2}}+o_n(1) .
\end{equation}
For large admissible $l$ the right-hand side is $\Theta_{k,C}\bigl(l^{-(k-2)\pi_k}\bigr)+o_n(1)$.
\end{lemma}

\begin{proof}
Writing $Z_l=\sum_{F\in\Ccal^{\Kcal}_l}\mathbf 1(F\subseteq\Hcal)$ gives $Z_l^{2}=\sum_{F,F'}\mathbf 1(F\cup F'\subseteq\Hcal)$. For two even covers with $|F\cap F'|=s$ we have $|F\cup F'|=2l-s$, hence $\E_\Hcal[\mathbf 1(F\cup F'\subseteq\Hcal)]=(m)_{2l-s}/(M)_{2l-s}$ and
\begin{equation}
   \frac{\E_\Hcal[Z_l^{2}]}{\E_\Hcal[Z_l]^{2}}
   =\sum_{s=0}^{l}\frac{n^{\Kcal}_s}{\E_\Hcal[Z_l]^{2}}\,
     \frac{(m)_{2l-s}}{(M)_{2l-s}} .
\end{equation}
By Lemma~\ref{lem:cover-intersect-bound}, for every fixed $s$,
\begin{equation}
   \frac{n^{\Kcal}_s}{\bigl(Z^{\Kcal}_l\bigr)^{2}}
   =\left(\frac{\alpha_{k,s}\alpha_{k,l-s}^{2}}{\alpha_{k,l}^{2}}+o_n(1)\right)
     n^{-ks/2},
\end{equation}
while, for fixed $l$ and $s$,
\begin{equation}
   \frac{(m)_{2l-s}}{(M)_{2l-s}}\left(\frac{(M)_l}{(m)_l}\right)^{2}
   =\bigl(1+o_n(1)\bigr)\frac{n^{ks/2}}{(C\,k!)^{s}} .
\end{equation}
Multiplying the two displays and using the definition of $\mu_q$ gives $\mu_s\mu_{l-s}^{2}/\mu_l^{2}+o_n(1)$ for the $s$-th summand; the $s=0$ term equals $1+o_n(1)$. As the sum has only $l+1$ terms, subtracting one from the normalized second moment proves~\eqref{app:eq:Zvar}.

For the large-$l$ estimate the summation must be uniform in $s$. Put
\[
   g_{k,t}:=\frac{(kt-1)!!}{t!\,(k!)^t}\qquad(kt\ \text{even}).
\]
The pairing argument in the proof of Theorem~\ref{thm:Enk} gives constants $0<c_k<C_k<\infty$ such that $c_kg_{k,t}\le\alpha_{k,t}\le C_kg_{k,t}$ for every sufficiently large admissible $t$; enlarging the constants covers the finitely many smaller admissible levels. We use the two-sided factorial bounds
\begin{equation}\label{eq:uniform-stirling}
 \sqrt{2\pi}\,t^{t+1/2}e^{-t}e^{1/(12t+1)}
 \le t!\le
 \sqrt{2\pi}\,t^{t+1/2}e^{-t}e^{1/(12t)},
 \qquad t\ge1.
\end{equation}
Applied to the factorials in $g_{k,t}$, these bounds give constants $\eta_k,c_{k,C},B_{k,C}>0$ such that, uniformly over admissible indices,
\[
 \frac{\mu_s\mu_{l-s}^2}{\mu_l^2}
 \le \left[B_{k,C}\left(\frac{s}{l}\right)^{k-2}\right]^s,
 \qquad 1\le s\le\eta_kl,
\]
and the same bound with $s$ replaced by $r=l-s$ holds for $1\le r\le\eta_kl$. On the interior $\eta_kl\le s\le(1-\eta_k)l$, the same substitution gives $e^{-c_{k,C}l\log l}$. For example, the power-carrying part of the logarithm at the lower endpoint is
\[
 \frac{k-2}{2}
 \bigl[s\log s+2(l-s)\log(l-s)-2l\log l\bigr]
 \le (k-2)s\log(s/l)+O_k(s),
\]
uniformly after choosing $\eta_k$ small; the square-root factors in \eqref{eq:uniform-stirling} are absorbed into $B_{k,C}^s$. The first nonzero lower-endpoint index is $s=\pi_k$, and its summand is $\Theta_{k,C}(l^{-(k-2)\pi_k})$. The opposite endpoint and the interior are smaller, while $s=l$ equals $1/\mu_l$ and is superpolynomially small at fixed $C$. Summing the three ranges proves the stated $\Theta_{k,C}(l^{-(k-2)\pi_k})$ estimate.
\end{proof}

\begin{corollary}[Concentration of the cover count]\label{cor:concentration}
Fix an admissible $l$ and $t>0$. Then
\begin{equation}
   \PP_\Hcal\left(\left|\frac{Z_l}{\E_\Hcal[Z_l]}-1\right|>t\right)
   \le\frac{1}{t^{2}}\left(\sum_{s=1}^{l}\frac{\mu_s\mu_{l-s}^{2}}{\mu_l^{2}}
     +o_n(1)\right)
   =O_{k,C}\bigl(t^{-2}l^{-(k-2)\pi_k}\bigr)+o_n(1)
\end{equation}
for large admissible $l$.
\end{corollary}
\begin{proof}
Immediate from Lemma~\ref{lem:Zvar} and Chebyshev's inequality.
\end{proof}

\begin{lemma}[Overlap sums at the detection scale]\label{lem:detection-overlaps}
Fix $k>2$, $\rho\in(0,1]$ and $\delta\in(0,1)$. If
\[
 C\ge\left(\frac{32}{\delta}\right)^{1/l}
         \frac{\rho^{-2}}{k!\,\alpha_{k,l}^{1/l}},
\]
then, for large admissible $l$,
\begin{align}
 \sum_{s=1}^{l}\frac{\mu_s\mu_{l-s}^2}{\mu_l^2}
    &=\frac1{\mu_l}
       +O_k\!\left(\rho^{2\pi_k}
          l^{-(k-2)\pi_k/2}\right),\label{eq:detection-overlap-unweighted}\\
 \sum_{s=1}^{l-1}\rho^{-2s}
          \frac{\mu_s\mu_{l-s}^2}{\mu_l^2}
    &=O_k\!\left(l^{-(k-2)\pi_k/2}\right).
    \label{eq:detection-overlap-weighted}
\end{align}
\end{lemma}

\begin{proof}
Put $p=(k-2)/2$. By~\eqref{app:eq:alpha-asymp}, after enlarging constants to cover the finitely many small admissible indices,
\[
   c_k\,t^{-1/2}(a_kt^p)^t
   \le\alpha_{k,t}\le
   C_k\,t^{-1/2}(a_kt^p)^t
\]
for every admissible $t$. The sample lower bound implies $\rho^{-2}(Ck!)^{-1}\le B_k\alpha_{k,l}^{1/l}$. Therefore the $s$th summand of~\eqref{eq:detection-overlap-weighted}, denoted $T_{l,s}$, satisfies
\[
 T_{l,s}
 \le B_k^s\alpha_{k,s}\alpha_{k,l-s}^2
           \alpha_{k,l}^{-2+s/l}.
\]
Applying~\eqref{eq:uniform-stirling} to this display gives constants $\eta_k,c_k,B_k>0$ such that
\begin{align*}
 T_{l,s}&\le[B_k(s/l)^p]^s,
    &&1\le s\le\eta_kl,\\
 T_{l,l-r}&\le[B_k(r/l)^{2p}]^r,
    &&1\le r\le\eta_kl,\\
 T_{l,s}&\le e^{-c_kl},
    &&\eta_kl\le s\le(1-\eta_k)l.
\end{align*}
Indeed, after writing $x=s/l$, the exponential part in the interior is
\[
 p l\,[x\log x+2(1-x)\log(1-x)],
\]
which is strictly negative on every compact subinterval of $(0,1)$. At the lower endpoint, the inequality $\log(1-s/l)\le-s/l$ yields the first bound uniformly for $s\le\eta_kl$; at the upper endpoint, setting $r=l-s$ gives the second. The square-root remainders in~\eqref{eq:uniform-stirling} are absorbed into $B_k^s$ or $B_k^r$.

Only admissible indices contribute, and the first positive one is $\pi_k$. Summing the lower endpoint gives $O_k(l^{-p\pi_k})$, the upper endpoint gives $O_k(l^{-2p\pi_k})$, and the interior is exponentially small. This proves \eqref{eq:detection-overlap-weighted}. The corresponding unweighted summand is $\rho^{2s}T_{l,s}$; hence its lower endpoint acquires $\rho^{2\pi_k}$, while the remaining ranges are no larger. Finally the omitted term $s=l$ is exactly $1/\mu_l$, proving \eqref{eq:detection-overlap-unweighted}.
\end{proof}

We now address the second heuristic assumption.

\begin{proposition}\label{prop:covariance}
Let $F,F'\in\Ccal_l$ with $|F|=|F'|=l$ and $|F\cap F'|=s$. Then, conditionally on $F,F'\subseteq\Hcal$, $\Cov_\rho(X_F,X_{F'})=\rho^{2l-2s}-\rho^{2l}$.
\end{proposition}
\begin{proof}
$X_FX_{F'}=X_{F\triangle F'}$, and since even covers form a closed cycle space, $F\triangle F'$ is an even cover of size $2l-2s$. Hence $\E_\rho[X_FX_{F'}]=\rho^{2l-2s}$, while $\E_\rho[X_F]\E_\rho[X_{F'}]=\rho^{2l}$.
\end{proof}

\begin{proof}[Proof of Theorem~\ref{thm:detection}]
Let $m=Cn^{k/2}$ and declare the model planted when $Z_l>0$ and $X(l)>\rho^{l}/2$. We first control the hypergraph randomness. Let $G$ be the event $\{Z_l\ge\E_\Hcal[Z_l]/2\}$. Corollary~\ref{cor:concentration} with $t=1/2$ bounds
\begin{equation}\label{app:eq:Gc}
   \PP_\Hcal(G^{c})\le 4\sum_{s=1}^{l}\frac{\mu_s\mu_{l-s}^{2}}{\mu_l^{2}}+o_n(1);
\end{equation}
we bound this after imposing the sample condition below, $l$ being fixed from this point onwards.

For $\Hcal\in G$ under the null model ($\rho=0$), $X(l)$ has mean zero and distinct covers have zero covariance, so $\Var_0(X(l))=1/Z_l$. Chebyshev's inequality, together with the contribution of $G^{c}$ and $\E_\Hcal[Z_l]=\mu_l+o_n(1)$, gives
\begin{equation}
   \PP_0(\text{false positive})
   \le\PP_\Hcal(G^{c})+\frac{8}{\rho^{2l}\mu_l}+o_n(1).
\end{equation}
Under the planted model the conditional mean is $\rho^{l}$, and Proposition~\ref{prop:covariance} gives
\begin{equation}
   \frac{\Var_\rho(X(l))}{\rho^{2l}}
   =\frac{\rho^{-2l}-1}{Z_l}
    +\frac{1}{Z_l^{2}}\sum_{s=1}^{l-1}\bigl(\rho^{-2s}-1\bigr)n_s ,
   \qquad
   \frac{\E_\Hcal[n_s]}{\mu_l^{2}}=\frac{\mu_s\mu_{l-s}^{2}}{\mu_l^{2}}+o_n(1),
\end{equation}
the second identity being Lemma~\ref{lem:cover-intersect-bound} together with the sampling step of Proposition~\ref{prop:EZ}. Chebyshev's inequality and averaging over $\Hcal$ then yield
\begin{equation}
   \PP_\rho(\text{false negative})
   \le\PP_\Hcal(G^{c})+\frac{8}{\rho^{2l}\mu_l}
     +16\sum_{s=1}^{l-1}\rho^{-2s}\frac{\mu_s\mu_{l-s}^{2}}{\mu_l^{2}}+o_n(1).
\end{equation}

It remains to choose $C$ and $l$. Requiring $8/(\rho^{2l}\mu_l)\le\delta/4$ is exactly
\begin{equation}
   C\ge\left(\frac{32}{\delta}\right)^{1/l}
   \frac{\rho^{-2}}{k!\,\alpha_{k,l}^{1/l}},
\end{equation}
which is~\eqref{eq:detexact} after substituting $C=m/n^{k/2}$, and which for large $l$ forces $C=\Omega(\rho^{-2}l^{1-k/2})$. Lemma~\ref{lem:detection-overlaps} gives the required uniform bounds on the two sums above. The sample condition gives $1/\mu_l\le\delta\rho^{2l}/32$, hence $4/\mu_l\le\delta/8$; since $\rho\le1$, the second remainder also controls the first. Choosing the implicit constant in $l=\Omega\bigl(\delta^{-2/((k-2)\pi_k)}\bigr)$ large enough makes four times the first remainder at most $\delta/8$ and sixteen times the second sum at most $\delta/4$. Then $\PP_\Hcal(G^{c})\le\delta/4+o_n(1)$ by~\eqref{app:eq:Gc}, and both error probabilities are below $\delta$ for all sufficiently large $n$.

Finally, inserting the large-$l$ form~\eqref{app:eq:alpha-asymp} of $\alpha_{k,l}$ into~\eqref{eq:detexact} gives
\begin{equation}
   m\ge\theta_{l,k,\delta}\,\rho^{-2}n\left(\frac nl\right)^{k/2-1},
   \qquad
   \theta_{l,k,\delta}=\bigl(1+o_l(1)\bigr)
   \left(\frac{32\sqrt l}{\delta L_k}\right)^{1/l}\gamma_k ,
\end{equation}
which is~\eqref{eq:detthreshold} up to a factor tending to one as $l$ grows.

It remains to verify the adaptive rule~\eqref{eq:adaptive-detector}. Under the null model, conditionally on every hypergraph with $Z_l>0$, $\E_0X(l)=0$ and $\Var_0(X(l))=1/Z_l$. Hence Chebyshev's inequality gives
\[
 \PP_0\!\left[X(l)>\sqrt{\frac{2}{\delta Z_l}}\,\middle|\,\Hcal\right]
 \le\frac\delta2.
\]
Under the planted model, on the event $G=\{Z_l\ge\E_\Hcal Z_l/2\}$, the sample condition implies
\[
 \rho^{2l}Z_l\ge\frac12\rho^{2l}\mu_l+o_n(1)
 \ge\frac{16}{\delta}+o_n(1).
\]
In particular, for all sufficiently large $n$ the left-hand side exceeds $8/\delta$, so the adaptive threshold is at most $\rho^l/2$. Its false-negative event on $G$ is contained in the one already bounded above, while $G^c$ has already been included in that bound. Thus both adaptive errors are at most $\delta$.
\end{proof}

\section{Counting Hypergraph Cycles}
\label{app:complete}

\subsection{Asymptotics for Complete Uniform Hypergraphs}

We count proper even $l$-covers (Definition~\ref{def:evencover}) on the complete $k$-uniform hypergraph $\Kcal=([n],\binom{[n]}{k})$, assuming $kl$ even. Such identities are classical in the analysis of random CSP refutation and appear e.g.\ in \citet{COCF2010,AOW2015,RRS2017,WAM2019,HKM2023}. We use the unnormalized Kravchuk polynomials
\begin{equation}\label{app:eq:kravchuk}
   K_j(x,M):=\sum_{r=0}^{j}(-1)^{r}\binom{x}{r}\binom{M-x}{j-r}.
\end{equation}

\begin{theorem}\label{thm:Enk}
Fix $k>2$ and $l\ge1$. The number of proper even $l$-covers (up to permutation) in $\Kcal$ is
\begin{equation}\label{app:eq:Enk-exact}
   Z^{\Kcal}_l=E_{n,k}(l)=\frac{1}{2^n}\sum_{w=0}^{n}\binom{n}{w}
      K_l\!\left(\frac{M-K_k(w,n)}{2},\,M\right),
   \qquad M=\binom{n}{k}.
\end{equation}
For fixed $k,l$, as $n\to\infty$,
\begin{equation}\label{app:eq:Enk-asymp}
   Z^{\Kcal}_l=\alpha_{k,l}\,n^{kl/2}+o_n\bigl(n^{kl/2}\bigr),
   \qquad
   \alpha_{k,l}=\frac{1}{l!}\left(\frac{1}{k!}\right)^{l/2}
     \E\left[H_l\!\left(\frac{H_k(Z)}{\sqrt{k!}}\right)\right],
   \quad Z\sim N(0,1),
\end{equation}
with $\alpha_{k,l}$ an $n$-independent constant, equal to zero when $l$ is inadmissible and equal to $1$ when $l=0$. For fixed $k>2$ and large admissible $l$, $\alpha_{k,l}$ obeys~\eqref{app:eq:alpha-asymp}. In the iterated limit $n\to\infty$ followed by admissible $l\to\infty$, this is equivalently
\begin{equation}\label{app:eq:Enk-gamma}
   Z^{\Kcal}_l\;=\;\bigl(1+o_l(1)\bigr)
      \frac{e^{-(k-1)/2}}{\sqrt\pi}\binom{M}{l}
      \left(\frac{2}{n}\right)^{kl/2}\Gamma\!\left(\frac{kl+1}{2}\right)
   \bigl(1+o_n(1)\bigr).
\end{equation}
\end{theorem}

\begin{proof}
Let $v_S\in\F_2^n$ be the characteristic vector of hyperedge $S$. Then
\begin{equation}
   E_{n,k}(l)=\sum_{\substack{T\subseteq\binom{[n]}{k}\\|T|=l}}
      \mathbf 1\left[\bigoplus_{S\in T}v_S=0\right]
   =\frac{1}{2^n}\sum_{y\in\F_2^{n}}\sum_{\substack{T\\|T|=l}}
      (-1)^{y\cdot\sum_{S\in T}v_S},
\end{equation}
using Boolean Fourier analysis. The sum over $T$ may be seen as a sum over $T\in\binom{[M]}{l}$; defining $X(y)\in\{0,1\}^{M}$ by $X_S(y)=y\cdot v_S$, the Kravchuk lemma gives $\sum_{T}(-1)^{X\cdot v_T}=K_l(|X(y)|,M)$. By symmetry $|X(y)|$ depends only on $|y|$:
\begin{equation}
   |X(y)|=\sum_{S}\frac{1-(-1)^{X_S(y)}}{2}
   =\frac{M}{2}-\frac12\sum_{S}(-1)^{y\cdot v_S}
   =\frac{M-K_k(|y|,n)}{2}.
\end{equation}
Grouping by Hamming weight yields~\eqref{app:eq:Enk-exact}.

For the fixed-$l$ asymptotics, let $W\sim\mathrm{Bin}(n,1/2)$ and $Z_n:=(n-2W)/\sqrt n$. For every fixed $j$ the central limit of the Kravchuk polynomial reads $M^{-j/2}K_j\bigl(\frac{M-\sqrt M z}{2},M\bigr)\to H_j(z)/j!$, where $H_j$ is the $j$th (probabilists') Hermite polynomial. The scaled expressions are fixed-degree polynomials whose coefficients converge, and $Z_n$ converges in moments to $Z\sim N(0,1)$, so applying this first to the inner and then to the outer polynomial gives
\[
   Z^{\Kcal}_l=\frac{M^{l/2}}{l!}
     \E\left[H_l\!\left(\frac{H_k(Z)}{\sqrt{k!}}\right)\right]+o_n(n^{kl/2})
   =\alpha_{k,l}n^{kl/2}+o_n(n^{kl/2}),
\]
using $M=\binom nk\sim n^{k}/k!$, which is~\eqref{app:eq:Enk-asymp}. Since $\alpha_{k,l}$ is a fixed multiple of $\E[H_l(H_k(Z)/\sqrt{k!})]$, it vanishes exactly when no proper even $l$-cover exists, i.e.\ when $l$ is inadmissible.

It remains to estimate $\alpha_{k,l}$ for large $l$. View the $kl$ factors in $\E[Z^{kl}]$ as $l$ boxes of size $k$. By Wick's formula $(kl-1)!!$ counts all pairings of these factors. Put $A_{k,l}:=\E[H_k(Z)^{l}]$; this counts those with no pair inside a box. The number of within-box pairs in a uniform pairing converges to a Poisson variable of mean $(k-1)/2$ --- its $r$th factorial moment tends to $((k-1)/2)^{r}$ --- whence $A_{k,l}=\bigl(e^{-(k-1)/2}+o_l(1)\bigr)(kl-1)!!$.

Let $P^\star_{k,l}$ count pairings with neither an internal box pair nor two boxes paired completely to one another. Intersections of complete-pairing events correspond exactly to matchings of disjoint pairs of boxes, so inclusion--exclusion gives
\begin{equation}\label{eq:outer-hermite-inclusion-exclusion}
 P^\star_{k,l}
 =\sum_{r=0}^{\lfloor l/2\rfloor}(-1)^r
    \frac{l!}{2^rr!(l-2r)!}(k!)^r A_{k,l-2r}.
\end{equation}
The probabilists' identity $H_l(y)=l!\sum_r(-1)^ry^{l-2r}/(2^rr!(l-2r)!)$ therefore yields exactly
\[
 \alpha_{k,l}=\frac{P^\star_{k,l}}{l!(k!)^l}.
\]
Moreover, a union bound over the $\binom l2$ possible completely paired box pairs gives
\[
 0\le A_{k,l}-P^\star_{k,l}
 \le\binom l2 k!\,A_{k,l-2},
 \qquad
 \frac{A_{k,l}-P^\star_{k,l}}{A_{k,l}}=O_k(l^{2-k})=o_l(1).
\]
Thus the outer Hermite polynomial changes the pairing count only by this relative error. Inserting $(kl-1)!!=(kl)!/(2^{kl/2}(kl/2)!)$ and Stirling's formula into \eqref{app:eq:Enk-asymp} gives
\[
   \alpha_{k,l}=\bigl(L_k+o_l(1)\bigr)l^{-1/2}
   \left(\frac{k^{k/2}e^{1-k/2}}{k!}\,l^{k/2-1}\right)^{l},
   \qquad L_k=\frac{e^{-(k-1)/2}}{\sqrt\pi},
\]
which is~\eqref{app:eq:alpha-asymp}. Combining this formula with Stirling's identity for $\Gamma((kl+1)/2)$ yields~\eqref{app:eq:Enk-gamma}. The factor $e^{-(k-1)/2}$ is essential: omitting the Hermite exclusion terms would, for example, give the wrong fixed-level constant already at $(k,l)=(4,3)$.
\end{proof}

\subsection{Expected Cycles in Random Uniform Hypergraphs}
\label{app:expected}

Now sample $m$ hyperedges uniformly from $\binom{[n]}{k}$ and let $Z_l$ be the number of proper even $l$-covers in the resulting random hypergraph. Its mean follows from Proposition~\ref{prop:EZ} together with Theorem~\ref{thm:Enk}.

\begin{theorem}\label{thm:EZell}
For fixed $k>2$ and admissible $l$, with $m=Cn^{k/2}$,
\begin{equation}\label{app:eq:EZ-asymp}
   \E_\Hcal[Z_l]
     =\binom{m}{l}\frac{Z^{\Kcal}_l}{\binom{M}{l}}
     \;=\;\bigl(\alpha_{k,l}+o_n(1)\bigr)\,(C\,k!)^{l}
     \;=\;\mu_l+o_n(1),
\end{equation}
so that, for large admissible $l$,
\begin{equation}\label{app:eq:EZ-large-l}
   \mu_l=\bigl(L_k+o_l(1)\bigr)\,l^{-1/2}
     \left(\frac{m\,l^{k/2-1}}{\gamma_k\,n^{k/2}}\right)^{l},
   \qquad \gamma_k=k^{-k/2}e^{k/2-1} .
\end{equation}
\end{theorem}

\begin{proof}
The first equality is Proposition~\ref{prop:EZ}. For the second, $l$ is fixed and $m\ll M$, so $\binom ml/\binom Ml=(1+o_n(1))m^{l}/M^{l}$; inserting $M=\binom nk=(1+o_n(1))n^{k}/k!$, $m=Cn^{k/2}$ and $Z^{\Kcal}_l=(\alpha_{k,l}+o_n(1))n^{kl/2}$ gives $\E_\Hcal[Z_l]=(\alpha_{k,l}+o_n(1))(Ck!)^{l}=\mu_l+o_n(1)$, which is the definition of $\mu_l$ used in~\eqref{eq:momentratio}. Equation \eqref{app:eq:EZ-large-l} then follows from~\eqref{app:eq:alpha-asymp} and $a_k=1/(\gamma_kk!)$.
\end{proof}

In particular $\mu_l\rho^{2l}\gg1$ is equivalent, to leading exponential order, to $m\gg\gamma_k\rho^{-2}n(n/l)^{k/2-1}$, which is~\eqref{eq:detthreshold}; the exact form of the resulting condition, with its constant, is \eqref{eq:detexact}.

\section{Proof of the Recovery Theorem}
\label{app:recovery}

We recall the notation needed for the recovery argument. The instance contains $m=Cn^{k/2}$ clauses on $n$ variables, each with a support of size $k$, and the HCE level $l$ is the number of clauses in a witness. The planted assignment is $x\in\{\pm1\}^n$, and $\rho$ is the mean of each independent noise sign. For a clause-index set $I\subseteq[m]$, its boundary is $\partial I:=\triangle_{a\in I}S_a$, the set of variables occurring an odd number of times among its supports. A level-$l$ $(i,j)$-witness is an $l$-element index set with boundary $\{i,j\}$; the recovery matrix $A^{(l)}$ sums its label products over all such witnesses and has zero diagonal, as in~\eqref{eq:Adef}.

\paragraph*{What recovery requires beyond detection.} Detection (Appendix~\ref{app:detection}) only needs the first two moments of a scalar statistic, the average $X(l)$ over even covers, whose overlaps are controlled by the counting estimate of Lemma~\ref{lem:cover-intersect-bound}. Recovery needs four additional ingredients. (i) \emph{A global matrix bound}: the statistic is the matrix $A^{(l)}$, so the residual must be controlled in operator norm; entrywise variances do not suffice, because witnesses in different entries share clauses and can create localized spectral contributions. (ii) \emph{Canonical centering and trace moments}: a Hoeffding decomposition of the residual (Subsection~\ref{app:rec-hoeffding}) ensures that only trace diagrams in which every clause index is repeated survive, and sequential exposure with pattern counting (Subsections~\ref{app:rec-exposure}--\ref{app:rec-connected}) bounds high trace moments, at the cost of the $(\log n)^{a}$ factor. (iii) \emph{Signal strength and alignment}: the witness count $N_l$ (Appendix~\ref{app:N_ell}) fixes the spike $\theta_l=\rho^l d_l$, and Weyl's and the Davis--Kahan inequalities turn spectral separation into overlap with $x$ (Subsection~\ref{app:rec-conclusion}). (iv) \emph{Boundary sectors}: witnesses whose endpoints lie in different support components vanish for even $k$ but not for odd $k$, where they are removed by support-based trimming (Subsection~\ref{app:rec-odd}). Recovery is also proved under i.i.d.\ clause sampling rather than the without-replacement model used for detection.

Here $W_l(i,j)$ denotes the number of level-$l$ $(i,j)$-witnesses, $N_l:=\E W_l(i,j)$ is its expectation, and $d_l:=nN_l$ is the corresponding degree scale. Thus $\theta_l=\rho^ld_l$ is the rank-one signal strength. To separate the planted signs from the fluctuations, recall $D_x:=\diag(x)$ and the gauged matrix $\widehat A^{(l)}:=D_xA^{(l)}D_x$. Its exactly centered residual is $\widetilde R^{(l)}:=\widehat A^{(l)}-\E\widehat A^{(l)}$. This differs from the residual $R^{(l)}:=A^{(l)}-\rho^lN_lxx^\top$ used in the recovery theorem only by conjugation and the deterministic diagonal shift in~\eqref{eq:shift}. Unless supports are explicitly conditioned on, expectations in this appendix average over both the sampled supports and the noise, with $x$ fixed.

This appendix proves Theorem~\ref{thm:recovery} and Corollary~\ref{cor:recovery-threshold}, together with the odd-$k$ trimming result Theorem~\ref{thm:recovery-odd} and Corollary~\ref{cor:recovery-threshold-odd}. Ground-variable labels (elements of $[n]$) and clause-index labels (elements of $[m]$) are kept strictly distinct throughout. The even-$k$ argument consists of the boundary, component, Hoeffding, diagram, exposure and connected-moment subsections \ref{app:rec-setup}--\ref{app:rec-conclusion}; the odd-$k$ contraction is isolated in Subsection~\ref{app:rec-odd}. Throughout, $B$ denotes a positive constant depending only on $k$ and $l$ whose value may change from line to line, $a_0=(k+1)l+2$, $a=a_0+1=(k+1)l+3$, and for a random matrix $M$ and an even integer $q\ge2$ we write
\begin{equation}\label{eq:app-moment-notation}
   \mathcal M_q(M):=\bigl(\E\tr|M|^{q}\bigr)^{1/q},
   \qquad
   \mathcal N_q(M):=\bigl(\E\opnorm{M}^{q}\bigr)^{1/q},
   \qquad
   \mathcal N_q(M)\le \mathcal M_q(M)\le n^{1/q}\mathcal N_q(M).
\end{equation}
In this notation, $|M|:=(M^\top M)^{1/2}$ is the matrix absolute value, $\tr$ denotes the trace, and $\opnorm{M}$ is the operator norm. Thus $\mathcal M_q$ measures a trace moment, while $\mathcal N_q$ measures an operator-norm moment; the displayed inequalities let us pass from the former to the latter.

\subsection{Setup, boundary probabilities and the exact entry variance}
\label{app:rec-setup}

We fix the planted vector $x\in\{\pm1\}^n$ and condition on it throughout. Recovery uses the i.i.d.\ sampling convention of Section~\ref{sec:prelim}:
\begin{enumerate}[nosep,leftmargin=1.6em]
\item the supports $S_1,\dots,S_m$ are drawn i.i.d.\ uniformly from $\binom{[n]}{k}$ --- repeated supports are retained and carry independent labels, so that clauses are indexed by $[m]$ rather than by a hypergraph;
\item the noise $\eta_1,\dots,\eta_m\in\{\pm1\}$ is independent of the supports and of itself, with $\E[\eta_a]=\rho\in[0,1]$ and $\eta_a^2=1$.
\end{enumerate}
Each observation is $b_a=\eta_a\prod_{v\in S_a}x_v$, and $A^{(l)}$ is as in~\eqref{eq:Adef}. The two sources of randomness are tracked separately: the witness \emph{structure} (the event $\partial I=\{i,j\}$) depends only on the supports, whereas the witness \emph{value} $\prod_{a\in I}b_a$ involves both supports and noise. This separation drives the entire argument.

For $I\subseteq[m]$ with $\partial I=\{i,j\}$ we have $\prod_{a\in I}b_a=x_ix_j\prod_{a\in I}\eta_a$, which gives the conditional mean~\eqref{eq:condmean}; averaging over the supports and using permutation symmetry gives the rank-one decomposition~\eqref{eq:decomposition} with the normalization~\eqref{eq:normalization}. It is convenient to write, for $w\in\F_2^n$ and $0\le r\le l$,
\begin{equation}\label{eq:app-pr}
   p_r(w):=\PP\bigl[\,\partial\{1,\dots,r\}=w\,\bigr],
   \qquad
   w_{ij}:=\mathbf 1_{\{i\}}+\mathbf 1_{\{j\}}.
\end{equation}
Here we identify a boundary set with its indicator vector in $\F_2^n$, so vector addition is symmetric difference, $\varnothing$ denotes the zero vector, and $|w|$ is the Hamming weight of $w$. The quantity $p_r(w)$ is therefore the probability that $r$ independent supports have the prescribed boundary $w$, and $w_{ij}$ represents the two-point boundary $\{i,j\}$. In particular, $N_l=\binom ml p_l(w_{ij})$, $d_l=nN_l$ and $\theta_l=\rho^ld_l$. Fixing $i\ne j$ and conditioning on the supports, the same computation as in Proposition~\ref{prop:covariance} gives
\begin{equation}\label{eq:app-condvar}
   \Var\bigl(A^{(l)}_{ij}\mid S_1,\dots,S_m\bigr)
   =(1-\rho^{2l})\,W_l(i,j)
    +\sum_{\substack{I,I'\,:\,\partial I=\partial I'=\{i,j\}\\ I\ne I'}}
      \bigl(\rho^{|I\triangle I'|}-\rho^{2l}\bigr),
\end{equation}
and averaging over the supports yields the exact overlap expansion in Proposition~\ref{prop:overlap}. For $l\ge4$, its endpoint scales include both $N_l=d_l/n$ and $n^{-1}\rho^{2l-4}d_l^{2-2/l}$, as summarized by Theorem~\ref{thm:second}. Such entrywise information is \emph{not} sufficient for an operator-norm bound because overlapping witnesses can localize on a few rows. The remainder of this appendix therefore bounds high trace moments of the exactly centered residual $\widetilde R^{(l)}$ of~\eqref{eq:shift}. We begin with the exact scale of the objects appearing in~\eqref{eq:app-condvar}.

\begin{lemma}[Prescribed boundary probability]\label{lem:boundary}
For fixed $r,k$ and prescribed $w$ of Hamming weight $h$,
\begin{equation}\label{eq:boundary-upper}
   p_r(w)=O_{k,r,h}\bigl(n^{-(kr+h)/2}\bigr).
\end{equation}
For feasible $h\in\{0,2\}$ this is a matching $\Theta$ estimate. The feasibility tests are
\begin{align}
   p_r(\varnothing)>0
     &\iff r=0\ \text{or}\ (r\ge2\ \text{and $kr$ is even}),
     \label{eq:zero-feasible}\\
   p_r(w)>0\quad(|w|=2)
       &\iff kr\ \text{is even and}\
          \bigl(r\ge2\ \text{or}\ (r=1\ \text{and}\ k=2)\bigr).
     \label{eq:two-feasible}
\end{align}
\end{lemma}

\begin{proof}
An ordered $r$-tuple of supports has $kr$ clause--ground-variable incidence slots. Each prescribed odd-degree variable consumes at least one slot and every other used variable consumes at least two; hence at most $(kr-h)/2$ unprescribed ground-variable labels occur. Counting labels and dividing by $\binom nk^{r}$ gives~\eqref{eq:boundary-upper}. For a feasible zero- or two-point boundary, reserve the prescribed slots. The remaining row counts have even total and no row count exceeds the sum of the others, so the slots can be paired so that every pair joins two distinct rows (for instance by repeatedly pairing slots from the two rows of largest positive count). Assigning a distinct fresh ground variable to each pair produces valid $k$-sets and the matching lower bound. Parity and the single-row exception give \eqref{eq:zero-feasible}--\eqref{eq:two-feasible}.
\end{proof}

\begin{definition}[Overlap probability with prescribed final boundary]
For $0\le u\le l$ and a final boundary $w\in\F_2^n$, set
\begin{equation}\label{eq:app-Gamma}
   \Gamma_u(w):=\sum_{v\in\F_2^n}p_u(v)\,p_{l-u}(w+v)^2 .
\end{equation}
Equivalently, $\Gamma_u(w)$ is the probability-weighted contribution of two $l$-clause witnesses that share exactly $u$ clauses and whose common final boundary is $w$.
\end{definition}

\begin{proposition}[Exact overlap formula]\label{prop:overlap}
For $i\ne j$,
\begin{equation}\label{eq:overlap}
   \E\bigl(\widetilde R^{(l)}_{ij}\bigr)^{2}
   =\sum_{u=0}^{l}\Xi_u
     \Bigl(\rho^{2(l-u)}\Gamma_u(w_{ij})-\rho^{2l}p_l(w_{ij})^2\Bigr),
   \qquad
   \Xi_u=\binom ml\binom lu\binom{m-l}{l-u},
\end{equation}
and the $u=0$ summand vanishes.
\end{proposition}

\begin{proof}
Expand the square and group ordered pairs of witnesses by $u=|I\cap I'|$. Conditionally on the boundary $v$ of the $u$ common clauses, both private parts must have boundary $v+w_{ij}$. Shared signs are squared to one and private signs have mean $\rho$, which produces $\rho^{2(l-u)}\Gamma_u(w_{ij})$; the centering subtracts $\rho^{2l}p_l(w_{ij})^2$. At $u=0$ the two witnesses are independent and the two terms cancel exactly.
\end{proof}

\begin{definition}[Active overlaps and variance proxy]
An overlap size $u\in\{1,\dots,l\}$ is \emph{active} if there is a $w\in\F_2^n$ with $|w|=2$ such that $p_u(w)>0$ and, in addition, $p_{l-u}(\varnothing)>0$. The associated variance proxy is
\begin{equation}\label{eq:variance-proxy}
   \mathcal V_l(d,\rho):=\sum_{\substack{u\in\{1,\dots,l\}\\ u\ \text{active}}}
   \rho^{2(l-u)}\,d^{\,2-u/l}.
\end{equation}
\end{definition}

\begin{theorem}[Sharp second moment]\label{thm:second}
Suppose $0\le d_l=nN_l\le(\log n)^{A}$ for some constant $A$. Then
\begin{equation}\label{eq:second}
   \E\tr\bigl[(\widetilde R^{(l)})^{2}\bigr]
   =\Theta_{k,l,\rho}\bigl(n\,\mathcal V_l(d_l,\rho)\bigr),
\end{equation}
and the smallest active overlap is
\begin{equation}\label{eq:classification}
   u^{\star}=
   \begin{cases}
      l, & l\in\{2,3\},\ (k,l)\ne(2,3),\\
      1, & k=2,\ l\ge3,\\
      2, & k\ge3,\ l\ge4 .
   \end{cases}
\end{equation}
\end{theorem}

\begin{proof}
For active $u$ the leading term of $\Gamma_u(w_{ij})$ is $v=w_{ij}$ and equals $p_u(w_{ij})p_{l-u}(\varnothing)^2$. Lemma~\ref{lem:boundary} together with $p_l(w_{ij})\asymp n^{-kl/2-1}$ gives the one-entry scale $n^{-1}\rho^{2(l-u)}d_l^{\,2-u/l}$. Writing $h=|v|$ and $\alpha=|w_{ij}\cap v|$, summing over all $v$ with this profile produces, apart from common factors, the exponent $(h-\alpha)-h/2-(h+2-2\alpha)=-h/2+\alpha-2\le-1$, with equality only at $v=w_{ij}$; every other profile loses at least a factor $n^{-1/2}$. If $u$ is inactive, the unique maximizing profile vanishes, so the whole class carries this fixed power loss and is negligible for polylogarithmic $d_l$. For $u\ge1$ the centering contribution obeys $\Xi_u p_l(w_{ij})^2=O(n^{-ku/2-2}d_l^{\,2-u/l})$, smaller than the active scale by a fixed power of $n$, and the $u=0$ term cancels exactly by Proposition~\ref{prop:overlap}. Summing the entry variances over the $n^2-n$ off-diagonal entries gives~\eqref{eq:second}, and the classification follows from~\eqref{eq:zero-feasible}--\eqref{eq:two-feasible}.
\end{proof}

For $k\ge3$ and $l\ge4$ the active terms form a subset of a finite geometric progression whose endpoints $u=2$ and $u=l$ are both present; since $k,l$ are fixed, the sum is comparable to its larger endpoint, so that uniformly in $\rho\le1$ and $d_l\ge1$
\begin{equation}\label{eq:variance-endpoints}
   \sqrt{\mathcal V_l(d_l,\rho)}
   \;\asymp\;\underbrace{d_l^{1/2}}_{\text{Wigner-like}}
   \;+\;\underbrace{\rho^{\,l-2}d_l^{\,1-1/l}}_{\text{correlations}} .
\end{equation}
The two terms of~\eqref{eq:recovery-moment} are exactly the trace-moment counterparts of these two endpoints.

\subsection{Support components and boundary sectors}
\label{app:rec-sectors}

For a witness $I$ let $\mathsf G(I)$ be the graph on its clauses that joins two clauses whose supports intersect; its connected components are called \emph{support components}.

\begin{lemma}[Boundary-component dichotomy]\label{lem:dichotomy}
If $\partial I=\{i,j\}$, then either one support component has boundary $\{i,j\}$, or two support components have boundaries $\{i\}$ and $\{j\}$. Every other component has empty boundary, and the two alternatives are disjoint.
\end{lemma}

\begin{proof}
Distinct support components use disjoint ground-variable sets, so their boundaries are disjoint. The statement lists the only partitions of the two-set $\{i,j\}$ into non-empty parts.
\end{proof}

We call the first alternative in Lemma~\ref{lem:dichotomy} \emph{boundary-connected} (bc) and the second \emph{boundary-disconnected} (bd). Let $\widehat A^{(l)}_{\mathrm{bc}}$ and $\widehat A^{(l)}_{\mathrm{bd}}$ be the corresponding sub-sums of the gauged matrix $\widehat A^{(l)}=D_xA^{(l)}D_x$, and set $R^{(l)}_{\sigma}=\widehat A^{(l)}_{\sigma}-\E\widehat A^{(l)}_{\sigma}$ for $\sigma\in\{\mathrm{bc},\mathrm{bd}\}$. Then
\begin{equation}\label{eq:sectors}
   \widetilde R^{(l)}=R^{(l)}_{\mathrm{bc}}+R^{(l)}_{\mathrm{bd}} .
\end{equation}
Boundary-connected does \emph{not} mean that the whole witness is connected: additional zero-boundary components are allowed.

\begin{proposition}[Disconnected-sector vanishing]\label{prop:vanishing}
If $k$ is even, or if $l<6$, then $R^{(l)}_{\mathrm{bd}}=0$ identically.
\end{proposition}

\begin{proof}
Let $J$ be a support component consisting of $r$ clauses. Its total number of clause--ground-variable incidences is $kr$, and if $d_J(v)$ denotes the number of clauses of $J$ containing $v$, then
\[
   kr=\sum_{v\in[n]}d_J(v)\equiv\#\{v:d_J(v)\ \text{odd}\}
   =|\partial J|\pmod 2 .
\]
Hence a component with singleton boundary has $kr$ odd, which forces both $k$ and $r$ odd; in particular no such component exists when $k$ is even. If $r=1$ then $\partial J=S_a$ has size $k\ge2$, so $r\ne1$, and since $r$ is odd every singleton-boundary component contains at least three clauses. By Lemma~\ref{lem:dichotomy} a boundary-disconnected witness contains two distinct singleton-boundary components, hence at least six clauses in total. Therefore $\widehat A^{(l)}_{\mathrm{bd}}=0$ whenever $k$ is even or $l<6$, and so is its centered version.
\end{proof}

For intuition, let $U_r(i)$ count connected $r$-clause sets of boundary $\{i\}$. At $(k,l)=(3,6)$ the matrix $\widehat A^{(6)}_{\mathrm{bd}}$ is, off the diagonal, $U_3U_3^{\top}$ minus intersecting component pairs: a low-rank contraction that is invisible to the entry-variance analysis of Theorem~\ref{thm:second}. This term is removed exactly by the trimming in Subsection~\ref{app:rec-odd}.

\subsection{Hoeffding decomposition and trace diagrams}
\label{app:rec-hoeffding}
For the boundary-connected sector, we use the abbreviation $Q_s:=Q_{\mathrm{bc},s}$ for the order-$s$ term defined in~\eqref{eq:hoeffding} below. The coefficient $\beta_s$ in~\eqref{eq:hoeffding-kernel} counts the number of $l$-element clause-index sets containing a given $s$-element set.
\paragraph*{Why a Hoeffding decomposition.} Each entry of the residual is a U-statistic of order $l$ in the i.i.d.\ clause variables $Z_a=(S_a,\eta_a)$: a sum over $l$-sets of clause indices of a fixed kernel. In the trace expansion of $\E\tr(R^{(l)})^q$, witnesses in different factors share clauses in arbitrary patterns, and the raw kernel is not centered given its other arguments, so a term in which some clause index appears in only one factor need not vanish. Bounding all such terms without cancellation would lose powers of $d_l$. The Hoeffding (ANOVA) decomposition rewrites the residual as a sum over orders $s\le l$ of \emph{canonical} U-statistics, whose kernels average to zero in each argument (Lemma~\ref{lem:hoeffding}). In the trace expansion every clause index must then occur in at least two factors (Proposition~\ref{prop:diagram-identity}), which reduces the moment bound to counting a finite family of equality diagrams. The lower orders $s<l$, weighted by $\beta_s$, capture the fluctuations created by witnesses that share only some of their clauses.

\begin{definition}[One-witness and sector kernels]\label{def:sector-kernels}
Write a clause variable as $z_t=(S_t,\eta_t)$, where $S_t\in\binom{[n]}k$ is its support and $\eta_t\in\{\pm1\}$ its noise sign. The symmetric \emph{one-witness kernel} is the matrix
\begin{equation}\label{eq:one-witness-kernel}
   H(z_1,\dots,z_l)_{ij}
   =\mathbf 1\{S_1\triangle\cdots\triangle S_l=\{i,j\}\}\prod_{t=1}^{l}\eta_t
   \quad(i\ne j),\qquad H_{ii}=0 ,
\end{equation}
i.e.\ the contribution of a single unordered set of $l$ clause samples before summing over $I\in\binom{[m]}{l}$. Form the support-intersection graph on the argument positions $[l]$, joining $\alpha$ and $\beta$ when $S_\alpha\cap S_\beta\ne\varnothing$. Define $H_{\mathrm{bc}}$ by retaining the entry in~\eqref{eq:one-witness-kernel} when $i$ and $j$ lie in the same support component, and $H_{\mathrm{bd}}$ by retaining it when they lie in two different components; all other entries are set to zero, so that $H=H_{\mathrm{bc}}+H_{\mathrm{bd}}$ pointwise.

For $1\le s\le l$ and $\Lambda\subseteq[s]$ --- a set of \emph{argument positions}, equivalently clause slots, not of ground variables --- put $w_t=z_t$ for $t\in\Lambda$ and $w_t=(S_t',\varepsilon_t)$ for $t\in[l]\setminus\Lambda$, and define the projection
\begin{align}\label{eq:sector-projection}
   (H_\sigma)_\Lambda(z_\Lambda)
   &:=\E_{(Z_t')_{t\in[l]\setminus\Lambda}}H_\sigma(w_1,\dots,w_l)\notag\\
   &=\frac{1}{\binom nk^{\,l-|\Lambda|}}
     \sum_{(S_t')_{t\notin\Lambda}}\ \sum_{(\varepsilon_t)_{t\notin\Lambda}}
     \Bigl(\prod_{t\notin\Lambda}\tfrac{1+\rho\varepsilon_t}{2}\Bigr)
     H_\sigma(w_1,\dots,w_l) .
\end{align}
Thus the arguments in $\Lambda$ are held fixed while every argument position outside $\Lambda$ is independently averaged over both its uniform $k$-support and its noise sign.
\end{definition}

With $Z_a=(S_a,\eta_a)$, define the order-$s$ canonical kernel and the multiplicity coefficient
\begin{equation}\label{eq:hoeffding-kernel}
   h_{\sigma,s}(z_1,\dots,z_s)
   =\sum_{\Lambda\subseteq[s]}(-1)^{s-|\Lambda|}(H_\sigma)_\Lambda(z_\Lambda),
   \qquad
   \beta_s=\binom{m-s}{l-s} .
\end{equation}

\begin{lemma}[Sectorwise Hoeffding decomposition]\label{lem:hoeffding}
Every $h_{\sigma,s}$ is canonical: for each argument position $r\in[s]$,
\begin{equation}\label{eq:canonicality}
   \E_{Z_r}\bigl[h_{\sigma,s}(z_1,\dots,z_{r-1},Z_r,z_{r+1},\dots,z_s)\bigr]=0,
\end{equation}
where $Z_r$ is a fresh clause variable and the remaining arguments are fixed. Moreover
\begin{equation}\label{eq:hoeffding}
   R^{(l)}_{\sigma}=\sum_{s=1}^{l}Q_{\sigma,s},
   \qquad
   Q_{\sigma,s}=\beta_s\sum_{J\in\binom{[m]}{s}}h_{\sigma,s}(Z_J).
\end{equation}
In a trace expansion, canonicality annihilates every clause-index equality block that occurs in only one matrix factor.
\end{lemma}

\begin{proof}
Fix $r\in[s]$ and average~\eqref{eq:hoeffding-kernel} over its $r$th argument. Group the subsets $\Lambda\subseteq[s]$ into pairs $\Lambda'$ and $\Lambda'\cup\{r\}$ with $\Lambda'\subseteq[s]\setminus\{r\}$. The projection $(H_\sigma)_{\Lambda'}$ does not depend on the $r$th argument, while averaging the fixed argument $Z_r$ in $(H_\sigma)_{\Lambda'\cup\{r\}}$ turns it back into a fresh clause variable, so $\E_{Z_r}(H_\sigma)_{\Lambda'\cup\{r\}}(z_{\Lambda'},Z_r) =(H_\sigma)_{\Lambda'}(z_{\Lambda'})$. The two members of each pair carry opposite signs in~\eqref{eq:hoeffding-kernel} and cancel, proving~\eqref{eq:canonicality}.

Next fix one $l$-clause witness with clause-index set $I$. M\"obius inversion on the subset lattice gives $H_\sigma(Z_I)-\E H_\sigma=\sum_{\varnothing\ne J\subseteq I}h_{\sigma,|J|}(Z_J)$. Summing over all $I\in\binom{[m]}{l}$, a fixed $s$-set $J$ occurs once for each $l$-set containing it, i.e.\ $\binom{m-s}{l-s}=\beta_s$ times; grouping by $s=|J|$ proves~\eqref{eq:hoeffding}. Finally, if a clause index occurs in only one matrix factor of a trace product, condition on all other clause variables and integrate that one: the factor vanishes by~\eqref{eq:canonicality}, hence so does the whole term.
\end{proof}

Fix an even $q\ge2$ and Hoeffding orders $\bm s=(s_1,\dots,s_q)\in[l]^{q}$, and let $\Omega_{\bm s}=\{(t,\alpha):1\le t\le q,\ 1\le\alpha\le s_t\}$ be the resulting slot set. Here $t$ labels a matrix factor in the trace product and $\alpha$ labels one of its clause arguments. A partition $\pi$ of $\Omega_{\bm s}$ records which slots carry the same clause index. It is \emph{admissible} if no block contains two slots with the same first coordinate and every block has size at least two; write $\mathfrak P(\bm s)$ for the set of admissible partitions and $[t,\alpha]_\pi$ for the block containing $(t,\alpha)$. An injection $\phi:\pi\hookrightarrow[m]$ assigns distinct clause indices to distinct blocks, hence realizes exactly the equality pattern encoded by $\pi$. For factor $t$ set
\begin{equation}\label{eq:J-phi}
   J_t(\phi):=\bigl(\phi([t,1]_\pi),\dots,\phi([t,s_t]_\pi)\bigr),
\end{equation}
whose entries are distinct by admissibility and injectivity, and define the diagram value
\begin{equation}\label{eq:diagram-value}
   D_\sigma(\bm s,\pi)
   =\frac{\prod_{t=1}^{q}\beta_{s_t}}{\prod_{t=1}^{q}s_t!}
    \sum_{\phi:\pi\hookrightarrow[m]}
    \E\tr\prod_{t=1}^{q}h_{\sigma,s_t}\bigl(Z_{J_t(\phi)}\bigr).
\end{equation}

\begin{proposition}[Exact sectorwise trace-diagram identity]
\label{prop:diagram-identity}
For $\sigma\in\{\mathrm{bc},\mathrm{bd}\}$ and every even $q\ge2$,
\begin{equation}\label{eq:diagram-identity}
   \E\tr\bigl[(R^{(l)}_{\sigma})^{q}\bigr]
   =\sum_{\bm s\in[l]^{q}}\ \sum_{\pi\in\mathfrak P(\bm s)}D_\sigma(\bm s,\pi).
\end{equation}
\end{proposition}

\begin{proof}
Insert~\eqref{eq:hoeffding} into each of the $q$ trace factors and choose $\bm s$; since $h_{\sigma,s_t}$ is symmetric, replace each unordered $J_t$ by an ordered tuple of distinct clause indices and divide by $s_t!$. Every term then assigns one clause index to each slot of $\Omega_{\bm s}$, and grouping slots with equal indices produces a unique partition $\pi$. Indices within one tuple are distinct, so no block contains two slots from the same factor; and a singleton block would make the term vanish by Lemma~\ref{lem:hoeffding}. Hence every surviving $\pi$ is admissible. Conversely each pair $(\pi,\phi)$ reconstructs one term through~\eqref{eq:J-phi}. Collecting terms with the same $\pi$ gives~\eqref{eq:diagram-value}, and summing over $\bm s$ and $\pi$ proves~\eqref{eq:diagram-identity}.
\end{proof}

Thus the exact trace expansion retains arbitrary multiway blocks: different trace factors may reuse clauses exposed by earlier factors. To count the ground-variable labels of such a pattern we expose factors sequentially and charge only the clauses appearing for the first time.

\subsection{Sequential exposure and pattern counting}
\label{app:rec-exposure}

\begin{lemma}[Sequential exposure]\label{lem:exposure}
Suppose some clauses have already been exposed and let $O$ contain every ground variable incident to them. Expose one further witness; let $f$ be the number of its clauses whose indices have not appeared before (equivalently, the number of fresh independent supports), and let $v_{\rm new}$ be the number of its ground variables outside $O$.
\begin{enumerate}[label=\textup{(\alph*)},nosep,leftmargin=2.0em]
\item If its boundary is $\{i,j\}$ with $i\in O$ and $i,j$ in one support component, then $v_{\rm new}\le kf/2$.
\item A singleton-boundary witness satisfies $v_{\rm new}\le(kf+1)/2$; if it is connected, its root is old and $f>0$, then $v_{\rm new}\le(kf-1)/2$.
\item A zero-boundary witness satisfies $v_{\rm new}\le kf/2$.
\end{enumerate}
\end{lemma}

\begin{proof}
The $f$ fresh clauses carry $kf$ incidences in total. Let $b$ be the number of these whose ground-variable endpoint lies in $O$; since previously exposed clauses cannot contain a variable outside $O$, the fresh clauses contribute exactly $kf-b$ incidences to variables outside $O$. Let $r_\partial$ be the number of boundary variables of the new witness lying outside $O$. Every outside non-boundary variable has positive even degree, hence degree at least two, and every outside boundary variable has degree at least one, so
\begin{equation}\label{eq:sequential-exposure-count}
   kf-b\ \ge\ 2(v_{\rm new}-r_\partial)+r_\partial=2v_{\rm new}-r_\partial,
   \qquad\text{i.e.}\qquad
   v_{\rm new}\le\frac{kf-b+r_\partial}{2}.
\end{equation}
For (a), $i\in O$ leaves at most $j$ outside. If $j\in O$ then $r_\partial=0$ and~\eqref{eq:sequential-exposure-count} gives $v_{\rm new}\le kf/2$. If $j\notin O$ then $r_\partial=1$, and because $i,j$ lie in one support component a path from $i$ to $j$ must cross from $O$ to its complement; at the first crossing a fresh clause meets a variable of $O$, so $b\ge1$ and the terms $-b$ and $+r_\partial$ cancel. For (b), $r_\partial\le1$ and $b\ge0$ give $v_{\rm new}\le(kf+1)/2$; in the rooted connected case, the root is the unique boundary variable, so $r_\partial=0$, and if $f>0$ connectivity forces some fresh clause to meet an old variable, so $b\ge1$. For (c), the boundary is empty, so $r_\partial=0$; together with $b\ge0$, equation~\eqref{eq:sequential-exposure-count} gives $v_{\rm new}\le kf/2$.
\end{proof}

\begin{corollary}[Completed connected trace patterns]\label{cor:trace-vertices}
A completed product of $q$ boundary-connected entries along a closed trace walk, using $N$ independent supports and $v$ ground variables, satisfies
\begin{equation}\label{eq:trace-vertices}
   v\le1+kN/2 .
\end{equation}
\end{corollary}

\begin{proof}
Expose the first trace variable, then the witnesses in cyclic order. Each left endpoint is then old, so Lemma~\ref{lem:exposure}(a) applies, and each support is new exactly once.
\end{proof}

\begin{lemma}[Pattern count]\label{lem:patterns}
There is a constant $C_k>0$, depending only on $k$, such that an expansion with $q$ factors has at most $(C_klq)^{a_0q}$ abstract clause- and ground-variable equality patterns, where $a_0=(k+1)l+2$. A pattern with $c$ sampled clause labels, $N$ actually used independent supports and $v$ ground-variable labels contributes at most
\begin{equation}\label{eq:pattern-cost}
   C_k^{lq}m^{c}n^{v}n^{-kN}
\end{equation}
before deterministic coefficients.
\end{lemma}

\begin{proof}
An abstract pattern records which of the at most $ql$ clause slots are equal and, separately, which of the at most $(kl+2)q$ incidence and trace positions represent the same ground variable, i.e.\ two set partitions of at most $a_0q$ positions; since the number of partitions of $r$ objects is at most $r^{r}$, there are at most $(C_klq)^{a_0q}$ of them. Once a pattern is fixed, its $c$ clause classes and $v$ ground-variable classes can be labeled in at most $m^{c}$ and $n^{v}$ ways, and each of the $N$ distinct independent supports carries probability $\binom nk^{-1}$; as $N\le ql$, $\binom nk^{-N}\le C_k^{lq}n^{-kN}$ for all sufficiently large $n$.
\end{proof}

\begin{lemma}[Tensorized completion of a canonical diagram]
\label{lem:tensorized-completion}
Fix $1\le s\le l$, an even $q\ge2$, and a surviving clause-index partition of the $qs$ core slots in the expansion of $\E\tr Q_s^q$. Let $c$ be its number of blocks. For factor $t$, choose $\Lambda_t\subseteq[s]$ in the expansion of the canonical kernel, and put
\[
   \Lambda:=\sum_{t=1}^q|\Lambda_t|,
   \qquad D:=qs-\Lambda.
\]
Let $c'$ be the number of clause-index blocks containing at least one retained slot and let $F:=c-c'$ be the number of completely discarded blocks. Then the corresponding projection term has the following properties.
\begin{enumerate}[label=\textup{(\roman*)},nosep,leftmargin=2.0em]
\item On an enlarged product probability space it is the expectation of a product of $q$ genuine boundary-connected $l$-clause kernels. Retained slots in the same block use the same sampled clause variable, and every omitted slot is filled by a fresh auxiliary clause variable; all auxiliary variables are mutually independent and independent of the sampled variables.
\item The number of independent support draws and the exponent used in the sampled clause-label upper bound are respectively
\begin{equation}\label{eq:completed-counts}
   N=c'+ql-\Lambda,
   \qquad E_0=c+q(l-s).
\end{equation}
Every block of a surviving partition has size at least two, so $D\ge2F$ and
\begin{equation}\label{eq:completed-surplus}
   N-E_0=D-F\ge F\ge0.
\end{equation}
\item Distinct independent draws remain distinct probability factors even if their realized $k$-supports coincide. After fixing a ground-variable equality pattern with $v$ labels, its absolute contribution, including the sum over sampled clause indices and the factors $\beta_s^q$, is at most
\begin{equation}\label{eq:completed-pattern-cost}
   C_k^{lq}m^{E_0}n^v n^{-kN}.
\end{equation}
The pattern may contain additional zero-boundary support components; they are included in the same $N$ draws and $v$ labels.
\end{enumerate}
The projection signs have absolute value one, there are at most $2^{sq}$ choices of $(\Lambda_1,\dots,\Lambda_q)$, and the abstract pattern count is bounded by Lemma~\ref{lem:patterns}.
\end{lemma}

\begin{proof}
For a fixed factor, Definition~\ref{def:sector-kernels} writes the selected projection as an expectation over its omitted clause arguments. Conditional on all retained sampled clauses, the product of these $q$ expectations equals one expectation over $q$ mutually independent collections of auxiliary arguments. This is the product-measure tensorization identity. It proves (i) and gives one independent draw for each of the $c'$ retained blocks and one for each of the $ql-\Lambda$ auxiliary arguments, proving the formula for $N$.

The injection assigning sampled clause indices to the $c$ blocks has at most $m^c$ choices, while $\beta_s^q=\binom{m-s}{l-s}^q\le m^{q(l-s)}$. This proves the formula for $E_0$. A completely discarded block contributes all of its slots to $D$; by canonicality every surviving block has size at least two. Hence $D\ge2F$, and direct substitution gives $N-E_0=D-F$, proving (ii).

The $N$ source variables just counted are independent draws from $\binom{[n]}k$. Equality of two realized support values is therefore an event in the ground-variable equality pattern, not an identification of their source variables; it still carries two factors $\binom nk^{-1}$. Once a pattern is fixed, label its $v$ ground-variable classes in at most $n^v$ ways and bound $\binom nk^{-N}\le C_k^{lq}n^{-kN}$. This proves \eqref{eq:completed-pattern-cost}. Each completed kernel is boundary-connected by construction, but its other support components may have zero boundary; no step of this count discards them. The remaining coefficient and pattern-count statements follow from the M\"obius expansion and Lemma~\ref{lem:patterns}.
\end{proof}

\subsection{The boundary-connected moment bound}
\label{app:rec-connected}

Write $Q_s=Q_{\mathrm{bc},s}$. Since $m=Cn^{k/2}$, the density $C$ plays the role of the natural expansion parameter, and $d_l\asymp C^{l}$ at fixed $k,l$.

\begin{theorem}[All connected Hoeffding orders]\label{thm:orders}
There is a constant $D_k>0$, depending only on $k$, such that, for $1\le s\le l$, $C>0$, $\rho\in[0,1]$, even $q\ge2$, and all sufficiently large $n$ (depending on $k,l$),
\begin{equation}\label{eq:order-bound}
   \mathcal M_q(Q_s)\le n^{1/q}(D_klq)^{a}\rho^{\,l-s}
   \begin{cases}
      C^{\,l-s+s/q}, & 0<C\le1,\\
      C^{\,l-s/2}, & C\ge1,
   \end{cases}
   \qquad a=(k+1)l+3 .
\end{equation}
If $\rho=0$ and $s<l$ then $Q_s=0$.
\end{theorem}

\begin{proof}
Since $q$ is even and $Q_s$ symmetric, $\tr|Q_s|^{q}=\tr Q_s^{q}$, and by~\eqref{eq:hoeffding}
\begin{equation}\label{eq:order-trace-expansion}
   \E\tr Q_s^{q}
   =\beta_s^{q}\sum_{J_1,\dots,J_q\in\binom{[m]}{s}}
     \E\tr\bigl[h_{\mathrm{bc},s}(Z_{J_1})\cdots h_{\mathrm{bc},s}(Z_{J_q})\bigr].
\end{equation}
Regard the $s$ arguments of each of the $q$ kernels as $qs$ \emph{core slots}. Grouping slots with equal clause index gives an equality partition with, say, $c$ blocks; singleton blocks vanish by canonicality~\eqref{eq:canonicality}, \emph{before} absolute values are taken, and the $s$ slots of one factor lie in distinct blocks. Hence
\begin{equation}\label{eq:order-core-blocks}
   s\le c\le qs/2 .
\end{equation}
Fix a surviving partition and expand each canonical kernel via~\eqref{eq:hoeffding-kernel}: a term chooses $\Lambda_t\subseteq[s]$, keeping those arguments and averaging the others out. Apply Lemma~\ref{lem:tensorized-completion}. Its notation gives $N=c'+ql-\Lambda$, $E_0=q(l-s)+c$, and $N-E_0\ge0$. Moreover,
\begin{equation}\label{eq:surplus}
   q(l-s)+s\le E_0\le q(l-s/2),
\end{equation}
by~\eqref{eq:order-core-blocks}.

Corollary~\ref{cor:trace-vertices} allows at most $1+kN/2$ ground-variable labels; this includes every additional zero-boundary component. By Lemma~\ref{lem:tensorized-completion}, a fixed completed pattern is bounded by
\[
   C_k^{lq}m^{E_0}n^{1-kN/2}
   =C_k^{lq}n^{1-k(N-E_0)/2}C^{E_0}
   \le C_k^{lq}nC^{E_0},
\]
using $m=Cn^{k/2}$ and $N\ge E_0$. The two bounds on $c$ give $C^{E_0}\le C^{q(l-s/2)}$ when $C\ge1$ and $C^{E_0}\le C^{q(l-s)+s}$ when $0<C\le1$. In factor $t$, averaging the $l-|\Lambda_t|\ge l-s$ omitted noise signs contributes at most $\rho^{\,l-s}$, so the product over factors is at most $\rho^{\,q(l-s)}$. Finally Lemma~\ref{lem:patterns}, together with the choices of the sets $\Lambda_t$, gives at most $(C_klq)^{a_0q}2^{lq}$ terms. Together with the factor $C_k^{lq}$ from a fixed pattern, the factor $2^{lq}$ is absorbed into the base by enlarging $D_k$, since $a_0\ge(k+1)l$. Thus the total is at most $(D_klq)^{a_0q}$ times the displayed density and noise factors. Taking $q$th roots and using $a=a_0+1$ yields~\eqref{eq:order-bound}. If $\rho=0$ and $s<l$, every projection averages at least one omitted noise sign, so $h_{\mathrm{bc},s}=0$.
\end{proof}

Canonical cancellation must precede completion: it removes singleton core blocks without taking absolute values.

We use the following standard consequence of the self-adjoint matrix Bernstein inequality \cite{Tropp2015}. If $X_1,\dots,X_m$ are independent centered self-adjoint $n\times n$ matrices with $\opnorm{X_a}\le\mathsf L$ almost surely and $\opnorm{\sum_a\E X_a^2}\le\mathsf v$, then, for every $q\ge2$,
\begin{equation}\label{eq:matrix-bernstein-moment}
   \bigl(\E\tr|\textstyle\sum_aX_a|^q\bigr)^{1/q}
   \le n^{1/q}B\bigl(\sqrt{q\mathsf v}+q\mathsf L\bigr).
\end{equation}
It follows by integrating the usual matrix Bernstein tail bound; the numerical constant $B$ is universal.

\begin{lemma}[Improved first projection for $k\ge3$]\label{lem:first}
For even $q\ge2$ and $k\ge3$, with the constant $D_k$ of Theorem~\ref{thm:orders},
\begin{equation}\label{eq:first}
   \mathcal M_q(Q_1)\le n^{1/q}(D_klq)^a\rho^{\,l-1}
   \bigl[\sqrt q\,n^{1/2-k/4}C^{\,l-1/2}+q\,n^{1-k/2}C^{\,l-1}\bigr].
\end{equation}
\end{lemma}

\begin{proof}
For a fixed support $S$ let $P_S(i,j)$ be the probability that $S$, completed by $l-1$ fresh supports, forms a boundary-connected witness of boundary $\{i,j\}$, and put $G_S=\binom{m-1}{l-1}P_S$. If $u=|S\cap\{i,j\}|$ then Lemma~\ref{lem:boundary} gives $G_S(i,j)\le(D_kl)^a n^{-k/2-1+u}C^{\,l-1}$. Indeed, its slot count uses at most $k(l-1)+2$ positions and hence at most $(C_kl)^{k(l-1)+2}$ equality patterns, which is bounded by $(D_kl)^a$ because $a=(k+1)l+3$. Block row sums therefore yield
\begin{equation}\label{eq:G-bounds}
   \opnorm{G_S}\le(D_kl)^a n^{1-k/2}C^{\,l-1},
   \qquad \norm{G_S\mathbf 1}_2^{2}\le(D_kl)^{2a}n^{2-k}C^{\,2l-2}.
\end{equation}
The law of $G_S$ is invariant under permutations of $[n]$, so the non-negative matrix $\E G_S^{2}$ has constant row sums and therefore
\begin{equation}\label{eq:G-variance}
   \opnorm{\E G_S^{2}}=\tfrac1n\E\norm{G_S\mathbf 1}_2^{2}
   \le(D_kl)^{2a}n^{1-k}C^{\,2l-2}.
\end{equation}
Now $Q_1=\sum_aX_a$ with independent centered summands $X_a=\rho^{\,l-1}\eta_aG_{S_a}-\E[\rho^{\,l-1}\eta_aG_{S_a}]$, and the matrix Bernstein moment inequality~\eqref{eq:matrix-bernstein-moment}, with the variance and norm proxies of~\eqref{eq:G-bounds}--\eqref{eq:G-variance} gives \eqref{eq:first}.
\end{proof}

\begin{theorem}[Connected-sector bound]\label{thm:connected}
For $k\ge3$, $\rho\in[0,1]$, even $q\ge2$, $C>0$, and all sufficiently large $n$ (depending on $k,l$),
\begin{equation}\label{eq:connected-kge3}
   \mathcal M_q\bigl(R^{(l)}_{\mathrm{bc}}\bigr)\le n^{1/q}(D_klq)^{a}
   \begin{cases}
      C^{\,l/q}+\rho^{\,l-2}C^{\,l-2+2/q}, & 0<C\le1,\\
      C^{\,l/2}+\rho^{\,l-2}C^{\,l-1}, & 1<C\le n^{k/2-1},\\
      C^{\,l/2}+\rho^{\,l-1}C^{\,l-1/2}, & C>n^{k/2-1}.
   \end{cases}
\end{equation}
\end{theorem}

\begin{proof}
By~\eqref{eq:hoeffding} and the triangle inequality $\mathcal M_q(R^{(l)}_{\mathrm{bc}})\le\sum_{s=1}^{l}\mathcal M_q(Q_s)$. Sum first the orders $2\le s\le l$. If $\rho=0$, the last assertion of Theorem~\ref{thm:orders} gives $Q_s=0$ for every $s<l$, so only the $s=l$ term remains and all three displayed bounds follow immediately. Assume $\rho>0$ below. For $0<C\le1$, Theorem~\ref{thm:orders} gives, suppressing the common prefactor $n^{1/q}(D_klq)^{a}$, the geometric sequence $u_s=\rho^{\,l-s}C^{\,l-s+s/q}$ with ratio $u_{s+1}/u_s=\rho^{-1}C^{-1+1/q}$, so that regardless of whether the ratio exceeds one
\begin{equation}\label{eq:connected-sparse-sum}
   \sum_{s=2}^{l}u_s\le(l-1)(u_2+u_l)
   =(l-1)\bigl(\rho^{\,l-2}C^{\,l-2+2/q}+C^{\,l/q}\bigr),
\end{equation}
the fixed factor $l-1$ being absorbed into $(D_klq)^a$. Likewise for $C\ge1$ with $w_s=\rho^{\,l-s}C^{\,l-s/2}$,
\begin{equation}\label{eq:connected-dense-sum}
   \sum_{s=2}^{l}w_s\le(l-1)\bigl(\rho^{\,l-2}C^{\,l-1}+C^{\,l/2}\bigr).
\end{equation}
It remains to absorb $Q_1$. If $C\le n^{k/2-1}$, dividing the first term of~\eqref{eq:first} by the relevant $s=2$ endpoint gives, up to a factor already present in $(D_klq)^a$, the quantity $n^{1/2-k/4}C^{\,3/2-2/q}\le1$ for $0<C\le1$ (using $q\ge2$, $k\ge3$) and $n^{1/2-k/4}C^{1/2}\le1$ for $1<C\le n^{k/2-1}$; for the second term the corresponding ratios are $n^{1-k/2}C^{\,1-2/q}\le1$ and $n^{1-k/2}\le1$. The factors $\rho\le1$, $q$ and $\sqrt q$ are absorbed into $(D_klq)^{a}$, so $\mathcal M_q(Q_1)$ is dominated by the $s=2$ term of~\eqref{eq:connected-sparse-sum} or~\eqref{eq:connected-dense-sum}. Finally, if $C>n^{k/2-1}$, apply the dense branch of Theorem~\ref{thm:orders} to all $1\le s\le l$ and repeat the endpoint estimate with endpoints $s=1$ and $s=l$.
\end{proof}

\begin{corollary}[Level-explicit dense deviation]\label{cor:level-explicit-radius}
Fix $k\ge3$, $l\ge2$, and $\gamma>0$, and let $a=(k+1)l+3$. There is a constant $D_{k,\gamma}>0$, independent of $l$, such that, whenever $1<C\le n^{k/2-1}$, for all sufficiently large $n$,
\begin{equation}\label{eq:level-explicit-radius}
 \PP\!\left[
    \opnorm{R^{(l)}_{\mathrm{bc}}}>
    D_{k,\gamma}^{a}(\log n)^a
    \bigl(C^{l/2}+\rho^{l-2}C^{l-1}\bigr)
 \right]\le n^{-\gamma}.
\end{equation}
\end{corollary}

\begin{proof}
For an even $q$, Theorem~\ref{thm:connected} and Markov's inequality give
\[
 \PP[\opnorm{R^{(l)}_{\mathrm{bc}}}>t]
 \le n\left[
    \frac{(D_klq)^a(C^{l/2}+\rho^{l-2}C^{l-1})}{t}
 \right]^q.
\]
Choose the least even integer $q\ge(\gamma+1)\log n/a$. For fixed $l$ and sufficiently large $n$, this choice satisfies $q\ge2$, $n^{(\gamma+1)/q}\le e^a$, and $lq\le C_{k,\gamma}\log n$. Taking
\[
 t=n^{(\gamma+1)/q}(D_klq)^a
    (C^{l/2}+\rho^{l-2}C^{l-1})
\]
proves~\eqref{eq:level-explicit-radius} after enlarging $D_{k,\gamma}$.
\end{proof}

\subsection{Even-\texorpdfstring{$k$}{k} residual, signal-to-noise and the threshold}
\label{app:rec-conclusion}

\begin{proposition}[Even-$k$ residual]\label{prop:app-even-recovery}
Fix even $k\ge4$, an admissible $l\ge2$, $0<C\le n^{k/2-1}$, $d_l\ge1$ and even $q\ge2$. For all sufficiently large $n$ depending on $k,l$, $R^{(l)}_{\mathrm{bd}}=0$ and
\begin{equation}\label{eq:app-even-residual}
   \mathcal M_q\bigl(\widetilde R^{(l)}\bigr)
   \le n^{1/q}(Bq)^{a}
   \Bigl[d_l^{1/2}+\rho^{\,l-2}d_l^{\,1-1/l}\Bigr].
\end{equation}
\end{proposition}

\begin{proof}
Proposition~\ref{prop:vanishing} gives $\widetilde R^{(l)}=R^{(l)}_{\mathrm{bc}}$ for even $k$. For $1<C\le n^{k/2-1}$, insert the dense branch of Theorem~\ref{thm:connected} and use the finite-$n$ calibration of Lemma~\ref{lem:finite-witness-calibration}. Its large-$m$ condition is automatic here for sufficiently large $n$: Lemma~\ref{lem:boundary} gives $d_l\le B_{k,l}C^l$, so $d_l\ge1$ bounds $C$ away from zero and hence $m=Cn^{k/2}\to\infty$. For $0<C\le1$, use the sparse branch and the facts that $d_l\ge1$ and $q\ge2$ imply $d_l^{1/q}\le d_l^{1/2}$ and $d_l^{1-2/l+2/(ql)}\le d_l^{1-1/l}$. Theorem~\ref{thm:second} and $\mathcal M_q\ge\mathcal N_q\ge\mathcal N_2\ge (n^{-1}\E\tr(\cdot)^2)^{1/2}$ show, when $d_l\le(\log n)^A$, that the same dependence on $d_l$ is sharp up to fixed constants (for $l\in\{2,3\}$ only the $d_l^{1/2}$ endpoint is active). At the corollary's threshold one may take any fixed $A\ge al$.
\end{proof}

\begin{proof}[Proof of Theorem~\ref{thm:recovery}]
Estimate~\eqref{eq:recovery-moment} is exactly~\eqref{eq:app-even-residual}. For the deviation bound, let $q_n=2\lceil\log n\rceil$ and abbreviate $\Delta_n:=d_l^{1/2}+\rho^{\,l-2}d_l^{\,1-1/l}$. Using $\opnorm{\widetilde R^{(l)}}^{q_n}\le\tr|\widetilde R^{(l)}|^{q_n}$ and Markov's inequality, for any $t>0$
\[
   \PP\bigl[\opnorm{\widetilde R^{(l)}}\ge t\bigr]
   \le t^{-q_n}\,\E\tr|\widetilde R^{(l)}|^{q_n}
   \le n\left(\frac{(Bq_n)^{a}\Delta_n}{t}\right)^{q_n},
\]
where the factor $n^{1/q_n}\le e^{1/2}$ from~\eqref{eq:app-even-residual} has been absorbed into $B$. Choosing $t=B_\gamma(\log n)^{a}\Delta_n$ with $B_\gamma$ large enough that the bracket is at most $e^{-(1+\gamma)}$ (possible because $q_n\le2\log n+2$, so $(Bq_n)^{a}\le B'(\log n)^{a}$) gives $\PP[\opnorm{\widetilde R^{(l)}}\ge t]\le n\,e^{-q_n(1+\gamma)}\le n^{-\gamma-1}$. Adding the deterministic diagonal shift of~\eqref{eq:shift}, which costs at most $\theta_l/n$, turns this into the bound~\eqref{eq:recovery-radius} for $\opnorm{R^{(l)}}$, with probability at least $1-n^{-\gamma}$.

For the correlation bound, work in the gauged frame: $D_xx=\mathbf 1$, so with $u=\mathbf 1/\sqrt n$ the decomposition~\eqref{eq:decomposition} reads
\[
   \widehat A^{(l)}
   =\theta_l\,uu^{\top}-\frac{\theta_l}{n}I+\widetilde R^{(l)} .
\]
The multiple of the identity shifts the whole spectrum and does not affect eigenvectors. On the event $\opnorm{\widetilde R^{(l)}}\le\mathfrak R_{n,\gamma}$, and under the assumption $\theta_l>(3+\varepsilon)\mathfrak R_{n,\gamma}$, Weyl's inequality separates the top eigenvalue of $\widehat A^{(l)}$ from the rest of its spectrum by a gap of at least $\theta_l-2\mathfrak R_{n,\gamma}>0$, so that eigenvalue is isolated. The Davis--Kahan $\sin\theta$ theorem \cite{DK1970,Yu2015} then gives, for a unit top eigenvector $\hat u$,
\[
   \sin\angle(\hat u,u)\le
   \frac{\mathfrak R_{n,\gamma}}{\theta_l-\mathfrak R_{n,\gamma}},
\]
and $|\langle\hat u,u\rangle|=\cos\angle(\hat u,u)= \sqrt{1-\sin^{2}\angle(\hat u,u)}$ gives~\eqref{eq:recovery-correlation}. Conjugating back by $D_x$ maps $u$ to $x/\sqrt n$ and leaves every norm involved unchanged. Choose the global sign of $\hat u$ so that $\langle\hat u,u\rangle\ge0$. With $\delta=\mathfrak R_{n,\gamma}/(\theta_l-\mathfrak R_{n,\gamma})$, the correlation bound gives $\|\hat u-u\|_2^2\le2\delta^2$. Each disagreement between $\sgn(\hat u)$ and $x$ contributes at least $1/n$ to this squared distance, proving~\eqref{eq:recovery-correlation}. Finally, $\theta_l>(3+\varepsilon)\mathfrak R_{n,\gamma}$ implies $\delta<1/2$, so the displayed binary overlap is strictly positive.
\end{proof}

The exact signal-to-noise ratio is $\SNR_l=\theta_l(1-1/n)/\opnorm{\widetilde R^{(l)}}$. Comparing $\theta_l=\rho^ld_l$ with the two endpoints in \eqref{eq:recovery-radius} gives respectively $\rho^{2l}d_l\gg(\log n)^{2a}$ and $\rho^{2l}d_l\gg(\log n)^{al}$. Since $l\ge2$, the latter is binding, which is \eqref{eq:snr}.

\begin{proof}[Proof of Corollary~\ref{cor:recovery-threshold}]
Let $D=D_{k,\gamma}$ be the constant in Corollary~\ref{cor:level-explicit-radius}, let $a_l=(k+1)l+3$, and put $\lambda=3+\varepsilon$. The witness asymptotic \eqref{app:eq:C-ell-k-scaling} implies
\[
 \kappa_{l,k}^{1/l}=c_k l^{k/2-1}(1+o_l(1))
 \quad\text{for some }c_k>0.
\]
Consequently the two suprema over admissible $l\ge2$,
\begin{align*}
 M_1&:=\sup_l D^{a_l}\kappa_{l,k}^{-(1-1/l)},\\
 M_2&:=\sup_l D^{2a_l/l}\kappa_{l,k}^{-1/l},
\end{align*}
are finite: their summands tend respectively to zero like $[D^{k+1}/(c_kl^{k/2-1})]^{l-1}$ and $D^{2(k+1)}/(c_kl^{k/2-1})$. Choose $K_{k,\gamma,\varepsilon}$ larger than a fixed multiple of $\lambda\max\{1,M_1,M_2\}$.

On the event of~\eqref{eq:level-explicit-radius}, write
\[
 t_l:=D^{a_l}(\log n)^{a_l}
       (C^{l/2}+\rho^{l-2}C^{l-1}).
\]
Since $\theta_l=(1+o_n(1))\rho^l\kappa_{l,k}C^l$, condition \eqref{eq:mainthreshold} and the definitions of $M_1,M_2$ give, for all sufficiently large $n$,
\[
 \theta_l>2\lambda t_l.
\]
Indeed, the correlated endpoint uses $K\ge\lambda D^{a_l}\kappa_{l,k}^{-(1-1/l)}$, while the Wigner endpoint uses $K\ge\lambda^{2/l}D^{2a_l/l}\kappa_{l,k}^{-1/l}$; the extra powers of $\log n$ are nonnegative because $l\ge2$.

For even $k$, $R^{(l)}_{\mathrm{bd}}=0$, so $\widetilde R^{(l)}=R^{(l)}_{\mathrm{bc}}$. Adding the deterministic diagonal shift costs $\theta_l/n$. For $n\ge2\lambda$, the preceding display yields
\[
 \theta_l>\lambda(t_l+\theta_l/n).
\]
Repeating the Weyl, Davis--Kahan, and sign-rounding argument of Theorem~\ref{thm:recovery} with radius $t_l+\theta_l/n$ gives the stated overlap. Corollary~\ref{cor:level-explicit-radius} gives failure probability at most $n^{-\gamma}$.
\end{proof}

The large-$l$ asymptotic~\eqref{app:eq:C-ell-k-scaling}, together with the $l$-independence of $K_{k,\gamma,\varepsilon}$, identifies the algebraic large-$l$ recovery curve without any hidden level-dependent prefactor. The lower bound on $n$ in the fixed-level theorem is still not uniform in growing $l$, and the exponent $a=(k+1)l+3$ grows with $l$; thus no theorem for $l=l(n)$ is claimed.

\subsection{Odd \texorpdfstring{$k$}{k}: trimming the boundary-disconnected sector}
\label{app:rec-odd}

Throughout this subsection $k\ge3$ is odd and $l$ is even. For $l\ge6$ the sector $R^{(l)}_{\mathrm{bd}}$ need not vanish. Instead of bounding its localized low-rank contribution, we remove it exactly by deleting a small, support-determined set of coordinates. For $l\in\{2,4\}$ this set is empty and no trimming is needed; for $l\ge6$, the resulting threshold contains the additional fixed-level factor $\kappa_{l,k}/\kappa^{\mathrm{bc}}_{l,k}$ (Corollary~\ref{cor:recovery-threshold-odd}).

\begin{remark}[Why the untrimmed matrix is not used]\label{rem:odd-obstruction}
This heuristic is not used below. At $l=6$, let $U_3(i)$ be the gauged signed count of connected three-clause sets with boundary $\{i\}$. Apart from sparse compatibility corrections, $\widehat A^{(6)}_{\mathrm{bd}}$ is $U_3U_3^{\top}$ off the diagonal. In the sparse-certificate regime $C^3=o(\sqrt n)$, which includes fixed and polylogarithmic $C$, Lemma~\ref{lem:boundary} gives $O(C^3\sqrt n)$ expected three-clause singleton-boundary certificates. Neglecting lower-order endpoint collisions then predicts $\E|\operatorname{supp}U_3|=\Theta(C^3\sqrt n)$ and a norm of the same order, whereas $\theta_6\asymp\rho^6C^6$. Thus the untrimmed top eigenvector can localize at polylogarithmic $C$.
\end{remark}

\begin{definition}[Singleton-boundary set and trimmed recovery matrix]
\label{def:trim}
For even $l$ define
\begin{equation}\label{eq:bad-set}
   \mathcal B_l:=\bigl\{v\in[n]:\ \partial P=\{v\}\ \text{for some }\
   r\in\{3,\dots,l-3\}\ \text{and }P\in\tbinom{[m]}r\bigr\},
   \qquad \mathcal G_l:=[n]\setminus\mathcal B_l .
\end{equation}
Let $\Pi_l:=\diag(\mathbf 1_{\mathcal G_l})$ and
\begin{equation}\label{eq:trimmed}
   \bar A^{(l)}:=\Pi_lA^{(l)}\Pi_l .
\end{equation}
We regard $\bar A^{(l)}$ as acting on $\operatorname{Ran}(\Pi_l)$, equivalently as the principal submatrix indexed by $\mathcal G_l$. The trimmed estimator takes the signs of its top eigenvector on $\mathcal G_l$ and assigns arbitrary signs on $\mathcal B_l$.
\end{definition}

Only odd values of $r$ can occur in~\eqref{eq:bad-set}; moreover $r\ne1$ because one $k$-support has boundary size $k\ge3$. Hence $\mathcal B_l=\varnothing$ for $l\in\{2,4\}$.

\begin{lemma}[Exact removal of the disconnected sector]\label{lem:trim-exact}
Every nonzero entry $(i,j)$ of $\widehat A^{(l)}_{\mathrm{bd}}$ satisfies $i,j\in\mathcal B_l$. Consequently
\begin{equation}\label{eq:trim-exact}
   \Pi_l\widehat A^{(l)}_{\mathrm{bd}}
   =\widehat A^{(l)}_{\mathrm{bd}}\Pi_l=0,
   \qquad
   \Pi_l\widehat A^{(l)}\Pi_l
   =\Pi_l\widehat A^{(l)}_{\mathrm{bc}}\Pi_l .
\end{equation}
\end{lemma}

\begin{proof}
If $I$ is a boundary-disconnected $(i,j)$-witness, Lemma~\ref{lem:dichotomy} gives disjoint support components $K_i,K_j\subseteq I$ with boundaries $\{i\}$ and $\{j\}$. Each has odd size at least three, so $|K_i|,|K_j|\le l-3$. Thus $i,j\in\mathcal B_l$, proving the claim.
\end{proof}

\begin{lemma}[Size of the singleton-boundary set]\label{lem:trim-size}
For even $l\ge6$ and $C>0$,
\begin{equation}\label{eq:trim-size}
   \E|\mathcal B_l|
   \le B\sqrt n\sum_{\substack{r\ \mathrm{odd}\\3\le r\le l-3}}C^r
   \le B\sqrt n\max\{C^3,C^{l-3}\}.
\end{equation}
Consequently, for every $\tau>0$,
\[
   \PP[|\mathcal B_l|>\tau n]
   \le\frac{B\max\{C^3,C^{l-3}\}}{\tau\sqrt n}.
\]
\end{lemma}

\begin{proof}
A union bound over the certifying index sets and Lemma~\ref{lem:boundary} with $h=1$ give
\[
 \E|\mathcal B_l|
 \le\sum_{r}n\binom mr Bn^{-(kr+1)/2}
 \le B\sqrt n\sum_r C^r.
\]
The second bound absorbs the fixed number of terms into $B$, and Markov's inequality gives the tail estimate.
\end{proof}

Let $W_{l,\mathrm{bc}}(i,j)$ count boundary-connected witnesses and set
\[
 N_{l,\mathrm{bc}}:=\E W_{l,\mathrm{bc}}(i,j),\quad
 d_{l,\mathrm{bc}}:=nN_{l,\mathrm{bc}},\quad
 \theta_{l,\mathrm{bc}}:=\rho^ld_{l,\mathrm{bc}}.
\]
Permutation symmetry gives
\begin{equation}\label{eq:bc-mean}
   \E\widehat A^{(l)}_{\mathrm{bc}}
   =\rho^lN_{l,\mathrm{bc}}(\mathbf1\mathbf1^{\top}-I).
\end{equation}
Proposition~\ref{prop:kappa-bc} shows that $N_{l,\mathrm{bc}}=(\kappa^{\mathrm{bc}}_{l,k}+o_n(1)) m^l/n^{kl/2+1}$ with $\kappa^{\mathrm{bc}}_{l,k}>0$.

\begin{theorem}[HCE weak recovery for odd $k$ after trimming]
\label{thm:recovery-odd}
Fix odd $k\ge3$, even $l\ge2$, and $\rho\in[0,1]$, and suppose $0<C\le n^{k/2-1}$ and $d_l\ge1$. Put $a=(k+1)l+3$ and $q_n=2\lceil\log n\rceil$. The conclusions below hold for all sufficiently large $n$ depending on $k,l$.
\begin{enumerate}[label=\textup{(\roman*)},nosep,leftmargin=2.0em]
\item For every even $q\ge2$,
\begin{equation}\label{eq:odd-moment}
   \mathcal M_q(R^{(l)}_{\mathrm{bc}})
   \le n^{1/q}(Bq)^a
   [d_l^{1/2}+\rho^{l-2}d_l^{1-1/l}].
\end{equation}
For every fixed $\gamma>0$, with probability at least $1-n^{-\gamma}$,
\begin{equation}\label{eq:odd-radius}
   \opnorm{R^{(l)}_{\mathrm{bc}}}+\frac{\theta_{l,\mathrm{bc}}}{n}
   \le\mathfrak R^{\mathrm{bc}}_{n,\gamma}
   :=B_\gamma(\log n)^a[d_l^{1/2}+\rho^{l-2}d_l^{1-1/l}]
     +\frac{\theta_{l,\mathrm{bc}}}{n}.
\end{equation}
\item If $l\ge6$, let $\tau\in(0,1)$; for $l\le4$ set $\tau=0$. On the event in~\eqref{eq:odd-radius} and $|\mathcal B_l|\le\tau n$, suppose
\begin{equation}\label{eq:odd-separation}
   (1-\tau)\theta_{l,\mathrm{bc}}
   >(3+\varepsilon)\mathfrak R^{\mathrm{bc}}_{n,\gamma}.
\end{equation}
Then the top eigenvalue of $\bar A^{(l)}$ is isolated. If $\hat u$ is a unit top eigenvector, then
\begin{equation}\label{eq:odd-correlation}
   \frac{|\langle\hat u,x\rangle|}{\sqrt n}
   \ge(1-\tau)^{1/2}(1-\delta^2)^{1/2},
   \qquad
   \delta:=\frac{\mathfrak R^{\mathrm{bc}}_{n,\gamma}}
   {(1-\tau)\theta_{l,\mathrm{bc}}-
      \mathfrak R^{\mathrm{bc}}_{n,\gamma}}.
\end{equation}
The rounded trimmed estimator satisfies
\begin{equation}\label{eq:odd-sign-correlation}
   \frac{|\langle\hat x,x\rangle|}{n}
   \ge1-4\delta^2-2\tau.
\end{equation}
In particular it achieves weak recovery whenever the right-hand side is positive.
\end{enumerate}
The probability of the joint event is at least
\[
   1-n^{-\gamma}-
   \begin{cases}
       0, & l\le4,\\
       \dfrac{B\max\{C^3,C^{l-3}\}}{\tau\sqrt n}, & l\ge6,
   \end{cases}
\]
with $\tau\in(0,1)$ in the second case.
\end{theorem}

\begin{proof}
Theorem~\ref{thm:connected} is valid for every $k\ge3$. Repeating the proof of Proposition~\ref{prop:app-even-recovery} for $R^{(l)}_{\mathrm{bc}}$ gives \eqref{eq:odd-moment}, and the same Markov argument gives \eqref{eq:odd-radius}.

Work in the gauged frame and write $g:=|\mathcal G_l|$ and $u_{\mathcal G}:=\mathbf1_{\mathcal G_l}/\sqrt g$. By Lemma~\ref{lem:trim-exact} and~\eqref{eq:bc-mean},
\begin{equation}\label{eq:trimmed-decomposition}
 \Pi_l\widehat A^{(l)}\Pi_l
 =\lambda_{\mathcal G}u_{\mathcal G}u_{\mathcal G}^{\top}+E_{\mathcal G},
 \quad
 \lambda_{\mathcal G}:=\frac gn\theta_{l,\mathrm{bc}},
 \quad
 E_{\mathcal G}:=\Pi_lR^{(l)}_{\mathrm{bc}}\Pi_l
 -\frac{\theta_{l,\mathrm{bc}}}{n}\Pi_l .
\end{equation}
Since $\opnorm{\Pi_l}\le1$, $\opnorm{E_{\mathcal G}}\le\mathfrak R^{\mathrm{bc}}_{n,\gamma}$ on the deviation event, while $g/n\ge1-\tau$. Weyl and Davis--Kahan therefore give the eigengap under~\eqref{eq:odd-separation} and $\sin\angle(\hat u^g,u_{\mathcal G})\le\delta$ for the gauged top eigenvector $\hat u^g=D_x\hat u$, viewed on $\operatorname{Ran}(\Pi_l)$ and extended by zero on $\mathcal B_l$. Hence
\[
 \frac{|\langle\hat u,x\rangle|}{\sqrt n}
 =\sqrt{\frac gn}\,|\langle\hat u^g,u_{\mathcal G}\rangle|,
\]
which proves~\eqref{eq:odd-correlation}.

For rounding, orient the gauged top eigenvector toward $u_{\mathcal G}$. Its squared distance from $u_{\mathcal G}$ is at most $2\delta^2$, so at most $2g\delta^2$ retained coordinates have the wrong sign. The discarded coordinates contribute at worst $-|\mathcal B_l|$ to the absolute overlap, which yields~\eqref{eq:odd-sign-correlation}. The probability statement follows from Lemma~\ref{lem:trim-size} and a union bound.
\end{proof}

\begin{corollary}[Smooth trade-off for recovery, odd $k$]
\label{cor:recovery-threshold-odd}
Fix odd $k\ge3$, even $l\ge2$, $0<\rho\le1$, $\gamma>0$, and $\varepsilon>0$. There is a constant $K^{\mathrm{odd}}_{k,\gamma,\varepsilon}>0$, independent of $l$, such that for every fixed even $l\ge2$, with $a=(k+1)l+3$, the following holds for all sufficiently large $n$ (where the lower bound on $n$ may depend on $l$). Let $C=C(n)\le n^{k/2-1}$ satisfy, if $l\ge6$, $\max\{C^3,C^{l-3}\}=o(\sqrt n)$, and suppose
\begin{equation}\label{eq:odd-threshold}
 C\ge K^{\mathrm{odd}}_{k,\gamma,\varepsilon}
 \rho^{-2}
 \frac{\kappa_{l,k}}{\kappa^{\mathrm{bc}}_{l,k}}
 \kappa_{l,k}^{-1/l}(\log n)^{(k+1)l+3}.
\end{equation}
Then, for all sufficiently large $n$, the trimmed estimator has overlap at least $1-4/(2+\varepsilon)^2-o(1)>0$ with probability $1-o(1)$.
\end{corollary}

\begin{proof}
For $l\ge6$ take $\tau_n=(\max\{C^3,C^{l-3}\}/\sqrt n)^{1/2}\to0$; for $l\le4$ take $\tau_n=0$. Proposition~\ref{prop:kappa-bc} gives $\theta_{l,\mathrm{bc}}=(\kappa^{\mathrm{bc}}_{l,k}/\kappa_{l,k}+o_n(1)) \theta_l$. Apply Corollary~\ref{cor:level-explicit-radius} to $R^{(l)}_{\mathrm{bc}}$. The two endpoint comparisons are the same as in the proof of Corollary~\ref{cor:recovery-threshold}: in the correlated endpoint the factor $\kappa^{\mathrm{bc}}_{l,k}$ cancels against the displayed ratio $\kappa_{l,k}/\kappa^{\mathrm{bc}}_{l,k}$, while the Wigner endpoint is no larger because $\kappa^{\mathrm{bc}}_{l,k}\le\kappa_{l,k}$. Thus the same finite suprema $M_1,M_2$ show that one may choose $K^{\mathrm{odd}}_{k,\gamma,\varepsilon}$ independently of $l$ so that $(1-\tau_n)\theta_{l,\mathrm{bc}}$ exceeds the level-explicit radius by a factor $3+\varepsilon$ for all sufficiently large $n$. Lemma~\ref{lem:trim-size} gives $\PP[|\mathcal B_l|>\tau_nn]=o(1)$, and the deviation event fails with probability at most $n^{-\gamma}$. Finally $\delta<1/(2+\varepsilon)$, so the proof of Theorem~\ref{thm:recovery-odd}, with this level-explicit radius, gives the stated overlap.
\end{proof}

For fixed $k$ followed by large even $l$, Proposition~\ref{prop:kappa-bc} and~\eqref{app:eq:C-ell-k-scaling} give
\[
 \frac{\kappa_{l,k}}{\kappa^{\mathrm{bc}}_{l,k}}
 \kappa_{l,k}^{-1/l}\sim\gamma_kl^{1-k/2}.
\]
Since $K^{\mathrm{odd}}_{k,\gamma,\varepsilon}$ is independent of $l$, this is the algebraic large-$l$ recovery curve without any hidden level-dependent prefactor. The lower bound on $n$ is still not uniform in growing $l$.

The side condition holds in the fixed-$\rho$ threshold regime, where $C$ is polylogarithmic. No claim is made here for arbitrary $n$-dependent $\rho$.

\paragraph{Computing $\mathcal B_l$.} The set $\mathcal B_l$ can be computed with the same meet-in-the-middle machinery used for HCE:
\begin{enumerate}[nosep,leftmargin=1.6em]
\item Fix an odd $r\in\{3,\dots,l-3\}$. Hash every $P_1\subseteq[m]$ with $|P_1|=(r+1)/2$ under the key $\partial P_1$.
\item For every $P_2\subseteq[m]$ with $|P_2|=(r-1)/2$ and every $v\in[n]$, look up the key $\partial P_2\triangle\{v\}$.
\item Declare $v\in\mathcal B_l$ as soon as a lookup succeeds.
\end{enumerate}
No disjointness check between $P_1$ and $P_2$ is needed. If $\partial P_1\triangle\partial P_2=\{v\}$, then $Q:=P_1\triangle P_2$ has $\partial Q=\{v\}$ and $3\le|Q|\le r\le l-3$; the lower bound follows because a singleton-boundary set has odd size and cannot contain one clause when $k\ge3$. Hence $v\in\mathcal B_l$. Conversely, every certificate $P$ of odd size $r$ has a disjoint split of the displayed sizes, so the procedure computes $\mathcal B_l$ exactly. Its expected cost is
\[
 \widetilde O_{k,l}(m^{(l-2)/2}+n\,m^{(l-4)/2})
\]
time, below the level-$l$ MITM cost. Trimming then only removes the corresponding rows and columns before the eigensolve; equivalently, witnesses with an endpoint in $\mathcal B_l$ need not be accumulated.

\subsection{Pure-noise refinement}
\label{app:rec-purenoise}

\begin{proposition}[Improved pure-noise bound]\label{prop:pure-noise}
If $\rho=0$ and $C\ge1$ then, for every even $q\ge2$,
\begin{equation}\label{eq:pure-noise}
   \mathcal M_q\bigl(R^{(l)}_{\mathrm{bc}}\bigr)\le n^{1/q}(Bq)^{a}C^{\,l/2},
   \qquad
   \mathcal M_q\bigl(\widetilde R^{(l)}\bigr)
   \le n^{1/q}(Bq)^{a}\sqrt n\,C^{\,l/2}.
\end{equation}
Consequently, at zero bias the full residual has high-probability upper scale $B(\log n)^{a}\sqrt{n\,d_l}$.
\end{proposition}

\begin{proof}
All proper noise projections vanish, so the first assertion is the $s=l$ case of Theorem~\ref{thm:orders}. In the full trace expansion a clause occurring an odd number of times has zero noise expectation, so the number $c$ of distinct clauses is at most $ql/2$. Exposing a general boundary-two witness along the trace with its left endpoint old, at most one outside ground variable can have odd degree, so a step with $f$ new clauses introduces at most $(kf+1)/2$ new ground variables; starting from one trace root gives $v\le1+kc/2+q/2$. By Lemma~\ref{lem:patterns} a pattern weighs at most $B^{q}n^{1+q/2}C^{c}\le B^{q}n^{1+q/2}C^{\,ql/2}$; count patterns and take $q$th roots.
\end{proof}

\section{Counting the Witnesses \texorpdfstring{$N_l$}{N ell}}
\label{app:N_ell}

It is useful to separate the purely combinatorial count from the random sampling of clauses. Fix a pair $\{i,j\}$ and let
\begin{equation}
   T_l(i,j):=\#\left\{\{D_1,\dots,D_l\}\subseteq\binom{[n]}{k}:
      D_r\text{ distinct},\ \triangle_{r=1}^{l}D_r=\{i,j\}\right\}.
\end{equation}
By symmetry this does not depend on $\{i,j\}$, so we write $T_l$. In the Poissonized inclusion model, where every $k$-set is included independently with probability $p=m/\binom{n}{k}$, the expected number of witnesses for a fixed pair is exactly
\begin{equation}\label{app:eq:N-from-T}
   N_l=T_l\,p^{l}=T_l\left(\frac{m}{\binom{n}{k}}\right)^{l}.
\end{equation}
For the i.i.d.\ clause model used in the main text, clause indices are distinct but their supports need not be. The distinct-support contribution is $(m)_lT_l/\binom nk^l$, while repeated-support tuples must be counted separately. They can contribute at the same power $n^{lk/2-1}$; their relative weight is small only in the subsequent large-$l$ comparison, not in the $n\to\infty$ limit at fixed $l$.

\subsection{The \texorpdfstring{$\F_2$}{F2} parity constraint}

Identify every support $D_r$ with its indicator vector in $\F_2^n$. The witness condition is $\sum_{r=1}^{l}\mathbf 1_{D_r}=e_i+e_j\pmod2$: every coordinate outside $\{i,j\}$ has even multiplicity, while $i$ and $j$ have odd multiplicity. Taking cardinalities modulo $2$ gives the necessary condition $2=|\{i,j\}|\equiv\sum_{r}|D_r|=l k\pmod2$. Hence $T_l=0$ unless $l k$ is even: for odd $k$ only even $l$ contributes, while for even $k$ every $l\ge2$ is admissible.

\subsection{The exact count for \texorpdfstring{$l=2$}{ell=2}}

For two supports, $D_1\triangle D_2=\{i,j\}$ if and only if the $k$-sets agree outside $\{i,j\}$ and differ on both distinguished coordinates. Thus there is a unique $(k-1)$-set $R\subseteq[n]\setminus\{i,j\}$ with $\{D_1,D_2\}=\{R\cup\{i\},R\cup\{j\}\}$, and for $n\ge k+1$,
\begin{equation}\label{app:eq:T2}
   T_2=\binom{n-2}{k-1}.
\end{equation}
Consequently, in the i.i.d.\ clause model,
\begin{equation}
   N_2=\E W_2(i,j)=\binom{m}{2}\frac{2T_2}{\binom{n}{k}^{2}}
   =\frac{(m)_2}{\binom{n}{k}^{2}}\binom{n-2}{k-1},
\end{equation}
and to leading order
\begin{equation}
   N_2=\left(1+O\!\left(\frac{k^2}{n}\right)+O\!\left(\frac1m\right)\right)
      \frac{(k!)^2}{(k-1)!}\frac{m^2}{n^{k+1}}
   =\left(1+O\!\left(\frac{k^2}{n}\right)+O\!\left(\frac1m\right)\right)
      k\,k!\,\frac{m^2}{n^{k+1}} .
\end{equation}

\subsection{The exact count for \texorpdfstring{$l=3$}{ell=3}}

The parity constraint forces $k$ even. Assume $k\ge4$ even and $n\ge3k/2+1$. Given two supports, the third is determined by
\begin{equation}\label{app:eq:third-support}
   D_3=D_1\triangle D_2\triangle\{i,j\},
\end{equation}
so an ordered triple is specified by $(D_1,D_2)$ subject to $|D_3|=k$; for $k\ge4$ the three supports are then automatically distinct. Let $s:=|D_1\setminus D_2|=|D_2\setminus D_1|$ and $t:=|\{i,j\}\cap(D_1\triangle D_2)|\in\{0,1,2\}$. Since $|D_1\triangle D_2|=2s$, \eqref{app:eq:third-support} gives $|D_3|=2s+2-2t$, so $|D_3|=k$ is equivalent to $s=s_t:=\frac k2-1+t$. Classify coordinates by membership,
\begin{equation}
   M_{\cap}=D_1\cap D_2,\quad A=D_1\setminus D_2,\quad B=D_2\setminus D_1,
   \quad O=[n]\setminus(D_1\cup D_2),
\end{equation}
with $|M_\cap|=k-s_t$, $|A|=|B|=s_t$ and $|O|=n-k-s_t$. If $\vec m=(m_M,m_A,m_B,m_O)$ records where $i,j$ fall, then $m_M+m_A+m_B+m_O=2$ and $m_A+m_B=t$. Filling the remaining slots with the other $n-2$ coordinates gives the exact count
\begin{align}\label{app:eq:T3-exact}
T_3=\frac{1}{6}\sum_{t=0}^{2}
\sum_{\substack{m_M,m_A,m_B,m_O\ge0\\ m_M+m_A+m_B+m_O=2,\ m_A+m_B=t}}
&\frac{2!}{m_M!\,m_A!\,m_B!\,m_O!}\notag\\
&\times\frac{(n-2)!}{(k-s_t-m_M)!\,(s_t-m_A)!\,(s_t-m_B)!\,(n-k-s_t-m_O)!},
\end{align}
terms with negative factorial arguments being omitted. The first factor in the summand counts the ways of choosing $\vec m$ summing to $2$; the second counts the ways of filling the remaining slots with the $n-2$ remaining variables. The factor $1/6$ converts ordered triples to unordered families. Hence $N_3=(m)_3T_3/\binom{n}{k}^{3}$ for even $k\ge4$, and $N_3=0$ for odd $k$. The leading part of~\eqref{app:eq:T3-exact} is
\begin{equation}\label{app:eq:T3-leading}
   T_3=\left(\kappa_3+O\!\left(\frac{k^2}{n}\right)\right)n^{3k/2-1},
   \qquad
   \kappa_3=\frac{1}{2(\frac k2+1)!(\frac k2-1)!^{2}}
           +\frac{1}{(\frac k2)!^{2}(\frac k2-1)!},
\end{equation}
so that $N_3=\bigl(\kappa_3+O(k^2/n)+O(1/m)\bigr)(k!)^{3}m^{3}/n^{3k/2+1}$.

\subsection{General fixed \texorpdfstring{$l$}{ell}: multigraph counting}

We now count the leading ordered i.i.d. support tuples entering $N_l$, for any fixed $l$ with $lk$ even. Support values may repeat, although the $l$ sampled clause indices are distinct. In a collapsing tuple, a generic coordinate $v\notin\{i,j\}$ must appear in an even number of positions, while $i$ and $j$ each appear an odd number. The largest power of $n$ comes from using as many distinct generic coordinates as possible: every generic coordinate appears exactly twice and each of $i,j$ exactly once. Any coordinate with multiplicity $\ge4$, or a distinguished coordinate with multiplicity $\ge3$, reduces the number of free generic coordinates and is lower order.

Label the tuple positions $1,\dots,l$. A generic coordinate appearing in supports $s$ and $t$ is an edge $\{s,t\}$ of a loopless multigraph on $[l]$; let $d_{st}\ge0$ be its multiplicity. Let $\alpha$ be the support containing $i$ and $\beta$ the support containing $j$, allowing $\alpha=\beta$. Since every support has size $k$,
\begin{equation}\label{app:eq:degree-constraint}
   \sum_{t\ne v}d_{vt}+\mathbf 1[v=\alpha]+\mathbf 1[v=\beta]=k,\qquad v=1,\dots,l,
\end{equation}
and the number of generic doubled coordinates is forced to be
\begin{equation}\label{app:eq:D-count}
   D:=\sum_{1\le s<t\le l}d_{st}=\frac{l k-2}{2}=\frac{l k}{2}-1 .
\end{equation}
For fixed $(\alpha,\beta)$ and feasible $(d_{st})$, choosing and assigning the $D$ generic coordinates contributes $(n-2)_D/\prod_{s<t}d_{st}!=n^{D}\prod_{s<t}(d_{st}!)^{-1}(1+O(D^2/n))$. Let $Q_l(i,j)$ be the number of ordered support tuples with boundary $\{i,j\}$. Summing over labeled configurations gives
\begin{equation}\label{app:eq:T-ell-general}
    Q_l(i,j)=l!\,\kappa_l\,n^{l k/2-1}
       \left(1+O_{k,l}\!\left(\frac1n\right)\right),\qquad
   \kappa_l=\frac{1}{l!}\sum_{\alpha,\beta\in[l]}
   \sum_{\substack{(d_{st})_{s<t}\in\mathbb Z_{\ge0}^{\binom{l}{2}}\\
       \text{satisfying }\eqref{app:eq:degree-constraint}}}
   \prod_{1\le s<t\le l}\frac{1}{d_{st}!} .
\end{equation}
This is valid for fixed $l$ with $l k=o(\sqrt n)$, the relative error being controlled by the birthday-type collision term $O((l k)^2/n)$. The sum does not require the incidence rows of two tuple positions to differ, so it already includes repeated support values. For example, at $l=4$ the choice $d_{12}=k$, $d_{34}=k-1$, $\alpha=3$, $\beta=4$ is precisely the leading configuration with the first two supports equal.

The checks for $l=2,3$ recover the formulas above. For $l=2$ the single edge has multiplicity $d_{12}=k-1$ and the two choices $(\alpha,\beta)=(1,2),(2,1)$ give $\kappa_2=1/(k-1)!$. For $l=3$, writing $g_v=\mathbf 1[v=\alpha]+\mathbf 1[v=\beta]$, the unique solution is $d_{12}=\frac k2-1+g_3$, $d_{13}=\frac k2-1+g_2$, $d_{23}=\frac k2-1+g_1$, which yields $\kappa_3$ in~\eqref{app:eq:T3-leading}.

\subsection{The sampled-witness constant}

For a fixed clause-index set, its ordered supports are independent uniform $k$-sets, so the unrestricted tuple count above gives
\begin{equation}
   N_l=\E W_l(i,j)
   =\binom ml\frac{Q_l(i,j)}{\binom nk^l}.
\end{equation}
Substituting~\eqref{app:eq:T-ell-general}, $\binom ml=m^l(1+O_l(1/m))/l!$, and $\binom nk=n^k(1+O_k(1/n))/k!$ yields
\begin{equation}\label{app:eq:N-ell-general}
   N_l=\kappa_{l,k}\frac{m^l}{n^{lk/2+1}}
      \left(1+O_{k,l}\!\left(\frac1n+\frac1m\right)\right),
   \qquad
   \kappa_{l,k}:=(k!)^l\kappa_l
   =\frac{(k!)^l}{l!}\sum_{\alpha,\beta\in[l]}
   \sum_{\substack{(d_{st})_{s<t}\in\mathbb Z_{\ge0}^{\binom{l}{2}}\\
       \text{satisfying }\eqref{app:eq:degree-constraint}}}
   \prod_{1\le s<t\le l}\frac{1}{d_{st}!} .
\end{equation}
Thus repeated supports enter the full i.i.d. coefficient exactly once, through the same unrestricted multigraph sum. The factor $(k!)^l$ comes from the sampling normalization $\binom{n}{k}^{-l}\sim(k!)^ln^{-kl}$.

\begin{lemma}[Finite-$n$ witness calibration]
\label{lem:finite-witness-calibration}
Fix admissible $k,l$. Uniformly over integers $m\ge l$, if $m=Cn^{k/2}$, then
\begin{equation}\label{eq:finite-witness-calibration}
 d_l=\kappa_{l,k}C^l
 \left(1+O_{k,l}\!\left(\frac1n+\frac1m\right)\right).
\end{equation}
Consequently there are $n_0(k,l)$ and $m_0(k,l)$ such that, whenever $n\ge n_0(k,l)$ and $m\ge m_0(k,l)$,
\[
 \frac12\kappa_{l,k}C^l\le d_l\le
 \frac32\kappa_{l,k}C^l.
\]
\end{lemma}

\begin{proof}
The displayed estimate for $Q_l(i,j)$ counts all leading incidence patterns. Every remaining pattern has at most $lk/2-2$ free generic ground-variable labels; there are only finitely many such patterns at fixed $k,l$, so their total is $O_{k,l}(n^{lk/2-2})$. Now use
\[
 \binom ml=\frac{m^l}{l!}\left(1+O_l(m^{-1})\right),
 \qquad
 \binom nk^{-l}=(k!)^ln^{-kl}\left(1+O_{k,l}(n^{-1})\right)
\]
in $N_l=\binom ml Q_l(i,j)/\binom nk^l$, and multiply by $n$. The two-sided bound follows after increasing $n_0(k,l)$ so that the relative error is at most $1/2$.
\end{proof}

\subsection{Asymptotics of \texorpdfstring{$\kappa_{l,k}$}{kappa ell k}}

We evaluate~\eqref{app:eq:N-ell-general} for large $l$ and fixed $k$. The multigraph sum has a generating-function form: let
\begin{equation}\label{app:eq:Phi}
   \Phi(x_1,\dots,x_l)=\exp\left(\sum_{1\le s<t\le l}x_sx_t\right),
   \qquad
   S(a):=\left[\prod_{v=1}^{l}x_v^{a_v}\right]\Phi(x_1,\dots,x_l).
\end{equation}
This coefficient is exactly $S(a)=\sum_{(d_{st}):\sum_{t\ne v}d_{vt}=a_v}\prod_{s<t}(d_{st}!)^{-1}$, because $\exp(x_sx_t)$ records how many generic variables are shared by supports $s$ and $t$. For a fixed ordered pair $(\alpha,\beta)$ the degree sequence is $a_v=k-\mathbf 1[v=\alpha]-\mathbf 1[v=\beta]$. There are two symmetry classes: if $\alpha=\beta$ there are $l$ choices and the degree sequence is $(k-2,k,\dots,k)$ up to permutation; if $\alpha\ne\beta$ there are $l(l-1)$ choices and it is $(k-1,k-1,k,\dots,k)$. Therefore
\begin{equation}\label{app:eq:C-two-terms}
   \kappa_l=\frac{S_2}{(l-1)!}+\frac{S_{11}}{(l-2)!},
   \qquad S_2=S(k-2,k,\dots,k),\quad S_{11}=S(k-1,k-1,k,\dots,k).
\end{equation}
To extract the exponential-in-$l$ scale, compare $S(a)$ with the configuration model in which loops are temporarily allowed; replacing \eqref{app:eq:Phi} by $\widetilde\Phi=\exp\bigl(\tfrac12(\sum_v x_v)^2\bigr)$ and writing $\sum_v a_v=2D$,
\begin{equation}\label{app:eq:config-coeff}
   \widetilde S(a)=\left[\prod_v x_v^{a_v}\right]\widetilde\Phi
   =\frac{(2D)!}{2^{D}D!\prod_v a_v!}=\frac{(2D-1)!!}{\prod_v a_v!} .
\end{equation}

\begin{lemma}[Loopless configuration coefficient]\label{lem:loopless-config}
Fix $k\ge3$. For either degree sequence $a=(k-2,k,\dots,k)$ or $a=(k-1,k-1,k,\dots,k)$,
\begin{equation}\label{eq:loopless-config}
   S(a)=\left(e^{-(k-1)/2}+o_l(1)\right)\widetilde S(a).
\end{equation}
In particular, $c_k\widetilde S(a)\le S(a)\le\widetilde S(a)$ for all sufficiently large $l$, for some $c_k>0$.
\end{lemma}

\begin{proof}
Attach $a_v$ labeled half-edges to each vertex $v$ and choose a uniform perfect matching of the $2D=kl-2$ half-edges. Dividing the number of all matchings by $\prod_va_v!$ gives $\widetilde S(a)$, while restricting to matchings with no pair internal to one vertex gives $S(a)$. Thus $S(a)/\widetilde S(a)=\PP[L=0]$, where $L$ is the number of internal pairs.

For every fixed $r$, the contribution to $\E(L)_r$ from choices of $r$ vertex-disjoint internal half-edge pairs is
\[
   \frac{r!\,e_r\bigl((\binom{a_v}{2})_{v\in[l]}\bigr)}
        {(2D-1)(2D-3)\cdots(2D-2r+1)},
\]
where $e_r$ is the $r$th elementary symmetric polynomial. Choices sharing a half-edge contribute $O_k(l^{r-1})/\Theta_k(l^r)=o_l(1)$. Since all but at most two degrees equal $k$, the display tends to $((k-1)/2)^r$. Hence the factorial moments of $L$ converge to those of $\operatorname{Poisson}((k-1)/2)$, so $\PP[L=0]\to e^{-(k-1)/2}$.
\end{proof}

For both degree sequences in~\eqref{app:eq:C-two-terms}, $2D=l k-2$ and
\begin{equation}
   \widetilde S_2=\frac{(2D-1)!!}{(k-2)!\,(k!)^{l-1}},
   \qquad
   \widetilde S_{11}=\frac{(2D-1)!!}{(k-1)!^{2}(k!)^{l-2}} .
\end{equation}
Their ratio is $(k-1)/k=\Theta_k(1)$, while the prefactors in~\eqref{app:eq:C-two-terms} differ by $(l-2)!/(l-1)!=1/(l-1)$. Thus the diagonal contribution is smaller by $\Theta_k(1/l)$ and
\begin{equation}\label{app:eq:Cell-asymp-step}
   \kappa_l^{1/l}\sim
   \left(\frac{(2D-1)!!}{(l-2)!\,(k-1)!^{2}(k!)^{l-2}}\right)^{1/l}.
\end{equation}
Using Stirling's formula, $((2D-1)!!)^{1/l}\sim(l k/e)^{k/2}$ and $((l-2)!)^{1/l}\sim l/e$, while the fixed factors $(k-1)!^{2}$ vanish under the $l$-th root. Hence
\begin{equation}\label{app:eq:C-scaling}
   \kappa_l^{1/l}\sim\frac{(l k/e)^{k/2}}{(l/e)\,k!}
   =\frac{k^{k/2}e^{1-k/2}}{k!}\,l^{k/2-1},
\end{equation}
and multiplying by the sampling factor gives the main-text constant
\begin{equation}\label{app:eq:C-ell-k-scaling}
   \kappa_{l,k}^{1/l}=k!\,\kappa_l^{1/l}
   \sim\frac{(l k/e)^{k/2}}{l/e}
   =\frac{e}{l}\left(\frac{kl}{e}\right)^{k/2}.
\end{equation}

\subsection{Boundary-connected witnesses for odd \texorpdfstring{$k$}{k}}
\label{app:N_ell-bc}

For a feasible configuration $(\alpha,\beta,(d_{st}))$ in \eqref{app:eq:N-ell-general}, let $G_d$ be the multigraph on $[l]$ with $d_{st}$ edges between $s$ and $t$. In a leading tuple, two support positions intersect exactly when they are adjacent in $G_d$, so the tuple is boundary-connected exactly when $\alpha$ and $\beta$ lie in the same connected component of $G_d$ (with $\alpha=\beta$ allowed). Define
\begin{equation}\label{eq:kappa-bc-definition}
 \kappa^{\mathrm{bc}}_{l,k}
 :=\frac{(k!)^l}{l!}
   \sum_{\alpha,\beta\in[l]}
   \sum_{\substack{(d_{st})\text{ satisfying }\eqref{app:eq:degree-constraint}\\
                    \alpha,\beta\text{ connected in }G_d}}
   \prod_{s<t}\frac1{d_{st}!}.
\end{equation}
This is a restriction of the same unrestricted ordered-tuple sum that defines $\kappa_{l,k}$; repeated support values are therefore included exactly once in both constants.

\begin{lemma}[Separated boundary vertices are rare]\label{lem:bd-component-tail}
Fix odd $k\ge3$ and even $l$. In the weighted multigraph sum defining $\kappa_{l,k}$, the fraction of configurations with $\alpha\ne\beta$ in different connected components is $O_k(l^{-(3k-5)/2})$.
\end{lemma}

\begin{proof}
Fix $\alpha\ne\beta$. The degree sequence is $a_\alpha=a_\beta=k-1$ and $a_v=k$ otherwise, with total degree $2D=kl-2$. Attach $a_v$ labeled half-edges to vertex $v$ and choose a uniform perfect matching. Conditional on the no-loop event $\mathsf L$, the resulting multiplicity vector has weight proportional to $\prod_{s<t}(d_{st}!)^{-1}$. Lemma~\ref{lem:loopless-config} gives $\PP(\mathsf L)\ge c_k>0$ uniformly in $l$.

If $\alpha$ and $\beta$ are disconnected, the component vertex set $V$ of $\alpha$ is closed under the pairing, contains $\alpha$ but not $\beta$, and has odd size $r$ with $3\le r\le l-3$. For fixed such $V$, put $2a_r=kr-1$. Then
\[
 \PP(V\text{ closed})
 =\frac{(2a_r-1)!!(2D-2a_r-1)!!}{(2D-1)!!}
 =\frac{\binom D{a_r}}{\binom{2D}{2a_r}}
 \le\binom D{a_r}^{-1}.
\]
There are $\binom{l-2}{r-1}$ choices for $V$. Pairing $V$ with its complement and applying a union bound gives
\begin{equation}\label{eq:bd-pairing-sum}
 \PP(\alpha\not\leftrightarrow\beta,\mathsf L)
 \le2\sum_{\substack{3\le r\le l/2\\r\ \mathrm{odd}}}
   \binom{l-2}{r-1}\binom D{a_r}^{-1}.
\end{equation}
Vandermonde gives
\[
 \binom D{a_r}
 \ge\binom{l-2}{r-1}\binom{D'}{a'_r},
 \quad D'=\frac{(k-2)l+2}{2},
 \quad a'_r=\frac{(k-2)r+1}{2}.
\]
The $r=3$ term is $O_k(l^{-(3k-5)/2})$. For $r\ge5$, convexity of $y\log(y/D')$ on $[a'_5,D'/2]$ bounds the remaining sum by $O_k(l^{1-(5k-9)/2})+l2^{-l/4}$, which is of the same or smaller order. Dividing~\eqref{eq:bd-pairing-sum} by $\PP(\mathsf L)\ge c_k$ proves the claim.
\end{proof}

\begin{proposition}[Boundary-connected witness constant]\label{prop:kappa-bc}
Fix odd $k\ge3$ and even $l\ge2$.
\begin{enumerate}[label=\textup{(\alph*)},nosep,leftmargin=2.0em]
\item
\[
 N_{l,\mathrm{bc}}
 =(\kappa^{\mathrm{bc}}_{l,k}+o_n(1))\frac{m^l}{n^{kl/2+1}},
 \qquad 0<\kappa^{\mathrm{bc}}_{l,k}\le\kappa_{l,k}.
\]
For $l\in\{2,4\}$ equality holds between the two constants.
\item As $l\to\infty$ through even values,
\[
 \frac{\kappa^{\mathrm{bc}}_{l,k}}{\kappa_{l,k}}
 =1-O_k\!\left(l^{-(3k-5)/2}\right).
\]
Consequently $(\kappa^{\mathrm{bc}}_{l,k})^{1/l}\sim\kappa_{l,k}^{1/l}$.
\end{enumerate}
\end{proposition}

\begin{proof}
Part (a) follows by repeating the leading-pattern count of \eqref{app:eq:T-ell-general} with the connectivity restriction in \eqref{eq:kappa-bc-definition}. At $l=2$, positivity follows from the exact count~\eqref{app:eq:T2}. For $l\ge4$, it follows from the feasible $\alpha=\beta$ class, in which both boundary variables lie in one support position. Equality for $l\in\{2,4\}$ follows from Proposition~\ref{prop:vanishing}.

For part (b), configurations with $\alpha=\beta$ are always boundary-connected. Lemma~\ref{lem:bd-component-tail}, applied to every $\alpha\ne\beta$ class, shows that the complementary weighted sum is at most $O_k(l^{-(3k-5)/2})\kappa_{l,k}$. This proves the ratio estimate, and the $l$th-root conclusion follows from~\eqref{app:eq:C-ell-k-scaling}.
\end{proof}

\section{Operator Comparison with Kikuchi and Square Constructions}
\label{app:operator-comparison}
This appendix shows that the HCE matrix, the Kikuchi matrix and the square construction are all restrictions of one operator on the Boolean cube. The methods differ only in whether the restriction to a slice is applied before or after the degree-$l$ polynomial is taken.

We work in the Hilbert space on the $n$-length Boolean cube $\ell_2(2^{[n]})$. Let $\mathsf{\hat{X}}_i$ be the operator that acts on the basis state corresponding to string $S$ as $\mathsf{\hat{X}}_ie_S = e_{S \triangle i}$. For a clause index $a \in [n]$ let
$\mathsf A_a = b_a\prod_{i \in S_a} \mathsf{\hat{X}}_i$, and similarly for a subset of clauses $I \subseteq [n]$ let $\mathsf A_I = \prod_{a\in I} b_a\prod_{i \in S_a} \mathsf{\hat{X}}_i$.
Then define the instance operator
\begin{equation}\label{eq:full-cube-clause-operator}
    \mathsf H :=\sum_{a=1}^m \mathsf A_a.
\end{equation}
In quantum terms, this is the Hadamard transform of the diagonal $k$XOR Hamiltonian corresponding to the instance.

Let $\mathsf \Pi_\ell$ be the orthogonal projection onto the span of Hamming weight $\ell$ basis vectors. Then the level-$\ell$ Kikuchi matrix is simply the Hamming weight $\ell$ slice
$$\mathsf K_\ell = \mathsf \Pi_\ell \mathsf H \mathsf \Pi_\ell.$$

Now define the elementary clause operator
\begin{equation}\label{eq:elementary-clause-operator}
   \mathsf E_l:=\sum_{I\in\binom{[m]}l}\mathsf A_I .
\end{equation}
The HCE side can be expressed as slices of this operator too. 
In the detection model, the discriminator can be written as
$$
 \sum_{F\in\Ccal_l}X_F
   =\langle e_\varnothing,\mathsf E_le_\varnothing\rangle,
      \label{eq:detection-compression}
$$
while the recovery matrix is
$$
 A^{(l)}
   =\operatorname{offdiag}(\mathsf \Pi_1\mathsf E_l\mathsf \Pi_1),
      \label{eq:hce-compression}
$$

The operators $\mathsf H$ and $\mathsf E_l$ can be related by a formal power series:
\begin{equation}\label{eq:full-cube-generating-function}
 \sum_{l=0}^m z^l\mathsf E_l
 =\prod_{a=1}^m(I+zA_a)
 =(1-z^2)^{m/2}
   \exp\bigl(\operatorname{artanh}(z)\,\mathsf H\bigr),
\end{equation}
On a simultaneous eigenvector of the commuting involutions $(\mathsf A_a)_{a\le m}$, suppose exactly $r$ eigenvalues equal $-1$. Then $\mathsf B$ has eigenvalue $m-2r$, while $\mathsf E_l$ has eigenvalue $\sum_j(-1)^j\binom rj\binom{m-r}{l-j}=K_l(r,m)$.
Consequently,
\begin{equation}\label{eq:full-cube-kravchuk}
   \mathsf E_l
   =K_l\!\left(\frac{mI-\mathsf H}{2},m\right),
\end{equation}
where $K_l(\cdot,m)$ is the unnormalized Kravchuk polynomial. 

Thus HCE is a matrix-valued degree-$l$ low-degree polynomial in $\mathsf H$, while Kikuchi is a constant Hamming weight slice of the same operator. HCE is not in general a power of one fixed Kikuchi slice: in $\mathsf \Pi_1K_l((mI-\mathsf H)/2,m)\mathsf \Pi_1$ the intermediate Boolean-cube boundaries are unrestricted, whereas powers of $\mathsf K_\ell = \mathsf \Pi_\ell\mathsf H\mathsf \Pi_\ell$ force every intermediate state to remain on the $\ell$-slice. For odd $k$ the balanced slice $\mathsf \Pi_\ell\mathsf H\mathsf \Pi_\ell$ vanishes because the symmetric difference of two $\ell$-sets has even cardinality, while odd-level HCE can be nonzero. 

\section{Meet-in-the-Middle Construction and Complexity}
\label{app:MITM}
A direct implementation of~\eqref{eq:Adef} enumerates all $l$-tuples of clauses and has runtime $O(m^{l})=O(n^{kl/2})$ when $m=Cn^{k/2}$. This naive cost is reduced by the standard meet-in-the-middle (MITM) strategy shown in Figure~\ref{fig:hce-schematic}. Put $r=\lfloor l/2\rfloor$ and $s=\lceil l/2\rceil$, and aggregate rather than list the hash buckets:
\[
   F_s(u):=\sum_{\substack{\Scal\in\binom{[m]}s\\
                            \triangle(\Scal)=u}}X_\Scal,
   \qquad u\in\F_2^n.
\]
The raw MITM matrix is then
\[
   G^{(l)}_{ij}:=\sum_{\Scal'\in\binom{[m]}r}
      X_{\Scal'}F_s\!\left(\triangle(\Scal')\triangle\{i,j\}\right)
   \quad(i\ne j),
\]
which takes $\widetilde O(m^s+n^2m^r)$ time if every pair $\{i,j\}$ is probed for each $r$-set. The key-scan construction works for any $l$: also build $F_r$, and write $\mathcal U_q:=\{u:F_q(u)\ne0\}$ for $q=r,s$. Each key in $\mathcal U_q$ has size at most $kq$. To find partners of the same size, for each $u\in\mathcal U_r$ and $i\in u$, $j\notin u$, probe $F_s$ at $(u\setminus\{i\})\cup\{j\}$. For partners of different sizes, delete every pair $\{i,j\}$ from each $u\in\mathcal U_r$ and probe $F_s$, and similarly delete pairs from keys in $\mathcal U_s$ and probe $F_r$. These exhaust all cross-table pairs of boundaries with $|u\triangle v|=2$. When $r=s$, the tables coincide: same-size partners are visited twice and different-size partners are visited twice by scanning both tables. When $r<s$, each cross-table match is visited once. Weighting accordingly recovers $G^{(l)}_{ij}$ exactly. There are $O_{k,l}(n)$ replacement probes per key in $\mathcal U_r$ and $O_{k,l}(1)$ deletion probes per key in either table. Thus building the tables and the raw matrix costs $\widetilde O_{k,l}(m^s+n|\mathcal U_r|+n^2)$ time and $\widetilde O_{k,l}(m^s+n^2)$ memory. For even $l=2h$ this is $\widetilde O_{k,l}(nm^h+n^2)$ time; for odd $l=2h+1$ it is $\widetilde O_{k,l}(m^{h+1}+nm^h+n^2)$ time. An ordered dictionary on the sparse keys gives these bounds deterministically (including logarithmic lookup overhead); hash lookups give the usual expected-time implementation. They hold for every input, not only random supports. Terms with $\Scal\cap\Scal'\ne\varnothing$ are removed algebraically. If $t=|\Scal\cap\Scal'|$, the common labels square to one and the remaining $l-2t$ labels form a lower-level witness. Counting the choices of the common labels and the split of the remaining labels gives the exact recurrence
\begin{equation}\label{eq:mitm-recurrence}
 G^{(l)}
 =\binom ls A^{(l)}
 +\sum_{t=1}^{r}
    \binom{m-l+2t}{t}
    \frac{(l-2t)!}{(s-t)!(r-t)!}\,A^{(l-2t)},
 \qquad A^{(0)}:=0.
\end{equation}
Here $A^{(1)}$ is defined by~\eqref{eq:Adef}: it vanishes for $k\ge3$, while for $k=2$ it records the observed clauses with support $\{i,j\}$. Computing the levels $l-2,l-4,\dots$ recursively therefore recovers the exact matrix~\eqref{eq:Adef}, without enumerating the possibly $\Theta(m^l)$ individual witnesses. The deterministic total time is $\widetilde O_{k,l}(n^3+m^s+nm^r)$ and memory $\widetilde O_{k,l}(n^2+m^s)$. For odd $l$, the $m^s$ table-construction term is not generally bounded by $nm^r$ when $m>n$. These are bounds for the stated probing procedures; the experiments of Section~\ref{sec:experiments} use the original all-pairs probe rather than the improved key scan. The resulting complexities are collected in Table~\ref{tab:complexity}.

The enumeration task underpinning HCE, given a list of $m$ vectors in $\F_2^{n}$, find size-$l$ subsets with a prescribed total XOR, is an instance of the $l$-XOR problem, a relative of generalized birthday and subset-sum problems \cite{Wagner2002}. The MITM procedure above is a standard exact generic baseline for this enumeration problem. We do not claim that its exponent is optimal for all-witness aggregation on random planted instances, where additional structure may be exploitable.

The structural difference is that HCE always performs spectral recovery on an $n\times n$ matrix and pays for clause-product enumeration, whereas Kikuchi avoids the enumeration but uses $\widetilde O((1+C)n^{\ell})$ memory for a lifted operator on a $\binom{n}{\ell}$-dimensional space. For $k=4$ and matched even levels $l=\ell$, the improved HCE bound and the stated Kikuchi bound have the same leading runtime exponent ($n^{l+1}$), not an asymptotic runtime separation. The representations nevertheless differ substantially, and this matters at the tested sizes. In our experiments, which use the original all-pairs HCE probe, HCE has substantially lower memory usage and runtime; for example at $n=50$, $k=4$, level-$6$ Kikuchi becomes computationally prohibitive (roughly $5.5$ hours per instance), whereas a level-$6$ HCE run with eight probe threads completes the densest tested instance ($C=0.25$) in roughly $300$ seconds. The two timings use different thread budgets (Appendix~\ref{app:protocol}) and are indicative rather than resource-matched.

\section{Decoding Pipelines Used in the Experiments}
\label{app:decoding}

Theorem~\ref{thm:recovery} concerns the bare spectral estimator of Figure~\ref{fig:hce-schematic}: $\hat x=\sgn(\hat u)$, where $\hat u$ is the top eigenvector of the single-level matrix $A^{(l)}$. The implementation used in Section~\ref{sec:experiments} wraps this estimator in data-dependent post-processing, and the Kikuchi baseline uses its own decoder. These steps are specified here to distinguish the practical pipelines from the estimators analyzed in Section~\ref{sec:theory}.

\paragraph{HCE.} For a target level $l$, the implementation retains the layers $A^{(2)},\dots,A^{(l)}$ produced during enumeration. The spectral step is applied to the combined matrix
\begin{equation}\label{eq:Atot}
   A_{\mathrm{tot}}=\sum_{l'=2}^{l}\hat w_{l'}A^{(l')} ,
\end{equation}
where $\hat w_{l'}\in[0,1]$ is a data-driven estimate of $\rho^{l'}$, so that each layer is weighted by an estimate of its signal strength. No knowledge of $\rho$ is used.

\emph{Layer weights.} Alongside $A^{(l')}$ the enumeration returns the witness counts $W_{l'}(i,j)$ of~\eqref{eq:condmean}; these are obtained by replacing every label $b_a$ by $1$ in the same hash computation. Write $\prod_{a\in I}b_a=x_ix_j\eta_I$ with $\eta_I:=\prod_{a\in I}\eta_a$. Then $(A^{(l')}_{ij})^2=\sum_{I,J}\eta_I\eta_J$, where the sum runs over pairs of level-$l'$ witnesses of $\{i,j\}$. The $W_{l'}(i,j)$ diagonal terms $I=J$ contribute exactly $1$ each. For $I\ne J$ the exact noise factor is $\E[\eta_I\eta_J]=\rho^{2l'-2|I\cap J|}$. The implemented weighting rule heuristically replaces every off-diagonal factor by its clause-disjoint value $\rho^{2l'}$, giving
\begin{equation}\label{eq:mom-weights}
\begin{aligned}
   \E\bigl[\|A^{(l')}\|_F^2\,\big|\,S_1,\dots,S_m\bigr]
   &\;\approx\;V_{l'}+\rho^{2l'}\bigl(Q_{l'}-V_{l'}\bigr),\\
   V_{l'}&:=\sum_{i\ne j}W_{l'}(i,j),
   &Q_{l'}&:=\sum_{i\ne j}W_{l'}(i,j)^2 .
\end{aligned}
\end{equation}
The term $V_{l'}$ is the exact conditional expectation of $\|A^{(l')}\|_F^2$ at $\rho=0$. Solving the heuristic relation \eqref{eq:mom-weights} for $\rho^{l'}$ gives
\[
   \hat w_{l'}\;=\;
   \begin{cases}
     \min\Bigl\{1,\ \bigl(\|A^{(l')}\|_F^2-V_{l'}\bigr)^{1/2}
        \big/\bigl(Q_{l'}-V_{l'}\bigr)^{1/2}\Bigr\},
       & \text{if } \|A^{(l')}\|_F^2>V_{l'} \text{ and } Q_{l'}>V_{l'},\\[3pt]
     0, & \text{otherwise.}
   \end{cases}
\]
A layer whose Frobenius mass does not exceed its noise floor is therefore discarded, as is a layer with $Q_{l'}=V_{l'}$. For odd $k$ and odd $l'$ the layer is identically zero and receives weight zero. The cap at $1$ reflects $\rho\le1$. This weighting rule is an experimental heuristic and is not used in the recovery theorem.

\emph{Candidate bitstrings.} We compute the full eigendecomposition of $A_{\mathrm{tot}}$ and retain the eigenvectors $\hat u_1,\hat u_2,\hat u_3$ of its three largest algebraic eigenvalues, together with an eigenvector $\hat u_{\min}$ of its smallest eigenvalue. Using $\sgn(0)=+1$, the candidate set consists of
\begin{enumerate}[leftmargin=1.5em,itemsep=1pt,topsep=2pt]
  \item the eight vectors $\pm\sgn(\hat u_1),\pm\sgn(\hat u_2), \pm\sgn(\hat u_3),\pm\sgn(\hat u_{\min})$;
  \item for $q\in\{2,3\}$, the rotated rounding $\sgn(\cos\theta_t\,\hat u_1+\sin\theta_t\,\hat u_q)$ with $\theta_t=\pi t/180$, $t=0,\dots,179$, that violates the fewest clauses among these $180$ angles.
\end{enumerate}
The rotations address cases where the planted direction is split across nearly degenerate top eigenvectors.

\emph{Final bitstring.} Each candidate $y\in\{\pm1\}^n$ is scored by
\[
   \mathrm{viol}(y):=
   \bigl|\{a\in[m]:\ b_a\ne\textstyle\prod_{v\in S_a}y_v\}\bigr|,
\]
and the candidate with the smallest score is returned as $\hat x$; ties are broken by the fixed order above. Selection uses only the observed clauses, never the planted $x$ or $\rho$. It is nevertheless in-sample post-processing, so the resulting curves need not equal those of the bare single-level estimator in Theorem~\ref{thm:recovery}.

\paragraph{Kikuchi.} A single level $\ell$ is used, with no multi-level combination, candidate selection or local search.

\emph{Eigenvector combination.} Let $v^{(1)},v^{(2)},v^{(3)}\in\R^{\binom{[n]}{\ell}}$ be unit eigenvectors of the three largest algebraic eigenvalues of $K$, with the signs returned by ARPACK. The implemented pipeline combines them with equal weights,
\[
   v=\frac{v^{(1)}+v^{(2)}+v^{(3)}}
           {\|v^{(1)}+v^{(2)}+v^{(3)}\|_2},
\]
using $v:=v^{(1)}$ if the sum vanishes. This implementation rule is not invariant under sign changes or basis rotations within a nearly degenerate eigenspace; the reported figures should therefore be understood as results for this specified numerical pipeline, not for a basis-invariant decoder.

\emph{Voting matrix.} In the planted model the informative Kikuchi eigenspace is correlated with $(\chi_T(x))_{T\in\binom{[n]}\ell}$, where $\chi_T(x):=\prod_{t\in T}x_t$. For an $(\ell-1)$-set $c$ and $u,u'\notin c$, the product $v_{c\cup\{u\}}v_{c\cup\{u'\}}$ is therefore a noisy estimate of $x_ux_{u'}$. Define $W\in\R^{n\times\binom{n}{\ell-1}}$ by $W_{u,c}:=v_{c\cup\{u\}}$ if $u\notin c$ and $W_{u,c}:=0$ otherwise, and set
\[
   V:=WW^\top-\diag(WW^\top),
   \qquad
   V_{uu'}=\sum_{\substack{c\in\binom{[n]}{\ell-1}\\u,u'\notin c}}
      v_{c\cup\{u\}}v_{c\cup\{u'\}}\quad(u\ne u'),
   \qquad V_{uu}=0 .
\]
The diagonal carries no pairwise sign information and is removed. The factorization avoids looping over pairs of Kikuchi coordinates and uses $O(n\binom{n}{\ell-1})$ memory for $W$.

\emph{Final bitstring.} We compute the top eigenvector $z$ of $V$ by dense \texttt{scipy.linalg.eigh} and return $\hat x=\sgn(z)$, with $\sgn(0)=+1$.

\section{Experimental Protocol}
\label{app:protocol}

\paragraph{Instances.} For each $(n,k,C,\rho)$ and seed we draw $x\in\{\pm1\}^n$ uniformly, set $m=\lfloor Cn^{k/2}\rfloor$, draw $S_1,\dots,S_m$ independently and uniformly from $\binom{[n]}k$ (the recovery model of Section~\ref{sec:theory}), and set $b_a=\eta_a\prod_{v\in S_a}x_v$ with $\PP[\eta_a=1]=(1+\rho)/2$. The same instances are used for every method.

\paragraph{Kikuchi baseline.} For even $k$ and $\ell\ge k/2$ we form $K_{T,U}=\sum_ab_a\mathbf 1[T\triangle U=S_a]$ on $\binom{[n]}{\ell}$. No degree normalization, row pruning or diagonal shift is applied; the diagonal is zero by construction. The $m\binom{k}{k/2}\binom{n-k}{\ell-k/2}$ generated entries are stored as COO records, using 32-bit row and column indices and 8-bit values, and the matrix is applied without conversion to a dense representation. Repeated clauses may generate duplicate coordinates, so this count is a count of stored records, not necessarily of distinct nonzero coordinates. The three largest algebraic eigenpairs are computed by ARPACK Lanczos (\texttt{scipy.sparse.linalg.eigsh}, \texttt{which=LA}, $\mathrm{ncv}=20$, tolerance set to machine precision). These eigenvectors are mapped to an estimate of $x$ by the voting procedure in Appendix~\ref{app:decoding}. No local search is applied.

\paragraph{HCE.} The layer matrices used by the implementation are computed exactly with hash tables of XOR boundaries and the original all-pairs probe, not the improved key scan of Appendix~\ref{app:MITM}. Products containing a repeated clause are removed algebraically; this is the implementation counterpart of the overlap recurrence~\eqref{eq:mitm-recurrence}. The resulting lower-level matrices are also retained for the multi-layer decoder of Appendix~\ref{app:decoding}. The spectral step uses dense \texttt{numpy.linalg.eigh} on an $n\times n$ matrix. No trimming is performed in the reported experiments: every odd-$k$ run has signal only at $l\in\{2,4\}$, where $\mathcal B_l=\varnothing$.

\paragraph{Resource metrics.} \emph{Runtime} is the wall-clock time of one solver call, from receiving the instance to returning $\hat x$, including matrix construction or enumeration, eigensolve and decoding, and excluding instance generation and I/O. \emph{Memory} is the recorded payload size of the largest persistent data structure: the COO records for Kikuchi and the size-$\lceil l/2\rceil$ hash table for HCE (all memory figures use $l=4$). This is a proxy for the dominant memory term of Table~\ref{tab:complexity}, not the process peak resident set size. It excludes the interpreter, transient buffers, allocator overhead and the Lanczos workspace ($\mathrm{ncv}\binom n\ell 8$ bytes for double-precision Kikuchi vectors). The proxy can therefore undercount either implementation and should be read only as a comparison of the selected persistent structures.

\paragraph{Hardware and software.} All timings were obtained on Intel CPUs with Python 3.11, NumPy 2.4.2 and SciPy 1.17.1. The Kikuchi baseline is single-process and uses no explicit parallelism; HCE uses one probe thread except for the level-$6$, $n=50$ timing quoted in Appendix~\ref{app:MITM}, which used eight probe threads. These threads independently compute the matrix element calculations for each $(i,j)$ pair.

\paragraph{Repetitions.} Heatmap cells are means over $10$ instances, and runtime curves are means over $3$ instances. No uncertainty intervals are reported, so the runtime curves are descriptive. The Kikuchi memory proxy is a deterministic function of $(n,k,m,\ell)$, namely $9m\binom{k}{k/2}\binom{n-k}{\ell-k/2}$ bytes before coalescing duplicate COO coordinates. The HCE proxy at $l=4$ is determined by $(n,k,m)$ up to coincidences of XOR keys, so additional repetitions would have little effect on the plotted memory curves.

\section{Additional Experimental Results}
\label{app:extra-experiments}

This appendix collects the recovery heatmaps for $k=5$ and $k=6$ referred to in Section~\ref{sec:experiments}. All settings (instance generation, seeds, number of repetitions) are identical to those of Figure~\ref{fig:k4-overlap}.

\begin{figure}[h]
  \centering
  \resultfig[0.92\linewidth]{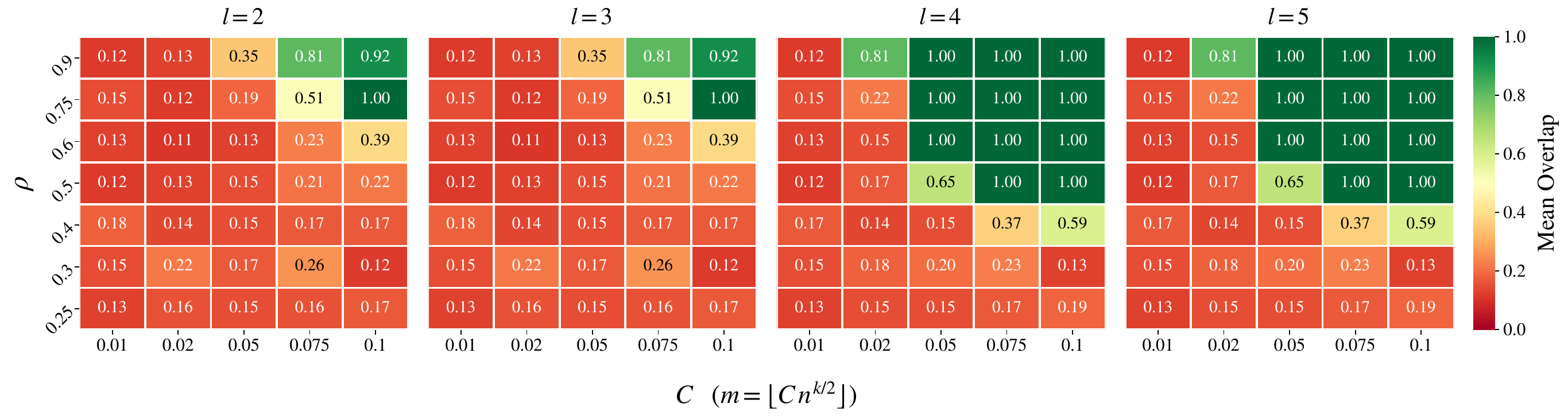}
  \caption{HCE overlap heatmap for $k=5$, $n=30$, levels $l=2,3,4,5$. Kikuchi does not directly apply to odd $k$. Odd levels carry no signal because $kl$ must be even: this can be seen by noting that the odd levels have the same recovery as the lower even levels (e.g. heatmap of level $3$ is exactly the same as that of level $2$)}
  \label{fig:k5-overlap}
\end{figure}

\begin{figure}[h]
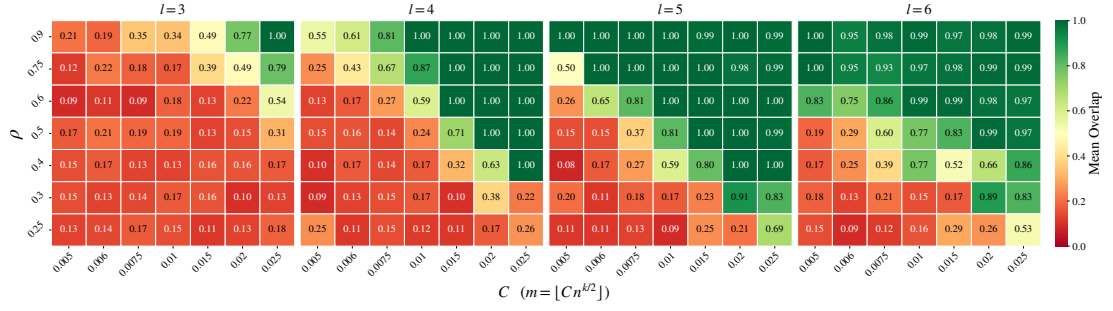
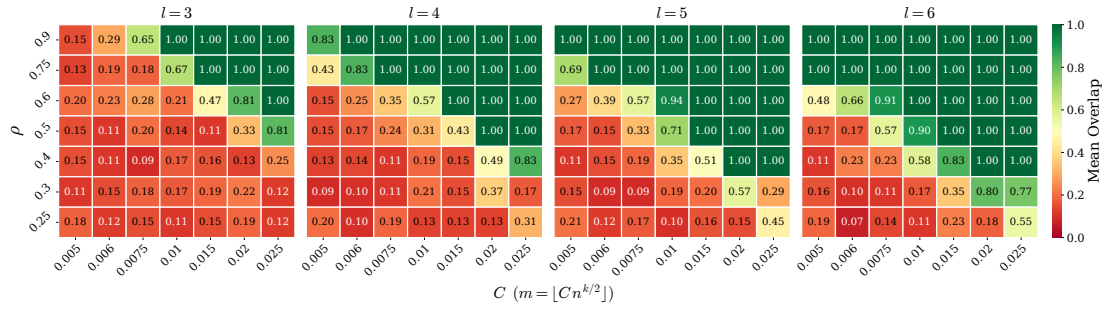

  \centering
  \begin{subfigure}{0.92\linewidth}
    \centering
    \resultfig{Kikuchi_n30_k6_heatmap_pub.pdf}
    \caption{Kikuchi spectral recovery.}
    \label{fig:k6-kikuchi}
  \end{subfigure}\\[0.6em]
  \begin{subfigure}{0.92\linewidth}
    \centering
    \resultfig{HCE_n30_k6_heatmap_pub.pdf}
    \caption{HCE recovery.}
    \label{fig:k6-hce}
  \end{subfigure}
  \caption{Recovery overlap for $n=30$, $k=6$, averaged over $10$ random instances with identical seeds across methods.}
  \label{fig:k6-overlap}
\end{figure}

Figure~\ref{fig:k5-overlap} shows HCE overlap for $k=5$ at levels $l=2,3,4,5$. Since $kl$ must be even, odd $l$ is inadmissible for odd $k$ and the corresponding recovery matrices vanish identically; the flat rows at $l=3$ and $l=5$ are exactly this parity obstruction, and they are a prediction of the theory rather than an artifact of the implementation. Symmetric Kikuchi does not apply directly to odd $k$, so no matched baseline is plotted. Figure~\ref{fig:k6-overlap} repeats the matched-level HCE/Kikuchi comparison of Figure~\ref{fig:k4-overlap} at $k=6$, where the two methods again track each other closely.

\end{document}